\documentclass[12pt,final,onecolumn]{IEEEtran}

\usepackage{latexsym}
 \usepackage{cite}    
\usepackage{graphicx}  
\usepackage[T1]{fontenc}
\usepackage[latin1]{inputenc}
\usepackage{dsfont}
\usepackage{amssymb}
\usepackage{amsmath}
\usepackage{amsfonts}
\usepackage{latexsym}
\usepackage{color}

\usepackage{algorithmic}

\begin{document}

\title{An impulsive dynamical systems framework  \\ for reset control systems}

\author{Alfonso~Ba\~nos, Juan I. Mulero, Antonio Barreiro, and Miguel A. Dav\'o
\thanks{This work has been supported by {\em FEDER- European Union} and {\em Ministerio de Ciencia e Innovaci\'on (Gobierno de Espa\~na)} under project DPI2013-47100-C2}
\thanks{A. Ba\~nos and M. A. Dav\' o, Dept. Inform\'atica y Sistemas, Univ. of Murcia, 30100 Murcia, Spain
        {\tt \small abanos@um.es}, {\tt \small mgdavo@um.es}}
\thanks{J. I. Mulero, Dept. Ing. Sistemas y Autom\'atica, Univ. Polit\'ecnica Cartagena, Cartagena, Spain
        {\tt \small juan.mulero@upct.es}}
\thanks{A. Barreiro, Dept. Ing. Sistemas y Autom\'atica, Univ. of Vigo, Pontevedra, Spain
        {\tt \small abarreiro@uvigo.es}}}

\maketitle

\begin{abstract}
Impulsive dynamical systems is a well-established area of dynamical systems theory, and it is used in this work to analyze several basic properties of reset control systems: existence and uniqueness of solutions, and continuous dependence on the initial condition (well-posedness). The work scope is about reset control systems with a linear and time-invariant base system, and a zero-crossing resetting law. A necessary and sufficient condition for existence and uniqueness of solutions, based on the  well-posedness of reset instants, is developed. As a result, it is shown that reset control systems (with strictly proper plants) do no have Zeno solutions. It is also shown that full reset and partial reset (with a special structure) always produce well-posed reset instants. Moreover, a definition of continuous dependence on the initial condition is developed, and also a sufficient condition for reset control systems to satisfy that property. Finally, this property is used to analyze sensitivity of reset control systems to sensor noise. This work also includes a number of illustrative examples motivating the key concepts and main results. 
\end{abstract}

\begin{keywords}
Reset control systems, impulsive dynamical systems, hybrid systems.
\end{keywords}

\section{INTRODUCTION}
Reset control systems trace back to the seminal work of Clegg \cite{clegg}, that introduced a nonlinear
integrator that sets its output to zero whenever its input is zero. Almost two decades later, the works by Horowitz and coworkers (\cite{horowitz-rosenbaum,krishman-horowitz}) propose design methods to incorporate a Clegg integrator (CI), and also a first order reset element (FORE), into a control loop. In the late 90s, the term {\em reset controller} is finally coined in the works by Hollot, Chait and coworkers (\cite{beker}), to describe a 'linear and time invariant system with mechanisms and laws to reset their states to zero', being the main motivation its use for overcoming fundamental limitations of linear and time invariant (LTI) control systems.

Impulsive and hybrid systems are active areas of dynamical systems theory that have been developed in the last three decades (\cite{bainov, laksh,branicky,antsaklis,lygeros,haddad,michel,goebel,lunze,yang}). Since reset controller dynamics is a combination of time and event based dynamics, it is not surprising that in the last decade different impulsive/hybrid dynamical system formulations were used for modeling and analysis of reset control systems. The survey \cite{schutter} emphasizes the diversity of hybrid systems formulations: hybrid automata, switched systems, piecewise models, complementary systems, hybrid inclusions, $\cdots$. There are two main frameworks that has been successfully used for modeling reset control systems: the framework of impulsive dynamical systems (IDS)\cite{haddad}, used in \cite{libro} and references therein; and the framework of hybrid inclusions (HI) developed in \cite{goebel}, used in \cite{steinbuch,nesic2,nesic}. Finally, another formulation of reset systems as hybrid automata has been investigated in \cite{polenkova}.

From a control practice point of view, an important issue in the different impulsive/hybrid systems formulations, directly related with their solution concept, is well-posedness. Historically, the term {\em well-posedness} comes from Hadamard \cite{hadamard}, who believed that mathematical models of physical phenomena should have these properties: i) a solution exists, ii) the solution is unique, and iii) the solution depends continuously on the initial condition (and, in general, on the  problem data). 
Regarding impulsive/hybrid systems, well-posedness (in the sense of Hadamard) has been recognized to be a very hard issue, and the term has been relaxed in different ways. In \cite{camlibel,lygeros,heemels,imura}, well-posedness is directly based on the existence and uniqueness of solutions, while in \cite{goebel} uniqueness of solutions is excluded and well-posedness is restricted to a relaxed sense of continuous dependence on the initial condition. On the other hand, although the term well-posedness is not explicitly used, existence and uniqueness of solutions, and also continuous dependence on the initial condition, are a main issue in the IDS framework \cite{bainov,laksh,haddad,michel}.   

In this work, both existence and uniqueness of solutions and continuous dependence on the initial condition will be investigated for reset control systems in the IDS framework, taking as a starting point the classical zero-crossing resetting law of Clegg and Horowitz. This formulation has been followed in several other recent works, for example \cite{bekerT}, \cite{beker} and references therein, and
also \cite{banos,banos2,libro,barreiro,carrasco,carrascoph,vidal}, including some successful experimental applications.  

In these precedent works, {\em existence and unicity of solutions} is simply assumed or is overtaken by using time regularization, that is modifying the resetting law definition by allowing reset actions to be performed only if some finite time has passed since the last reset action. However, the problem is simply avoided and thus an in-depth analysis of the
problem is missed. In addition, it is generally assumed that time regularization poses intrinsic difficulties in analysis and implementation of reset control systems \cite{nesic2,haddad2014}, and thus it is desirable to remove that restriction. To the knowledge of authors, the first work about existence and uniqueness of solutions, without including time regularization, is \cite{banosmulero}. A related recent work considers reset systems with a different resetting law, based on a reset band that includes a type of spatial regularization \cite{barreiro2}. 
 In \cite{banosmulero}, a sufficient condition is developed, given by the non-existence of after-reset states that are elements of the unobservable subspace of the base system.  This work will follow this research direction to investigate existence and uniqueness of solutions, searching necessary and sufficient conditions.  

In addition, the IDS framework will also be used for investigating {\em continuous dependence of solutions on the initial condition}, and in contrast with the HI framework, without losing uniqueness of solutions. It will be shown how for reset control systems, with a base LTI system, and with exogenous signals modeled by Bohl functions, continuous dependence on the initial condition can be characterized without introducing nondeterminism. 
In general, pointwise continuous dependence 
 on the initial condition is not a common property of impulsive and hybrid systems \cite{haddad, goebel, lygeros}. This is mainly due to the fact that for two solutions corresponding to an initial condition and some small perturbation on it, the discontinuity instants are in general different. More specifically, in the IDS framework, a quasi-continuous property has been introduced in \cite{had dad}, that is based on the continuity of the maps relating the initial condition to the resetting instants, but it has been shown that this property is not satisfied for reset control systems \cite{banosmulero,libro}. This fact makes the continuous dependence problem challenging, and in fact to the knowledge of authors it is unexplored for reset control systems in the IDS framework.  
 This has been a main motivation for this work, where a notion of continuous dependence inspired in \cite{dishliev}, that uses the Hausdorff metric, will be used for the characterization of this fundamental property. Some preliminary related recent work, developed in the HI framework  appears in \cite{copp}. Here, it should be emphasized that the well-posedness concept of the HI framework (see \cite{GSTbook}, Ch. 6, p. 126) 
is based on a relaxed sense of continuous dependence (outer semicontinuous dependence to be precise), and should not be confused with the continuous dependence concept to be developed in this work, which is a much stronger property.

Summarizing, this work will elaborate a rigorous IDS framework for reset control systems, approaching two basic problems regarding well-posedness: existence and uniqueness of solutions, and continuous dependence on the initial condition. Although it is only investigated the zero-crossing resetting law, it is believed that the different concepts and methods to be developed will provide a solid framework to analyze most of the resetting laws that has been found useful in practice. The main contributions of this work are:

\begin{itemize}
\item Existence and uniqueness of reset control systems solutions on forward time (excluding pathological behaviors like deadlock and existence of Zeno solutions), is shown to be equivalent to the well-posedness of reset instants (they are well defined and distinct). 

\item For reset systems, not necessarily reset control systems, well-posedness of reset instants is shown to be equivalent to the invariance of a subspace which is a subset of the base system unobservable subspace. 

\item For the significative class of reset compensators with full reset, 
reset control systems have always well-posed reset instants, as far as the exogenous inputs are generated by exosystems (Bohl functions). In the case of reset compensation with partial reset, reset control systems are guarantied to have well-posed reset instants only in some cases: the reset compensator has a special structure, or some zero/pole cancellations of a particular structure are present.
As a result, the time regularization restriction may be removed in the reset compensator definition, since it is not necessary for avoiding Zeno solutions and deadlock. 

\item A sufficient condition for continuous dependence on the initial condition, using a new elaborated concept based on the Hausdorff distance between trajectories. 
\item An analysis of reset control system sensitivity to sensor noise based on the developed property of  continuous dependence on the initial condition. Again, it is shown how full reset/partial reset compensators produce reset control systems that are not sensitive to sensor noise, where the noise signal is an arbitrary Bohl function.
\end{itemize}

 In Section II, besides notation, IDS and also reset control system are formally defined, the solution concept is elaborated and some basic properties are also stated. Section III is devoted to the existence and uniqueness of solutions for reset systems, based on the equivalent property of reset instants well-posedness. In addition, 
necessary and sufficient conditions are developed for well-posedness of reset instants; it is also shown how reset control systems based on full reset compensators (or partial reset with a particular structure -right reset-) always have well-posed reset instants. In Section IV, a concept of continuous dependence is developed based on the Hausdorff distance between a trajectory and a perturbed trajectory.  It is shown with some simple counterexamples that functions mapping initial conditions to reset instants have jump discontinuities, and thus previous IDS continuous dependence results are useless to approach the problem. Finally, a sufficient condition for continuous dependence of reset control systems on the initial condition is given, including several illustrative examples; moreover, for a class of sensor noise signals, modeled as Bohl functions, the characterization of sensitivity with respect to sensor noise is also investigated using the continuous dependence property.  

\section{Impulsive dynamical systems and Reset systems}

\subsection{Notation and Background}
 $\mathds{N}$ is the natural numbers set, $\mathds{R}^{+}$ is the set of nonnegative real numbers, $\mathds{C}$ is the complex numbers set, $\mathds{R}^n$ is the n-dimensional euclidean space,  and $(\mathbf{x},\mathbf{y})$, with column vectors $\mathbf{x} \in \mathds{R}^n$ and $\mathbf{y} \in \mathds{R}^m$, denotes the column
vector  $\left(
\begin{array}
[c]{c}%
\mathbf{x}\\
\mathbf{y}%
\end{array}
\right)  $.
$(t_k)_{k=1,2,\cdots}$ is a sequence of real numbers. $I_{n}$ is the $n\times n$ identity matrix, $O_{n_1 \times n_2}$  is the $n_1\times n_2$ zero matrix (if it is clear from the context subscripts are eliminated), and ${\bf 0}$ is a column vector of zeros. $\mathcal{N}(A)$, for a matrix $A \in \mathds{R}^{n\times m}$, stands for the null space of $A$. $\varnothing$ denotes the empty set, and $\setminus$ denotes sets difference; when used with a sequence, $\mathds{R}\setminus (t_k)_{k=1,2\cdots} := \mathds{R}\setminus \{t_1, t_2, \cdots\}$. For a set $\mathcal{X} \subset \mathds{R}$, $min \mathcal{X}$ is an element of $\mathcal{X}$ that is its greatest lower bound, and $min \varnothing = \infty$. For a set $\mathcal{X} \subset \mathds{R}^n$, $\bar{\mathcal{X}}$ denotes its closure. $\mathcal{S}^{n-1}$ is a $(n-1)$-sphere, $\mathcal{S}^{n-1} := \{ \mathbf{x} \in \mathds{R}^n :  \| x \|= 1 \}$; $\mathcal{HS}^{n-1}$ is a unit $(n-1)$-hemisphere, $\mathcal{HS}^{n-1} := \{(x_1,x_2, \cdots,x_n) \in  \mathds{R}^n: x_1 = \cos(\theta_1),  x_2 = \sin(\theta_1)  \cos(\theta_2), \cdots,  x_n = \sin(\theta_1) \sin(\theta_2) \cdots  \sin(\theta_{n-1});  \theta_1, \cdots, \theta_{n-2}  \in  [0,\pi],  \theta_{n-1} \in [0, \pi) \}$.   
$\leftarrow$ is an assignment. $\lor$ and $\land$ are the logical disjunction and conjunction, respectively.

A Bohl function (\cite{steinbuch,trentelman}) is defined as a linear combination of functions of the form $t^ke^{\lambda t}$, where $k$ is a nonnegative integer and $\lambda \in \mathds{C}$. Given a matrix $A \in \mathds{R}^{n\times n}$ and a linear subspace $\mathcal{V} \subset \mathds{R}^n$,  $\mathcal{V}$ is $A$-invariant (\cite{inv}) if $A\mathbf{x} \in \mathcal{V}$ for any $\mathbf{x} \in \mathcal{V}$. 
With some abuse of notation, $I:= [0,T) \subseteq \mathds{R}^+$ is an interval where the endpoint $T$ may be finite or infinite (if the interval is related with data like an initial condition ${\mathbf{x}_0}$, $I_{\mathbf{x}_0}$ is used). 
A function $f:I \rightarrow \mathds{R}^n$ is {\em left continuous with right limits} (or simply {\em left-continuous}) in $I$ if the left limit $\mathbf{x}(a^-):= \lim\limits_{t \rightarrow a, t < a} \mathbf{x}(t)$ exists and $\mathbf{x}(a) = \mathbf{x}(a^-)$ for any  $a \in I\setminus\{0\}$, and the right limit $\mathbf{x}(a^+):= \lim\limits_{t \rightarrow a, t > a} \mathbf{x}(t)$ exists for any $a \in I$.

A (linear) state-dependent impulsive dynamical system (IDS) is given by 
\begin{equation}
\left\{  \begin{aligned}
\mathbf{\dot{x}}(t) &= A \mathbf{x}(t), \hspace{1cm} & \mathbf{x}(t)  \notin \mathcal{M} \\
\mathbf{x}(t^+) & = A_R\mathbf{x}(t), \hspace{1cm} & \mathbf{x}(t) \in \mathcal{M} \\
\mathbf{x}(0) &=  \mathbf{x}_0 
\end{aligned}
\right.  \label{1}
\end{equation}
\noindent {where $\mathbf{x}(t) \in {\mathds{R}}^{n}$, $t \geq 0$, is the system state at the instant $t$, $\mathcal{M} \subset{\mathds{R}}^{n}$ is the reset set, and $A, A_R \in \mathds{R}^{n \times n}$. State-dependent IDS has been developed in the monograph \cite{haddad} and references therein, and will be used in this work as the framework to represent reset control systems and to investigate well-posedness.


The first equation in \eqref{1} will be referred to as the \textit{base system}, while the
second equation in \eqref{1} will be referred to as the \textit{resetting law}. For this IDS, there exists a unique solution $\psi(t)=e^{At}\psi_0$ of the (continuous) base system with initial condition $\psi(0) =\psi_0$ on $[0,\infty)$, for any $ \psi_{0} \in \mathds{R}^n$. 
When at some instant $t \geq 0$, referred to as {\em reset instant},  $\mathbf{x}(t) \in \mathcal{M}$ is true (a \textit{crossing} is performed) the state $\mathbf{x}(t)$ jumps to $\mathbf{x}(t^+) = A_R\mathbf{x}(t) \in \mathcal{M_{R}}$, where $\mathcal{M_{R}} := A(\mathcal{M})$ is the after-reset set. Otherwise, the state $\mathbf{x}(t)$ evolves with the base system dynamics.
Here, the term {\em crossing} is used in a relaxed sense, it has to be understood that a crossing is performed when the solution intersects the reset set. 
 
For a given initial condition $\mathbf{x}_0$, reset instants are denoted by $t_k$, $k=1, 2, \cdots$. A function $\mathbf{x}:I_{ \mathbf{x}_{0}} \rightarrow
\mathds{R}^{n}$ is a \textit{solution} of the IDS \eqref{1} on the interval $I_{ \mathbf{x}_{0}}$, with initial condition $\mathbf{x}_0$,  if (\cite{bainov,laksh,haddad})

\begin{itemize}
\item $\mathbf{x}(0) = \mathbf{x}_0$
\item $\mathbf{x}$ is differentiable and $\mathbf{\dot{x}}(t)= A\mathbf{{x}}(t)$ for any $t\in I_{ \mathbf{x}_{0}}$, $t> 0$, $t \neq t_k$, $k = 1, 2, \cdots$
\item  $\mathbf{{x}}(t)$ is left-continuous in $I_{ \mathbf{x}_{0}}$ and  $\mathbf{x}(t^+)= A_R\mathbf{x}(t)$ if $t = t_k \in I_{ \mathbf{x}_{0}}$, $k = 1, 2, \cdots$ 
\end{itemize}

Note that for a particular solution there may exist no crossings, a finite or a infinite number of crossings, and in a finite or infinite time interval $I_{ \mathbf{x}_{0}}$. And that, if a crossing is performed at the instant $a \in I_{ \mathbf{x}_{0}}$, then the solution has a jump discontinuity at that instant, that is the limits $\mathbf{x}(a^+)$ and $\mathbf{x}(a^-)$ exist, and $\mathbf{x}(a) = \mathbf{x}(a^-)$. A IDS solution is a {\em c\`agl\`ad} function (French "continue \`a gauche limite \`a droite"), not to be confused with {\em c\`adl\`ag} functions ("continue \`a droite limite \`a gauche"), mentioned for example in \cite{goebel}.

On the other hand, note that if the sets $\mathcal{M}$ and $\mathcal{M_{R}}$ are not disjoint, then for $\mathbf{x}_0 \in \mathcal{M_{R}} \cap\mathcal{M}$ an infinity number of jumps are performed without continuous evolution between them (this is usually referred to as livelock \cite{haddad}) and no solution of the IDS (1) would exist according to the above definition. Thus, the following standing assumption must be satisfied: $\mathcal{M_{R}} \cap\mathcal{M} = \varnothing$.

\subsection{Reset systems}
In this work, reset systems are defined as a particular class of  the IDS (1),  in which the resetting law is based on the crossing of an hyperplane given by $ \mathcal{H}_C :=   \{ \mathbf{x} \in \mathbb{R}^n : C\mathbf{x} = 0 \}$, that is $ \mathcal{H}_C =  \mathcal{N}(C)$, where $C$ is a row vector; and, in addition, $A_R$ is an orthogonal projector such as the last $n_r$ components of $A_R\mathbf{x}$ are zero, and the first $\bar{n}_r:=n -n_r$ components of $A_R\mathbf{x}$  remain unchanged after a crossing at $\mathbf{x} \in \mathcal{M}$. Note that the subspace $ \mathcal{H}_C $ can not be directly used as the reset set, at least the origin $\mathbf{0}$ belongs both to $ \mathcal{H}_C $ and $A_R(\mathcal{H}_C) $, and thus it always true that $\mathcal{H}_C \cap A_R(\mathcal{H}_C) \neq \varnothing$.  Thus, for building the reset set from $ \mathcal{H}_C $, it is necessary to remove from $\mathcal{H}_C $ all the fixed points of $A_R$ as a map. In the following, this set of fixed points will be denoted by ${\mathcal F}_R$.         

\vspace{0.125cm}

{\bf Definition II.1 (reset system)}:  {\em A reset system $(A,C,n_r)$, where $A\in \mathds{R}^{n \times n}$, $C\in \mathds{R}^{1 \times n}$, $n \geq 2$, and $1\leq n_{r} < n$, is an IDS as given by (1), with $A_R$ given by
\begin{equation}
A_R = \left(
\begin{array}{cc}
I_{n_{\bar{r}}} &   O_{n_{\bar{r}} \times n_{r}}
\\
O_{n_{r} \times n_{\bar{r}} } & O_{n_{r} \times n_{r} }\\
\end{array}
\right)  \label{(2.3)}
\end{equation}\

\noindent and reset set $\mathcal{M}$ given by 
\begin{equation}
\mathcal{M} = \mathcal{N}(C) \setminus\mathcal{F_{R}}  \label{(2.1b)}
\end{equation}
\noindent where $\mathcal{F_{R}} = \mathcal{N} ( \left(
\begin{array}{c}
I - A_R\\
   C         
\end{array}
\right) )$ and $n_{\bar{r}}=n-n_r$. }

\vspace{0.25cm}

Note that $n_r = n$ would be the case in which the state is fully reset to zero at a crossing; therefore, either the reset system evolves as the base system if no crossings are performed, or the system reaches the origin at the first crossing. This trivial case has been removed from the above definition, and thus no first order reset systems may exist. In addition,  (3) is consistent with \cite{beker}, that is a reset is performed at the instant $t$ if  $C\mathbf{x}(t) = 0$ and $(I-A_R)\mathbf{x}(t) \neq 0$.

It is also convenient to introduce the hyperplane $\mathcal{H}_R := \mathcal{N}(I -A_R)$. 
Note that the definition of the reset set $\mathcal{M}$, according to (3), only depends on $n_r$ and $C$. Since $A_R$ is a projector, then $ \mathcal{F}_R = \mathcal{N}(I -A_R)\cap \mathcal{N}(C) = \mathcal{H}_C \cap \mathcal{H}_R$; and thus $\mathcal{M}$ contains all the states of the hyperplane $\mathcal{H}_C$ that are not fixed points of $A_R$. Therefore, it is clear that the after-reset set $\mathcal{M}_R =  A_R(\mathcal{M})$ does not contain states of the reset set $\mathcal{M}$, and thus $\mathcal{M_{R}} \cap\mathcal{M} = \varnothing$. In addition, note that the set of fixed points $\mathcal{F}_R$ 
is  a subspace while the reset set $\mathcal{M}$ is not ($\mathbf{0} \notin \mathcal{M}$), and that ${\bar{\mathcal M}} = \mathcal{M} \cup \mathcal{F}_R = \mathcal{H}_C$. 
On the other hand, for reset systems with $C = (C_{n_{\bar r}}, O_{1 \times n_{r}})$ (for example the reset control systems in Section III.B), it results that $C(I-A_R) = O$ (that is the hyperplanes $\mathcal{H}_C$ and $\mathcal{H}_R$ are orthogonal), and thus it easily follows that $\mathcal{M}_R = \mathcal{F}_R$.
 Some examples:\\

\subsubsection{Second order reset systems (n = 2)}
note that the simplest reset system is second order; for a second order reset system $(A,C,1)$ ($n_r = 1$ is the only possible value), three subclasses are possible (assuming that $C\neq O$):   

\begin{itemize}
\item (Fig. 1.a) If $C= \left(
\begin{array}{cc}
  c_1 & 0     
\end{array}
\right)
 $ for some $c_1 \in \mathds{R}\setminus \{0\}$, then $\mathcal{H}_C = \{ ( x_1,x_2) \in \mathds{R}^2 : x_1 = 0 \}$, $\mathcal{M} = \mathcal{H}_C \setminus \{\mathbf{0}\}$, and $\mathcal{M}_R = \mathcal{F}_R = \mathcal{H}_R = \{\mathbf{0}\}$, 
 
 \item  If 
$C= \left(
\begin{array}{cc}
  0 & c_2     
\end{array}
\right)
$ for some $c_2 \in \mathds{R}\setminus \{0\}$,
then $\mathcal{H}_C = \mathcal{H}_R = \{ ( x_1,x_2) \in \mathds{R}^2 : x_2 = 0 \}$, $\mathcal{F}_R = \mathcal{H}_C$, and $\mathcal{M} =  \mathcal{M}_R = \varnothing$ (this is a trivial case, no reset action may be perfomed since $\mathcal{M} =  \varnothing$),

\item (Fig. 1.b) Otherwise, $ \mathcal{H}_C = \mathcal{N}(C)$, $\mathcal{F}_R =   \{\mathbf{0}\}$, $\mathcal{M} = \mathcal{H}_C \setminus \{\mathbf{0}\}$,  $\mathcal{H}_R = \{ ( x_1,x_2) \in \mathds{R}^2 : x_2 = 0 \}$, and $\mathcal{M}_R = \mathcal{H}_R \setminus  \{\mathbf{0}\}$. 

\end{itemize}

\begin{figure}[htbp] 
   \centering
   \begin{tabular}{c}
   	\includegraphics[width=12cm]{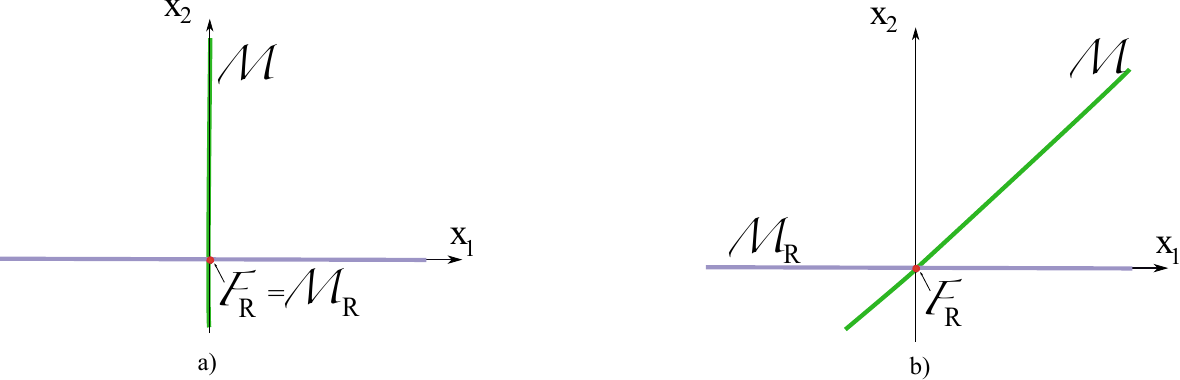} 
   \end{tabular}
   \caption{$\mathcal{M}$, $\mathcal{M}_R$, and $\mathcal{F}_R$ for second order reset systems}
   \label{fig:example31}
\end{figure}

\subsubsection{Third order reset systems (n=3)}
In this case, $n_r \in \{1,2\}$ . For $n_r = 1$, the hyperplane $\mathcal {H}_R$ is the plane $x_1-x_2$, and several subclasses are possible depending on whether the hyperplanes ${\mathcal H}_C$ and $\mathcal {H}_R$ are orthogonal or not. Assume that they are not identical (if they are identical then a trivial case with $\mathcal{M} = \varnothing$ is obtained), then $\mathcal{M}_R = \mathcal{F}_R$ if they are orthogonal (Fig. 2.b), and $\mathcal{M}_R = \mathcal {H}_R \setminus \mathcal{F}_R$, otherwise (Fig. 2.a). For $n_r = 2$ the hyperplane $\mathcal {H}_R$ is the $x_1$-axis, and again several subclasses are possible, for example Fig. 2.c shows the case corresponding to ${\mathcal H}_C\cap \mathcal {H}_R = \{\mathbf{0}\}$, and then $\mathcal{F}_R =   \{\mathbf{0}\}$, $\mathcal{M} = \mathcal{H}_C \setminus \{\mathbf{0}\}$, and $\mathcal{M}_R = \mathcal {H}_R\setminus  \{\mathbf{0}\}$. 

\begin{figure}
\centering
\begin{tabular}{c}
 \includegraphics[width=12cm]{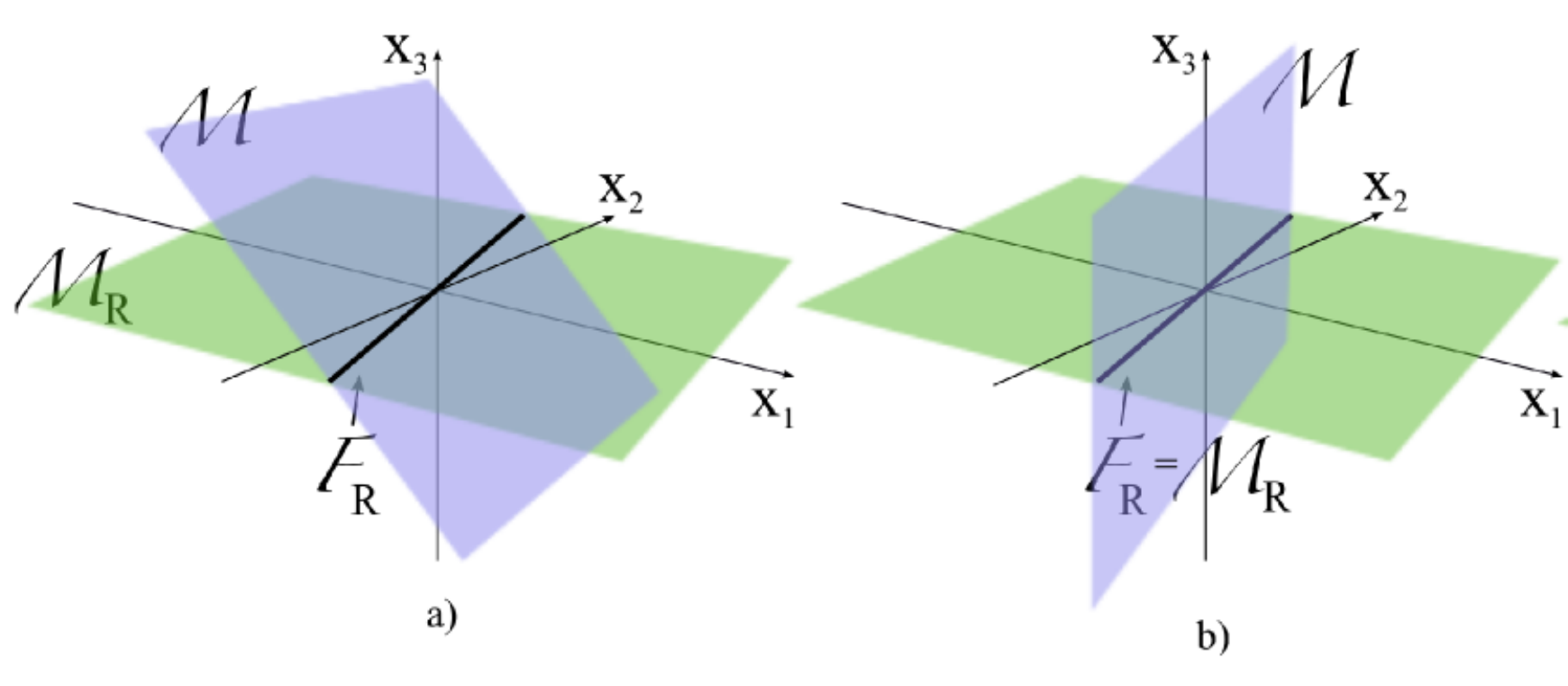}  \\ 
\includegraphics[width=6cm]{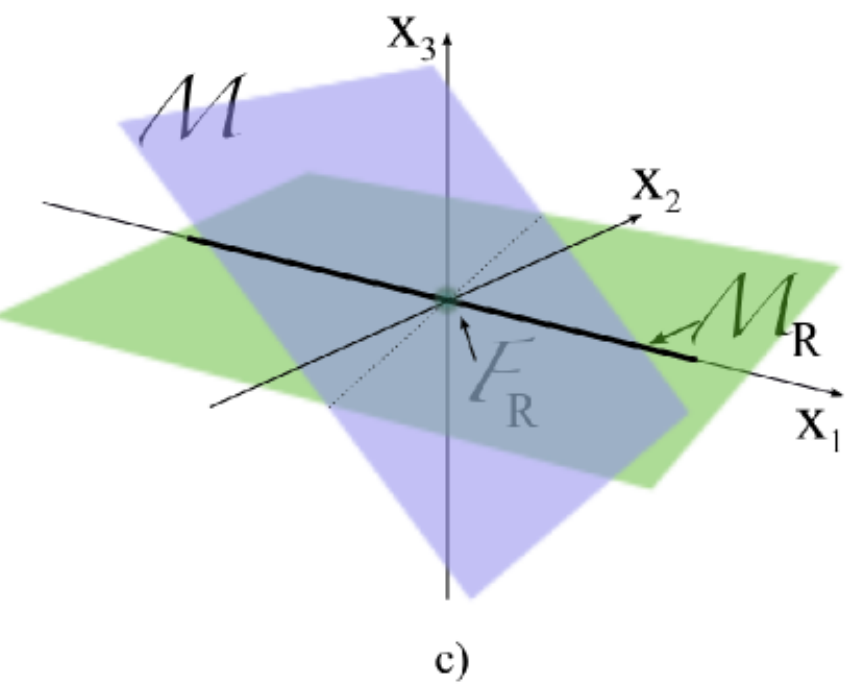}  
\end{tabular}
\caption{$\mathcal{M}$, $\mathcal{M}_R$, and $\mathcal{F}_R$ for third order reset systems}
   \label{fig:example31}
\end{figure}

\vspace{0.125cm}
\subsubsection {Reset systems with $C = (C_{n_{\bar r}}, O_{n_{r}})$} this is an important case, corresponding for example to the reset control systems to be analyzed in Section IV.  In this case, $C(I-A_R) = O$ (that is the hyperplanes $\mathcal{H}_C$ and $\mathcal{H}_R$ are orthogonal), and thus it easily follows that $\mathcal{M}_R = \mathcal{F}_R$ (see Fig. 1.a and Fig. 2.b).

\section{Existence and uniqueness of solutions}

A key topic in IDS analysis is the existence and uniqueness of solutions. In this Section, the concept  of reset instants well-posedness will be elaborated, and it will be shown to be equivalent to the existence and uniqueness of reset control systems solutions on forward time, and for any arbitrary initial condition. 

\subsection{Reset systems with well-posed reset instants}

{\bf Definition III.1 (well-posed reset instants)}: {\em A reset system $(A,C,n_r)$ has {well-posed reset instants} if for any initial condition $\mathbf{{x}}_0 \in \mathds{R}^n$ there exists a sequence $\mathds{T}:\{1,2,\cdots\} \rightarrow \mathds{R}^+ \cup \{\infty\}$, denoted by $\mathds{T}= (t_1,t_2,\cdots)$,  and given by the following procedure, being  $t_0 = 0$:

\begin{itemize}
\item 
$\small
	t_1 = \left \{ 
		\begin{array}{ccc}
		0  &,\mathbf{{x}}_{0} \in \mathcal{M}    \\
		\min \{ \Delta \in \mathds{R}^+ : e^{A\Delta}\mathbf{{x}}_{0} \in \mathcal{M} \}  &,\mathbf{{x}}_{0}  \in \mathds{R}^n\setminus\mathcal{M} 
  	\end{array} \right. \\
\nonumber
\nonumber
$

\item 
\begin{algorithmic}
\STATE $i = 1$
\WHILE {$t_i \neq \infty$}  
	\STATE 
	$\begin{array}{l}
	\mathbf{{x}}_i = A_Re^{A(t_{i}-t_{i-1})}\mathbf{{x}}_{i-1} \\
	t_{i+1} = t_{i} + \min \{ \Delta \in \mathds{R}^+ : e^{A\Delta}\mathbf{{x}}_{i} \in \mathcal{M} \}\\
	i \leftarrow i+1\\
	N = i  
	\end{array}$
\ENDWHILE
\end{algorithmic}

\end{itemize}
}

\vspace{0.125cm}

Here $\mathds{T}$ is the finite or infinite sequence of reset instants corresponding to an initial condition. Note that, for a reset system with well-posed reset instants, 
all the reset instants are distinct, since for $i \geq 1$ after-reset states $\mathbf{{x}}_i$ satisfy $\mathbf{{x}}_i = A_Re^{A(t_{i}-t_{i-1})}\mathbf{{x}}_{i-1} \notin \mathcal{M}$ ($ \mathcal{M}$ and $ \mathcal{M}_R$ are disjoint). That is, $\mathds{T}= (t_1, t_2, \cdots)$ satisfies $0  \leq t_1 < t_2 < \cdots <$ for any initial condition, the three possible cases are: i) $\mathds{T}= (\infty)$ (there is no reset actions), ii) $\mathds{T}= (t_i)_{i=1}^ N = (t_1, t_2, \cdots, t_{N-1}, \infty)$ (a finite number, $N-1$, of reset actions), and iii) $\mathds{T}= (t_i)_{i=1}^ \infty$ (an infinity number of reset actions). On the other hand, a system that does not have well-posed reset instants is said to have {\em ill-posed} reset instants. 

\vspace{0.125cm}

{\em Example III.1 (reset system with well-posed reset instants)}:  Consider a second order reset system $(A,C,1)$ with
\begin{equation}
A=
\left(
\begin{array}{cc}
  0 &  -\omega   \\
  \omega & 0   
\end{array}
\right), 
C = 
\left(
\begin{array}{cc}
  1 & -1     
\end{array}
\right)
 \label{(2.1c)}
\end{equation}
for some constant $\omega > 0$. By definition, $\mathcal{H}_R$ is the $x_1$-axis, $\mathcal{H}_C = \{ (x_1,x_2) \in \mathds{R}^2 : x_1 = x_2 \} $, $\mathcal{F}_R = \{\mathbf{0}\}$, $\mathcal{M} =\mathcal{H}_C \setminus \{\mathbf{0}\}$, and $\mathcal{M}_R =\mathcal{H}_R \setminus \{\mathbf{0}\}$ (see Fig. 1.a). It is not difficult to see that this reset system has reset instants given by\\

\begin{itemize}
\item $
	t_1 = \left \{ 
		\begin{array}{cc}
		\infty &  x_{20}= x_{10} = 0\\
		\frac{1}{\omega}(\frac{\pi}{4}- atan(\frac{x_{20}}{x_{10}}))   &{otherwise}  
	  	\end{array} \right. \\
$

\item For $i \geq 2$,  $t_i =  t_{i-1} + \frac{\pi}{4 \omega}
  	 \\
$
\end{itemize}
That is, $\mathds{T}=(\infty)$ for $\mathbf{x}_0 = \mathbf{0}$ (since $\mathbf{0} \notin \mathcal{M}$ then no crossings are performed), and $\mathds{T} = (\frac{1}{\omega}(\frac{\pi}{4}- atan(\frac{x_{20}}{x_{10}})) + (i-1)ï¿½ \frac{\pi}{4\omega})_{i=1}^{\infty}$ for any $\mathbf{x}_0 \neq \mathbf{0}$ (a periodic sequence after the second reset instant, with fundamental period $\frac{\pi}{4\omega}$). Fig. 3 shows a solution of the reset system with initial condition $\mathbf{x}_0$, and after-reset states $\mathbf{x}_1, \mathbf{x}_2,\cdots$.

\begin{figure}
\centering
\begin{tabular}{c}
\includegraphics[width=5cm]{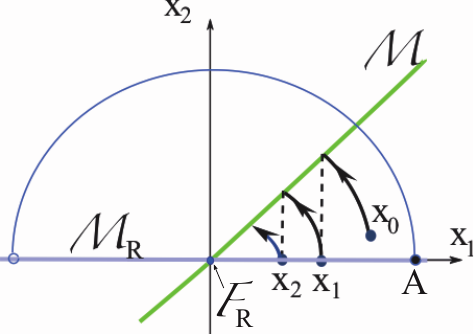}  \\ 
\includegraphics[width=10cm]{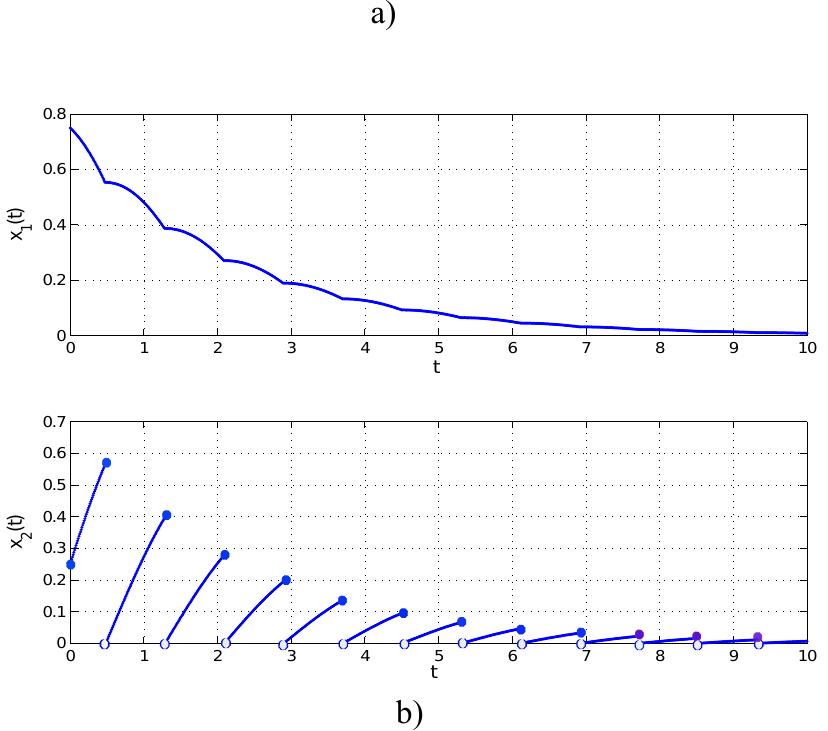}  
\end{tabular}
\caption{a) $\mathcal{M}$, $\mathcal{M}_R$, and $\mathcal{F}_R$ for Example III.1, b) Time simulation for $\omega=1$ and $\mathbf{x}_0 = (0.75,0.25)$}
   \label{fig:example31}
\end{figure}

\vspace{0.125cm}

{\em Example III.2 (reset system with ill-posed reset instants)}: 
This system is used in \cite{nesic2} for analyzing some weak points in the
definition of reset systems given in \cite{beker}. Consider a reset system
$(A,C,1)$ with%
\begin{equation}
A=\left(
\begin{array}
[c]{ccc}%
-1 & 0 & 0\\
0 & -1 & -1\\
0 & 1 & -1
\end{array}
\right)  , \hspace{0.1cm} 
C = \left(
\begin{array}
[c]{ccc}%
1 & 0 & 0
\end{array}
\right)  
\end{equation}
\noindent where $\mathcal{H}_R$ is the $x_1-x_2$ plane, $\mathcal{H}_C$ is the $x_2-x_3$ plane, $\mathcal{F_{R}} = \mathcal{M_{R}} = span\{(0,1,0)\}$ (the $x_2$-axis), and $\mathcal{M} = \mathcal{H}_C \setminus \mathcal{F}_R$  (see Fig. 2). In \cite{nesic2} it is correctly argued that for any initial condition
$\mathbf{x}_{0} = (0,a,0) \in\mathcal{M_{R}} $, the solution is
ill-defined. In fact, the problem is that the reset system has ill-posed reset instants, since for any $\mathbf{x}_{0} = (0,a,0) \in\mathcal{M_{R}} $, with $a\neq 0$,  
\begin{equation}
 \min \{ \Delta \in \mathds{R}^+ : e^{A\Delta}\mathbf{{x}}_{0} \in \mathcal{M} \} = \min \hspace{0.1cm} (0,T_1)
\nonumber
\end{equation}
\noindent does not exist since the interval $(0,T_1)$ is open and thus $t_1$ (and $\mathds{T}$) does not exist. Here $ T_1>0 $ is some instant prior to $t^* \approx 3.11$, the first non-zero instant in which  $e^{At^*}\mathbf{{x}}_{0} \in \mathcal{M}_R$. Note that the trajectory of the base system is a stable focus in the $x_2-x_3$ plane (Fig. 2).

\begin{figure}[t] 
   \centering
   \includegraphics[width=4.5cm,keepaspectratio]{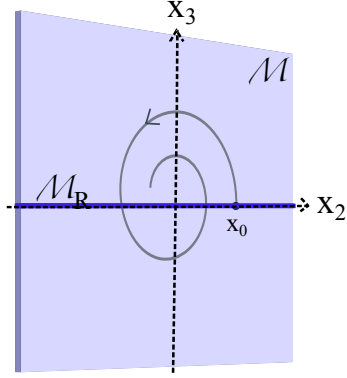}
   \caption{Reset system with ill-posed reset instants (it is shown the trajectory of the base system): 
   $\mathbf{x}(\epsilon) \in \mathcal{M}$ for any $\mathbf{x}_0 \in \mathcal{M}_R$ and $\epsilon \in (0,t^\ast)$. }
   \label{fig:illposed}
\end{figure}

{\bf Proposition III.1}: {\em The reset system $(A,C,n_r)$ has well-posed reset instants if and only if for any $\mathbf{{x}}_0 \in \mathcal{HS}^{n-1} \cap\mathcal{F}_R $ there exists a number $t_1 \in \mathds{R}^+ \cup \{\infty\} $ such that 
\begin{equation}
t_1 = \min \{ \Delta \in \mathds{R}^+: e^{A\Delta}\mathbf{{x}}_0 \in \mathcal{M} \}  
\end{equation}
}
{\bf Proof}:  Since by definition ${\mathcal F}_R = \mathcal{H}_C \cap \mathcal{H}_R$ and  
${\mathcal M} = \mathcal{H}_C\setminus {\mathcal F}_R$, then $\bar{\mathcal{M}}\setminus \mathcal{M} = {\mathcal F}_R$. 
For $ \mathbf{{x}}_{0} = \mathbf{0}$ ($\notin \mathcal{M}$), $t_1 = \min \{ \Delta \in \mathds{R}^+ : e^{A\Delta}\mathbf{{x}}_{0} \in \mathcal{M} \} = \min \varnothing = \infty$ and thus $\mathds{T} = (\infty)$. On the other hand, since $\bar{\mathcal{M}}\setminus \mathcal{M} = {\mathcal F}_R$ the minimum in (6) exists for any $\mathbf{{x}}_{0}  \in \mathds{R}^n\setminus\mathcal{M} $ if and only it exists for any $\mathbf{{x}}_{0}  \in \mathcal{F}_R\setminus\{\mathbf{0}\}$. In addition,  $e^{A\Delta}\mathbf{{x}}_{0} \in \mathcal{M}$ for some  $\Delta \in \mathds{R}^+$ if and only if $e^{A\Delta}(\alpha\mathbf{{x}}_{0}) \in \mathcal{M}$ for any $\alpha \in \mathds{R}\setminus \{0\}$. As a result,
$
\min \{ \Delta \in \mathds{R}^+ : e^{A\Delta}\mathbf{{x}}_{0} \in \mathcal{M} \}$ exists for any $ \mathbf{{x}}_{0}  \in \mathds{R}^n\setminus\mathcal{M}$ if and only if the minimum exists for any $ \mathbf{{x}}_{0}  \in \mathcal{HS}^{n-1} \cap\mathcal{F}_R$, where $\mathcal{HS}^{n-1} $ is the unit $(n-1)$-hemisphere in $\mathds{R}^n$ centered at the origin $\mathbf{{0}}$. Therefore, it is true that $t_1 \geq 0$, in fact $t_1 = 0$ if $\mathbf{{x}}_{0} \in \mathcal{M}$, and   $t_1 > 0$ otherwise.

For the following reset instants the reasoning is similar. The first after-reset state is $\mathbf{x}_1 = A_Re^{At_1}\mathbf{x}_0 \notin {\mathcal M}$ (note that $\mathbf{x}_1 \in {\mathcal M}_R$), and the second reset instant $t_2$ is given by 
$t_2 = t_1+ \min \{ \Delta \in \mathds{R}^+ : e^{A\Delta}\mathbf{{x}}_{1} \in \mathcal{M} \}$, for some $\mathbf{{x}}_{1}  \in \mathcal{M}_R 
$. For $\mathbf{{x}}_{1} = \mathbf{0}$,  $t_2 = \min \{ \Delta \in \mathds{R}^+ : e^{A\Delta}\mathbf{{x}}_{1}  \in \mathcal{M} \} = \min \varnothing = \infty$, and for $\mathbf{{x}}_{1}\in \mathcal{M}_R\setminus \{\mathbf{0}\}$ the minimum exists if and only if it exist for 
$\mathbf{{x}}_{1}\in \mathcal{HS}^{n-1} \cap\mathcal{F}_R$ by using the above argument. 
The same reasoning is again applied for the rest of the reset instants. 
$\square$ 

\vspace{0.25cm}
This Proposition reduces the dimensionality when checking whether a reset system has well-posed reset instants or not, simply by checking if a minimum exists for a reduced number of states that are elements of the hemisphere $\mathcal{HS}^{n-1} \cap\mathcal{F}_R$. This is particularly simple for low-order reset systems, as shown in the next examples. 
Moreover, for reset systems with well-posed reset instants, a function $\tau_{\mathcal{HS}}:\mathcal{HS}^{n-1} \cap \bar{\mathcal{M}}_R \rightarrow \mathds{R}^+$ is defined (note that  $\bar{\mathcal{M}}_R = \mathcal{M}_R \cup \mathcal{F}_R$):
\begin{equation}
\tau_{\mathcal{HS}}(\mathbf{x}) := \min \{ \Delta \in \mathds{R}^+ : e^{A\Delta}\mathbf{{x}} \in \mathcal{M} \} 
\end{equation}

For simplicity of notation, it is also convenient to define a function $\tau_{\mathcal{S}}:\mathcal{S}^{n-1} \cap \bar{\mathcal{M}}_R \rightarrow \mathds{R}^+$  such that
\begin{equation}
\tau_{\mathcal{S}}(\mathbf{x}):=  \left \{
\begin{array}{ccc}
 \tau_{\mathcal{HS}} (\mathbf{x})&  ,\mathbf{x} \in \mathcal{HS}^{n-1} \cap\bar{\mathcal{M}}_R  \\
  \tau_{\mathcal{HS}} (-\mathbf{x})& , \tiny{otherwise}    
\end{array}
\right.
\end{equation}

Thus, given the first reset instant $t_1$, the sequence $\mathds{T}= (t_1,t_2, \cdots)$ may be obtained as 
\begin{equation}
 \left \{
\begin{array}{c}
  t_{k+1} = t_{k} + \tau_{\mathcal{S}}(\frac{\mathbf{x}_{k}}{\| \mathbf{x}_{k} \|  } )\\
   \mathbf{x}_k = A_Re^{At_k}\mathbf{x}_{k-1}   
\end{array}  \right. 
\end{equation}
for $k = 1, 2, \cdots$. And the first reset instant $t_1$ is  

\begin{equation}
t_1 = \left \{ 
		\begin{array}{ccc}
		0  &,\mathbf{{x}}_{0} \in \mathcal{M}    \\
		\tau_{\mathcal{S}}(\frac{\mathbf{x}_{0}}{\| \mathbf{x}_{0} \|  } ) &,\mathbf{{x}}_{0} \in \bar{\mathcal{M}}_R\setminus \{\mathbf{0}\}\\
		\min \{ \Delta \in \mathds{R}^+ : e^{A\Delta}\mathbf{{x}}_{0} \in \mathcal{M} \}  &,\tiny{otherwise} 
  	\end{array} \right. 
\end{equation}
 
In addition, (7-10) may be used to define functions $\tau_i:\mathds{R}^n \rightarrow \mathds{R}$, for $i = 1, 2, \cdots, N$, such that $t_i = \tau_i(\mathbf{x}_0)$ is the $i^{th}$ element of the sequence of reset instants $\mathds{T}$. These functions take an important role in the impulsive systems literature (\cite{bainov, haddad,laksh}).

\vspace{0.125cm}

{\em Example III.3}: For the reset system of example III.1, $\mathcal{HS}^{n-1} \cap{\mathcal{F}}_R = \varnothing$, and thus it has directly well-posed reset instants. In addition, $\mathcal{HS}^{n-1} \cap\bar{\mathcal{M}}_R = \{ (1,0) \}$ (the point A in Fig. 1a), and $\tau_{\mathcal{HS}}((1,0)) = \frac{\pi}{4\omega}$, thus the reset instants may be easily computed since they are periodic with period  $\frac{\pi}{4\omega}$ after the second reset instant.  On the other hand, for the reset system of Example III.2, for $\mathbf{x}_0 \in \mathcal{HS}^{n-1} \cap{\mathcal{F}}_R = \{ (0,1,0) \}$ the first reset instant $t_1 = \min (0,T_1)$ is not well defined since $(0,T_1)$ is open ant thus the minimum does not exist; therefore, the reset system has not well-posed reset instants.

\vspace{0.125cm}

{\em Example III.4}:  Consider a reset system $(A,C,1)$ with 
\begin{equation}
A =\left(
\begin{array}
[c]{cccc}%
0 & 0 &  -0.35 &3\\
1 & 0 & -2.40 & 1\\
0 & 1 & -4.35 &0\\
0& 0  & -1 & -1
\end{array}
\right),   
C =\left(
\begin{array}
[c]{cccc}%
0 & 0 & 1 & 0
\end{array}
\right)
\end{equation}

\noindent In this case, the set $ \mathcal{S}^{n-1} \cap \bar{\mathcal{M}}_R = \mathcal{S}^{n-1} \cap {\mathcal{F}}_R $ is the unit circumference centered at the origin of the plane $x_1-x_2$. It may be parameterized by 
\begin{equation}
\mathbf{x}(\theta) = \left(
 \cos (\theta),
 \sin (\theta), 
0,
0
\right) 
\end{equation}
\noindent for $\theta \in [0,2\pi)$ and $\tau_{\mathcal{S}}$ may be computed as a function of $\theta$, that is $\tau_{\mathcal{S}}(\theta) := \tau_{\mathcal{S}}(\mathbf{x}(\theta))$ (see Fig. 3), by using (7)-(8) (note that $\tau_{\mathcal{S}}(\theta + \pi) = \tau_{\mathcal{S}}(\theta)$ for $\theta \in [0, \pi)$). This is equivalent to solve for $t$ the implicit equation
\begin{equation}
Ce^{At}\left(
 \cos (\theta),
 \sin (\theta), 
0,
0
\right) = 0 
\end{equation}

\noindent The result is that the domain of the map $\tau_{\mathcal{HS}}$ is $\mathcal{HS}^{n-1} \cap {\mathcal{F}}_R$ (and thus the domain of  $\tau_{\mathcal{S}}$ is $\mathcal{S}^{n-1} \cap {\mathcal{F}}_R$), and then by Prop. III.1 
the reset system has well-posed reset instants. On the other hand,  $\tau_{\mathcal{S}}$ has a discontinuity at $\theta = \pi$. As a result, functions $\tau_i$, $i=1,\cdots,N$  are also discontinuous. It is worthwhile to mention that the continuity of these functions is a common assumption in most of the work done about IDS \cite{haddad}, and thus it is not directly applicable to reset systems. 
\begin{figure}[htbp]
\centering
\includegraphics[width=12cm,keepaspectratio]{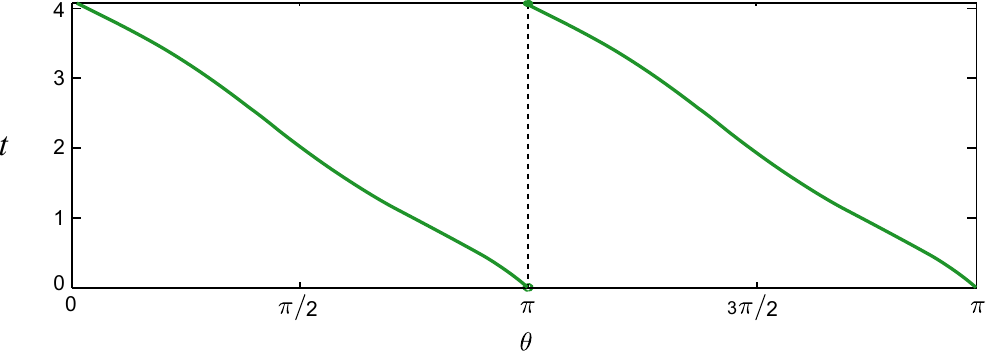}
\caption{Reset instants corresponding to states in $\mathcal{S}^{n-1} \cap\mathcal{M}_R$ as a function of the parameter $\theta$, $t = \tau_S(\theta)$ (note that $\tau_S(\pi^-) = 0$, and $\tau_S(\pi) = \tau_S(\pi^+) \approx 4.13$).}
\label{2:f_dim_2_ex_1}
\end{figure}
In the following, a geometric condition based on the system matrices $A$, $A_R$, and $C$ will be developed for the reset system  $(A,C,n_r)$ to have well-posed reset instants. 
\vspace{0.25cm}

{\bf Proposition III.2}: {\em The reset system $(A,C,n_r)$ has well-posed reset instants if and only if the subspace 
\begin{equation}
\mathcal{F}_{RU} := {\cal N}(
\left(
\begin{array}{c}
  I - A_R   \\
  {\cal O}_{base}   
\end{array}
\right)
 )
\end{equation}
 is $A$-invariant, where $\mathcal{O}_{base}$ is the observability matrix of the base system. }

\vspace{0.125cm}

{\bf Proof}: First note that  $\mathcal{F}_{RU} = \mathcal{F}_R\cap\mathcal{N}( {\cal O}_{base})$ and thus $\mathcal{F}_{RU}$ is the subspace of fixed points of $A_R$ that are unobservable; on the other hand, for a given $\mathbf{x}_0 \in \mathds{R}^n$, the function $f:\mathds{R} \rightarrow \mathds{R}$ defined as $f(\Delta) = Ce^{A\Delta}\mathbf{x}_0$ is an analytical function (it is a Bohl function). A property of $f$ to be used below is that either $f(\Delta) = 0$ for any $\Delta \in \mathds{R}$ or $f$ has isolated zeros.

By Prop. III.1, that $(A,C,n_r)$ has well-posed reset instants is equivalent to the existence of $\min \{ \Delta \in \mathds{R}^+ : e^{A\Delta}\mathbf{{x}}_0 \in \mathcal{M} \}  $ for any $\mathbf{{x}}_0 \in \mathcal{HS}^{n-1} \cap {\mathcal{F}}_R $, or equivalently for 
any $\mathbf{{x}}_0 \in \mathcal{F}_R$. 

\vspace{0.15cm}
{\em if}) For $\mathbf{{x}}_0 \in \mathcal{F}_R$, either $f(\Delta) = Ce^{A\Delta}\mathbf{x}_0= 0$ for any $\Delta \in \mathds{R}$ or $f$ has isolated zeros. If $f(\Delta) = 0$ for any $\Delta \in \mathds{R}$, then $\mathbf{{x}}_0 \in \mathcal{F}_R\cap\mathcal{N}( {\cal O}_{base}) = \mathcal{F}_{RU} $, and since $\mathcal{F}_{RU} $ is $A$-invariant then $e^{A\Delta}\mathbf{x}_0 \notin \mathcal{M}$. Thus,  
$\min \{ \Delta \in \mathds{R}^+ : e^{A\Delta}\mathbf{{x}}_0 \in \mathcal{M} \} = \min \varnothing = + \infty $. On the other hand, if $f$ has isolated zeros then $Ce^{A\Delta}\mathbf{x}_0\neq 0$ and thus $e^{A\Delta}\mathbf{x}_0 \notin \mathcal{M}$ for $\Delta \in (0,\epsilon)$ and some constant $\epsilon >0$. The result is that $\min \{ \Delta \in \mathds{R}^+ : e^{A\Delta}\mathbf{{x}}_0 \in \mathcal{M} \}  $  does also exist in this case.

\vspace{0.125cm}
{\em only if}) By contradiction, if $\mathbf{{x}}_0 \in \mathcal{F}_{RU}$  and $\mathcal{F}_{RU} $ is not $A$-invariant then $f(\Delta) = Ce^{A\Delta}\mathbf{x}_0 = 0$ for any $\Delta \in \mathds{R}$,  and $e^{A\Delta}\mathbf{x}_0 \in \mathcal{N}(\mathcal{O}_{base})\setminus \mathcal{F}_{RU} \subset \mathcal{M} $, for $\Delta \in (0, \epsilon)$ and some constant $\epsilon >0$. Thus $\min \{ \Delta \in \mathds{R}^+ : e^{A\Delta}\mathbf{{x}}_0 \in \mathcal{M} \}$ does not exist for $\mathbf{{x}}_0 \in \mathcal{F}_{RU}$, wich is a contradiction. 
$\Box$

\vspace{0.125cm}
{\em Example III.5: } For the reset system of Example III.1, the after reset set is $\mathcal{ F}_R=\{ \mathbf{0} \}$. In this case, $\mathcal{ F}_{RU}=\{ \mathbf{0} \}$ is trivially $A$-invariant, and thus the reset system has well-posed reset instants according to Prop. III.2. In the case of Example III.2, the after reset set $\mathcal{ F}_R $ is the $x_2$-axis and the unobservable subspace of the base system is $\mathcal{N}( {\cal O}_{base}) = span\{(0,0,1),(0,1,0)\}$ (the $x_2-x_3$ plane); in addition, $\mathcal{ F}_{RU}$  is the $x_2$-axis and $A(\mathcal{ F}_{RU})= span\{(0,-1,1)\}
\nsubseteq \mathcal{ F}_{RU} $; as a result, $\mathcal{F}_{RU}$ is not $A$-invariant and thus the reset system has ill-posed reset instants.


By definition,  a \textit{Zeno solution} $\mathbf{x}$ of the IDS (1) exists for some initial condition $\mathbf{x}_{0} \in \mathds{R}^n$ if there exists an infinite sequence of reset instants $\mathds{T} =( t_k)_{k = 1}^\infty$, and $t_\infty \in \mathds{R}^+$ such as  $t_k \rightarrow t_\infty$ as $k \rightarrow \infty$. Note that for a reset system with well-posed reset instants, a solution $\mathbf{x}$ exists on $I_{\mathbf{x}_{0}}$ for any $\mathbf{x}_{0} \in \mathds{R}^n$, where   $I_{\mathbf{x}_{0}} = [0,t_\infty)$ if it is a Zeno solution, and $I_{\mathbf{x}_{0}} =  [0,\infty)$ otherwise. A simple counterexample (\cite{joaquintesis, libro}) shows that in general Zeno solutions may exist for reset systems with well-posed reset instants.

\subsection{Reset control systems}
In this work, a reset control system (Fig. \ref{controlsetup}) refers to a feedback interconnection of a LTI system $P$ 
and a reset compensator $R$ with base system $R_{base}$. 
$P$ is described by:%
\begin{equation}
 \label{(2.1)}
\ \left\{
\begin{array}{l}%
\mathbf{ \dot{x}}_{p}(t) = A_{p} \mathbf{ x}_p(t) + B_{p} u(t)\\
y(t) = C_{p} \mathbf{x}_{p}(t)%
\end{array}
\right. 
\end{equation}
with $\mathbf{x}_{p}(0) =\mathbf{x}_{p0}$, and the reset compensator $R$ is given by
\begin{equation}\small
 \ \left\{
\begin{array}{ll}%
\mathbf{\dot{x}}_{r}(t) = A_{r} \mathbf{x}_{r}(t) + B_{r} e(t),&  \hspace{2mm} e(t) \neq0 \lor (I-A_\rho)\mathbf{x}_r(t)=\mathbf{0} \\
\mathbf{{x}}_{r}(t^{+}) = A_{\rho}\mathbf{x}_{r}(t), &\hspace{2mm} e(t) = 0 \land (I-A_\rho)\mathbf{x}_r(t)\neq \mathbf{0} \\
v(t)  =  C_{r} \mathbf{x}_{r}(t) + D_r e(t) 
\end{array}
\right.  \label{(2.2)}
\end{equation}
\noindent with $\mathbf{x}_{r}(0) = \mathbf{x}_{r0}$.
Here $\mathbf{x}_p \in \mathds{R}^{n_p}$, $\mathbf{x}_r \in \mathds{R}^{n_r}$, and $e,u \in \mathds{R}$. 
It is assumed that the
last $n_{\rho}$ compensator states are set to zero at the reset instants, then
$A_{\rho}$ is partitioned in blocks as%
\begin{equation}
A_{\rho}= \left(
\begin{array}{cc}
I_{n_{\bar{\rho}} } & O_{n_{\bar{\rho}} \times {n_{\rho}}%
}\\
O_{n_{\rho} \times n_{\bar{\rho}} } & O_{n_{\rho} \times n_{\rho} }\\
\end{array}
\right)  \label{(2.3)}
\end{equation}
\noindent where $n_{\bar{\rho}}=n_r-n_\rho$, and in addition $A_r$, $B_r$, and $C_r$ are partitioned into blocks with appropriate block dimensions:
\begin{equation}
   A_r =  \left (
   \begin{array}{cc}
      A_{r_{11}} &\hspace{-0.12cm} A_{r_{12}}\\
      A_{{r_{21}}} & \hspace{-0.12cm}A_{r_{22}}
   \end{array}
   \right), 
   B_r =  \left (
   \begin{array}{c}
      B_{r_{1}}\\
      B_{r_{2}}
   \end{array}
   \right), 
   C_r =  \left (
   \begin{array}{cc}
      C_{r_{1}} & \hspace{-0.12cm}  C_{r_{2}}
   \end{array}
   \right) 
\end{equation}
\noindent In the case of a {\em full reset} compensator, all the elements of $A_{\rho}$ are 0; otherwise, $R$ is a {\em partial reset} compensator.
\begin{figure}[t] 
   \centering
   \includegraphics[width=8cm,keepaspectratio]{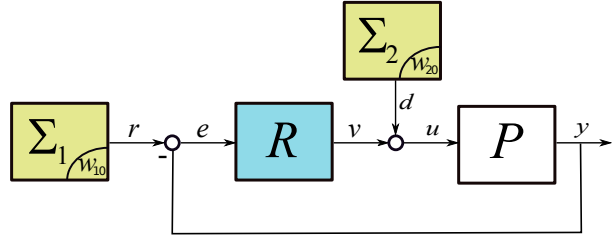}
   \caption{Reset control system (exosystems have not inputs, their initial conditions are explicitly shown in the bottom right corner).}
   \label{controlsetup}
\end{figure}

As it is usual in control practice, reset control systems are driven by external or exogenous inputs such as reference or disturbance signals (note that output measurement noise may be included in the reference signal for analysis of existence and uniqueness of solutions). It will be assumed that the reference input $r$ and the disturbance input $d$ are generated by exosystems $\Sigma_1$ and $\Sigma_2$ respectively, with state space models 
\begin{equation}
  \Sigma_1: \left \{
   \begin{array}{llll}
     \mathbf{\dot{w}}_1(t) &= A_{1}\mathbf{w}_1(t), \hspace{0.5cm} \mathbf{w}_1(0)=\mathbf{w}_{10} \\
     r(t) & = C_{1}\mathbf{w}_1(t), \hspace{0.5cm}  \\
    \end{array}
   \right.
    \label{(5.1)}
\end{equation}
\noindent with $\mathbf{w}_1 \in \mathbb{R}^{m_1}$, and
\begin{equation}
  \Sigma_2: \left \{
   \begin{array}{llll}
     \mathbf{\dot{w}}_2(t) &= A_{2}\mathbf{w}_2(t), \hspace{0.5cm} \mathbf{w}_2(0)=\mathbf{w}_{20} \\
    d(t) & = C_{2}\mathbf{w}_2(t), \hspace{0.5cm}  \\
    \end{array}
   \right.
    \label{(5.2)}
\end{equation}
\noindent with $\mathbf{w}_2 \in \mathbb{R}^{m_2}$. These exosystems allow to generate signals like steps, ramps, sinusoids, etc. (Bohl functions). Now,  the feedback connection is obtained by making $e = r -y$ and $u = v + d$. Define the closed-loop state as 
$\mathbf{x} = 
\left(\mathbf{w}_1, \mathbf{w}_2, \mathbf{x}_p,\mathbf{x}_r \right)$,  then the reset control system of Fig. 4 can be represented as a reset system $(A,C,n_\rho)$ with 
\begin{equation}
\begin{array}{l}
 {A}=\left(
\begin{array}[c]{cccc}%
A_1 & O & O & O\\
O & A_2 & O & O\\
B_pD_rC_1 & B_pC_2 & A_p - B_pD_rC_p & B_pC_r\\
B_rC_1 & O & -B_rC_p & A_r  %
\end{array}
\right) \\
\\
{C} =  \left( \begin{array}
[c]{cccc}%
C_1  & O & -C_p & O
\end{array} \right)
 \end{array} \label{A-C}
 \end{equation}

Note that, according to Def. II.1, the  reset set is $\mathcal{M}=\mathcal{N}(C)\setminus \mathcal{F}_R$, and  thus $R$ performs reset actions at the instant $t$ when $e(t) = C\mathbf{x}(t) = 0$ only if $\mathbf{x}(t) \notin  \mathcal{F}_R$, that is when  $e(t) = 0$ and $(I-A_\rho)\mathbf{x}_r(t) \neq \mathbf{0}$. 
 In addition, since the last $n_\rho$ values of $C$ are zero, then $\mathcal{M}_R = \mathcal{F}_R$, and therefore  $\mathcal{M}_{RU} := \mathcal{M}_R \cap \mathcal{N}(\mathcal{O}_{base}) = \mathcal{F}_{RU}$. 


For a block diagram representation of the reset compensator $R$ as given by (16)-(18), it is sufficient to employ an extension of the Clegg integrator as shown in Fig. \ref{CI}, in which $s:\mathds{R}^+ \rightarrow \{0,1\}$ is a boolean-valued trigger function that takes values $0$ (false) and $1$ (true): 

\begin{equation}
   CI: \left \{
   \begin{array}{llll}
     \dot{v}(t) = e(t), \hspace{1cm} s(t) = 0\\
           v(t^+) = 0, \hspace{1.15cm} s(t) = 1 
    \end{array}
   \right.
  \label{(5.10)}
\end{equation}

\begin{figure}[t]
\centerline{\includegraphics[width=9cm,keepaspectratio]{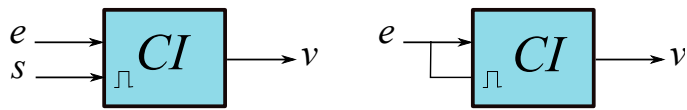}}
\caption{({\em left}) Two-inputs Clegg integrator, ({\em right}) Clegg integrator (by simplicity, the connection of the signal $e$ to the trigger input means that $s$ is the bolean expression $e = 0 \land v \neq 0$).} 
\label{CI}
\end{figure}

A block diagram of $R$ that allows a direct practical implementation is given in Fig. \ref{bloques}. 
On the other hand, if $A_{r_{21}} = O$ ($A_{r_{12}} = O$) then $R$ will be referred to as a {\em right reset compensator} ({\em left reset compensator}); 
the name is related with the right (left) triangular block structure of the matrix $A_r$. It is worthwhile to mention that some of  the reset compensator with partial reset (see Fig. \ref{RRC}) that has been found useful in practice \cite{libro} are right reset compensators.

In the following, a necessary and sufficient condition for existence and uniqueness of solutions will be developed; and, in addition, this result will be applied to reset control systems with a full/partial reset compensation structure. 

\begin{figure}[t]
\centerline{\includegraphics[width=10cm,keepaspectratio]{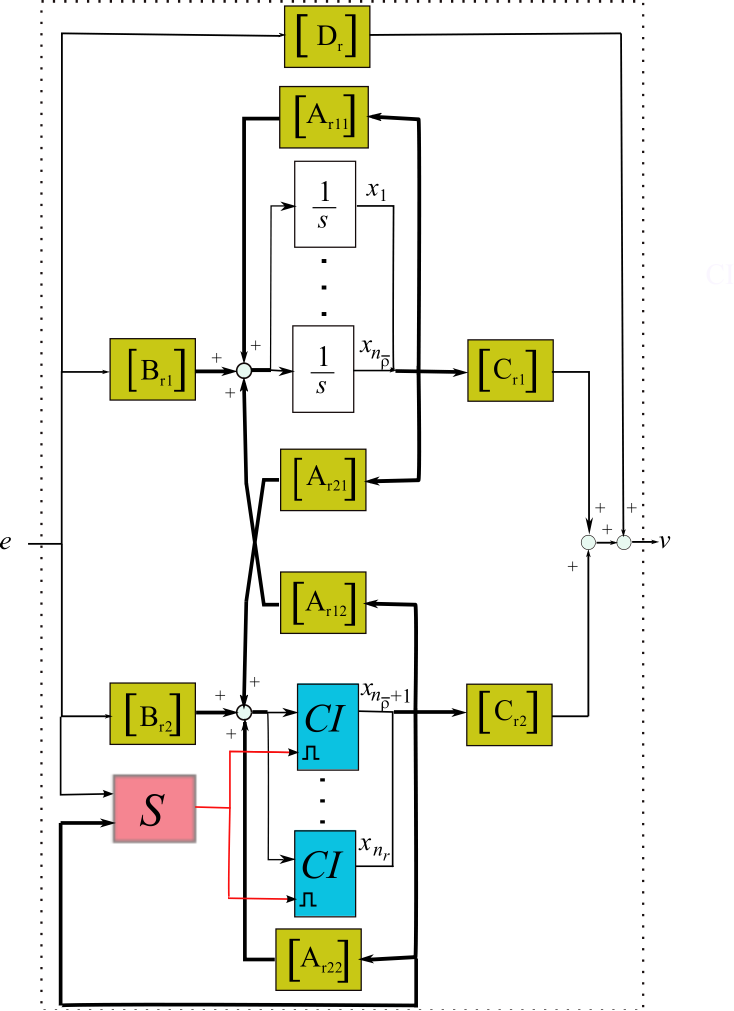}}

\caption{Blocks diagram of the reset compensator $R$ (for a block $[ H ]$ the output vector is the matrix multiplication of $H$ with the input vector); the compensator state is ${\mathbf x}_r =  (x_1, \cdots, x_{n_{\bar \rho}},x_{{n_{\bar \rho}}+1},\cdots,x_{n_r})$, and the block $S$ produces the boolean-valued function $S(e,x_{{n_{\bar \rho}}+1},\cdots,x_{n_r}) (t) = 0$ if $e(t) \neq 0 \lor (x_{{n_{\bar \rho}}+1},\cdots,x_{n_r})(t) = (0,\cdots,0)$, and $S(e,x_{{n_{\bar \rho}}+1},\cdots,x_{n_r}) (t) = 1$ otherwise.
}
\label{bloques}
\end{figure}

\begin{figure}[t]
\centerline{\includegraphics[width=8cm,keepaspectratio]{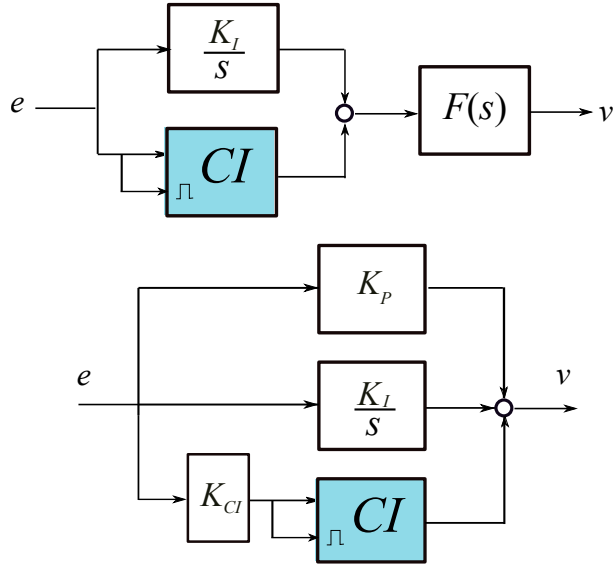}}
\caption{Right-reset compensators: ({\em top}) Horowitz reset compensator (\cite{krishman-horowitz}, $K_I$ is a parameter to be tuned and $F(s)$ a transfer function to be designed), ({\em botton}) PI+CI compensator (\cite{libro,vidal}, $K_P$, $K_I$ and $K_{CI}$ are compensator parameters to be tuned).}
\label{RRC}
\end{figure}

{\bf Proposition III.3}: {\em The reset control system $(A,C,n_\rho)$, with $A$ and $C$ given by (21), has a unique solution on $[0,\infty)$, for any $\mathbf{x}_0 \in \mathds{R}^n$,  if and only if it has well-posed reset instants.
}

\vspace{0.125cm}
{\bf Proof}:  {\em only if}) 
By contradiction, if $(A,C,n_\rho)$ does not have well-posed reset instants, then for some initial condition $\mathbf{x}_0 \in \mathds{R}^n$ the reset instant sequence $\mathds{T}$ is not well-defined, that is for some integer $k$, $1 \leq k < \infty$,
$\min \{ \Delta \in \mathds{R}^+ : e^{A\Delta}x_{k-1} \in \mathcal{M} \}$ does not exist and thus the solution is not defined for $t>t_{k-1}$ and some finite $t_{k-1}$. This is in contradiction with the solution to be defined on $[0,\infty)$.

{\em if})  By well-posedness of the reset instants, $\mathds{T}$ is well-defined for any $\mathbf{x}_0 \in \mathds{R}^n$. If the sequence $\mathds{T}$ is finite then the result directly follows by existence and uniqueness of solutions of the LTI base system; otherwise, it will be shown that if $\mathds{T} = ( t_k)_{k = 1}^\infty$ then $t_k \rightarrow \infty$ as $k\rightarrow \infty$.  Since for $k\geq2$ it is true that $\mathbf{x}(t_k^+) \in \mathcal{M}_R$, to complete the proof it is enough with showing that $t_k \rightarrow \infty$ as $k\rightarrow \infty$ for any $\mathbf{x}_0 \in \mathcal{M}_R$ with an infinite sequence $\mathds{T}$. This directly follows from the following result: an initial condition in the after-reset set $\mathcal{M}_R$, with dimension $m$, will have sequences of decreasing reset intervals with length at most $m-1$ (a detailed proof is given in \cite{banosmulero,libro}, note that $\mathcal{F}_R = \mathcal{M}_R$ ). Thus, since for the base system, solutions exists and are unique for any initial condition, it directly follows that the reset system is well-posed.
\
$\Box$

\vspace{0.05cm}

Since the only way in which Zeno solutions may exist is that reset instants be well-posed, from Prop. III.3 it may be concluded that reset control systems do not have Zeno solutions, since the solution is defined on $[0,\infty)$ when reset instants are well-posed (note that this is only true for reset control systems in which $P$ has a strictly proper transfer function as given by (15)). Thus:

\begin{itemize}
\item Ill-posed reset instants implies the existence of deadlock for some initial condition, but not the existence of Zeno solutions. 

\item Well-posed reset instants implies that neither deadlock nor Zeno solutions do exist. 

\end{itemize}

\subsection{Full reset and right reset compensation}

A natural question to ask is whether existence and uniqueness of reset control system solutions can be checked in a simple manner for a given system $P$ and a reset compensator $R$, and for any exogenous inputs, that is with independence of the exosystems. By using Prop. III.2 and III.3, this is about to derive simple conditions for the subspace $\mathcal{M}_{RU} = \mathcal{N}({I -A_R}) \cap \mathcal{N}(\mathcal{O}_{base})$ to be $A$-invariant. 

Note that if the base system is observable, that is ${\mathcal O}_{base}$ is full rank, then ${\mathcal M}_{RU} = \{\mathbf{0}\}$  and the condition is trivially satisfied; but this is also the case if $dim({\mathcal N}({\mathcal O}_{base})) = 1$, that is if there exists only one unobservable mode, since $dim({\mathcal M}_{RU}) \leq dim({\mathcal N}({\mathcal O}_{base})) = 1$ then ${\mathcal M}_{RU} = \{\mathbf{0}\}$ or  ${\mathcal M}_{RU} = {\mathcal N}({\mathcal O}_{base})$ (in both cases ${\mathcal M}_{RU}$ is $A$-invariant). The general case is much more involved; in the following, the cases of full reset and right reset compensation are analyzed.

\vspace{0.125cm}

{\bf Proposition III.4}: {\em The reset control system $(A,C,n_\rho)$, with $A$ and $C$ given by (21), has well-posed reset instants if the reset compensator is full reset or partial reset with right reset.}

{\bf Proof:}  
Regroup states as $\bar{\mathbf{x}}_p= (\mathbf{w}_1, \mathbf{w}_2,\mathbf{x}_p)$, and split the compensator state $\mathbf{x}_r$ into two parts, $\mathbf{x}_r = (\mathbf{x}_{\bar{\rho}},\mathbf{x}_\rho)$, where $\mathbf{x}_{\bar{\rho}}$ and $\mathbf{x}_\rho$, corresponding to non-resetting and resetting states respectively. Thus the closed-loop state is $\mathbf{x} = (\bar{\mathbf{x}}_p,\mathbf{x}_{\bar{\rho}}, \mathbf{x}_\rho)$. Moreover, by using submatrices with appropriate dimensions, matrices $A$, $A_R$ and $C$ are partitioned as 
\begin{equation}
\small
\begin{array}{ll}
 A = 
\left(
\begin{array}{ccc}
  \bar{A}_p & \bar{B}_p C_{r_1} &  \bar{B}_p C_{r_2}\\
  -B_{r_1} \bar{C}_p & A_{r_{11}}   &   A_{r_{12}} \\
  B_{r_2} \bar{C}_p &A_{r_{21}}   &   A_{r_{22}}
\end{array}
\right),   &  \hspace{-0.2cm}  A_R = 
\left(
\begin{array}{ccc}
  I& O  &O   \\
  O&I   & O  \\
  O& O & O   
\end{array}
\right)
\\
 \\
 C = 
\left(
\begin{array}{ccc}
\bar{C}_p  & O & O   
\end{array}
\right)   &   
 \end{array}
\end{equation}

Now, using (3) the subspace of after-reset and unobservable states ${{\mathcal M}}_{RU} = {{\mathcal F}}_{RU}  =
{\cal N}(
\left(
\begin{array}{c}
  I - {A}_R   \\
  {{\cal O}}_{base}   
\end{array}
\right)
 )$. For any state $\mathbf{x} \in {{\mathcal M}}_{RU}$, it is true that $\mathbf{x}_\rho = \mathbf{0}$, and $\bar{C}_p \bar{\mathbf{x}}_p = {C}\mathbf{x}  =  0$, and then 
\begin{equation}
\small
(I-{A}_R){A} \mathbf{x} = 
\nonumber
\end{equation}
\begin{equation}
\left(
\begin{array}{ccc}
  O & O &O \\
  O & O &O \\  
  O & O & I 
\end{array}
\right) \hspace{-0.15cm}
 \left(
\begin{array}{ccc}
  \bar{A}_p & \bar{B}_p C_{r_1} &  \bar{B}_p C_{r_2}\\
  -B_{r_1} \bar{C}_p & A_{r_{11}}   &   A_{r_{12}} \\
  B_{r_2} \bar{C}_p &A_{r_{21}}   &   A_{r_{22}} 
\end{array} 
\right)   \hspace{-0.15cm}
\left(
\begin{array}{c}
  \bar{\mathbf{x}}_p  \\
  \mathbf{x}_{\bar{\rho}}   \\
  \mathbf{0}   
\end{array}
\right) = 
\nonumber
\end{equation}
\begin{equation}
\left(
\begin{array}{c}
  \mathbf{0}  \\
  \mathbf{0}  \\
  B_{r_2}\bar{C}_p \bar{\mathbf{x}}_p +  A_{r_{21}}   \mathbf{x}_{\bar{\rho}}   
\end{array}
\right) = 
\label{(2.8a)}
\left(
\begin{array}{c}
  \mathbf{0}  \\
  \mathbf{0}  \\
    A_{r_{21}}   \mathbf{x}_{\bar{\rho}}   
\end{array}
\right)  
\end{equation}

\noindent As a result, ${{\mathcal M}}_{RU}$ is ${A}$-invariant if and only if $A_{r_{21}}   \mathbf{x}_{\bar{\rho}}   = \mathbf{0}$ for any $\left(
\begin{array}{ccc}
  \bar{\mathbf{x}}_p &
  \mathbf{x}_{\bar{\rho}}  &
  \mathbf{0}   
\end{array}
\right) \in {\mathcal N}({{\mathcal O}}_{base}) $. The result follows since for a full reset compensator $\mathbf{x}_{\bar{\rho}} = \mathbf{0}$, and for a right reset compensator $A_{r_{21}}= 0$. $\Box$

{\em Example III.6 (full reset compensation with ill-posed reset instants)}: Example III.2 describes a full reset system with ill-posed reset instants; note that the reset system does not correspond to a reset control system as given in Fig. \ref{controlsetup}. 

 
  {\em Example III.7 (reset control system with reference input and well-posed reset instants)}:
   Consider the reset control system of Fig. \ref{controlsetup}, where $R$ is a CI and $P$ is an integrator. In addition, consider a sinusoidal reference $r(t)= A \sin(\omega t +  \phi)$, for some given constants $A$, $\omega >0$, and $\phi$; it is given by the exosystem
\begin{equation}
   \left \{
   \begin{array}{l}
     \mathbf{\dot{w}}_1(t) = \left( 
\begin{array}{cc}
 0 & \omega   \\
 -\omega & 0 
\end{array}
\right) \mathbf{w}_1(t),  \ \ \mathbf{w}_1(0) = \left( 
\begin{array}{c}
 A \sin \phi \\
 A \cos\phi 
\end{array} \right)\\
     r(t)  = 
\left(
\begin{array}{cc}
  1&0        
\end{array}
\right) {\mathbf w}_1
    \end{array}
   \right.
\label{(5.11)}
\end{equation}
Since a disturbance signal is not considered in this example, Prop. III.3 can be used by simply eliminating the row and column blocks corresponding to the disturbance exosystem in the matrices ${A}$, ${A}_R$, and ${C}$. The result is 
\begin{equation}
{A}=\left(
\begin{array}
[c]{cccc}%
0 & \omega & 0 & 0\\
  -\omega  & 0&0 & 0\\
  0 & 0 & 0 & 1\\
  1 & 0 &-1 & 0
\end{array}
\right), \ \
{C}=\left(
\begin{array}
[c]{cccc}%
 1 & 0 & -1 & 0 
\end{array}
\right)
\label{(5.12)}
\end{equation}
\noindent Now, the observability matrix of the (closed-loop) base system is 
\begin{equation}
{{ O}}_{base}=\left(
\begin{array}
[c]{cccc}%
1 & 0 & -1 & 0\\
0 & \omega &  0 & -1\\
-(1+\omega^2) & 0 & 1 & 0\\
  0 & -\omega(1+\omega^2) & 0 & 1
\end{array}
\right
)
\label{(5.13)}
\end{equation}
\noindent which is full rank for any $\omega > 0$. As a result ${\mathcal M}_{RU} = \{ {\mathbf 0} \}$ is trivially $A$-invariant and thus the system is well-posed for arbitrary sinusoidal reference inputs. Note that, since Proposition III.4 applies (the reset compensator $R$ is a Clegg integrator and thus it is full reset), in this case it is not necessary to check the $A$-invariance of the subspace ${\mathcal M}_{RU}$; and moreover, it is possible to assure a much more general result: the reset control system has well-posed reset instants not only for sinusoidal references but for any exogenous inputs   generated by exosystems.


\subsection{Partial reset compensation (left reset compensators)}
The general case of partial reset compensation is much more involved; in the following, existence and uniqueness of solutions will be analyzed for a type of left reset compensators, in particular for reset compensators $R$ that are a series interconnection of a LTI compensator $G_1 = (A_{r_1},B_{r_1},C_{r_1})$ with state $\mathbf{x}_{\bar{\rho}} \in \mathds{R}^{n_{\bar{\rho}}}$,  and a full reset compensator $R_2$ with a base system $R_{2,b}= (A_{r_2},B_{r_2},C_{r_2})$, and with state $\mathbf{x}_{{\rho}} \in \mathds{R}^{n_{{\rho}}}$ (Fig. \ref{leftcompensator}). 

\begin{figure}[t]
\centerline{\includegraphics[width=6cm,keepaspectratio]{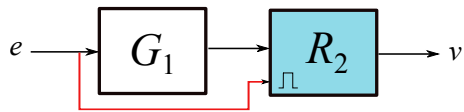}}
\caption{A left reset compensator} 
\label{leftcompensator}
\end{figure}

Thus, $R$ has a base system $R_{b} = (A_{r},B_{r},C_{r})$ with 
\begin{equation}
\begin{array}{ll}
 A_r = 
\left(
\begin{array}{cc}
  {A}_{r_1} & O\\
  B_{r_2} C_{r_1} & A_{r_{2}} 
\end{array}
\right)   &   B_r = 
\left(
\begin{array}{c}
 B_{r_1}  \\
  O     
\end{array}
\right)
\\
 \\
 C_r = 
\left(
\begin{array}{cc}
 O & C_{r_2}   
\end{array}
\right)   &   
 \end{array}
\end{equation}

\noindent and thus it is a left reset compensator. Now, consider $R$ as the compensator of the reset control system $(A,C,n_\rho)$ of Fig. 7, with $A$ and $C$ given by (23). In addition, by defining again the closed loop state as $\mathbf{x} = (\bar{\mathbf{x}}_p,\mathbf{x}_{\bar{\rho}}, \mathbf{x}_\rho)$, with  $\bar{\mathbf{x}}_p = ({\mathbf{w}}_1,\mathbf{w}_2, \mathbf{x}_p)$, $A$ and $C$ can be partitioned as: 
\begin{equation}
\small
\begin{array}{ll}
 A = 
\left(
\begin{array}{ccc}
  \bar{A}_p & O &  \bar{B}_p C_{r_2}\\
 - B_{r_1} \bar{C}_p & A_{r_{1}}   &  O \\
 O &B_{r_2}C_{r_1}   &   A_{r_{2}}
\end{array}
\right)   &   
 C = 
\left(
\begin{array}{ccc}
\bar{C}_p  & O & O   
\end{array}
\right)      
 \end{array}
\end{equation}

It will be assumed that the realizations $(A_{r},B_{r},C_{r})$ and $(A_{p},B_{p},C_{p})$ are minimal, and that $(\bar{A}_p,\bar{B}_p,\bar{C}_p)$ is observable\footnote{Using interconnection properties, note that $(\bar{A}_p,\bar{B}_p,\bar{C}_p)$ is simply a parallel connection of a system $(A_1,O,C_1)$ and the series connection of $(A_2,O,C_2)$ and $(A_{p},B_{p},C_{p})$ (see Fig. 7), thus it directly follows that unobservable modes of $(\bar{A}_p, \bar{B}_p, \bar{C}_p)$ are given by common eigenvalues of $A_1$ and $A_2$, common values of eigenvalues of $A_1$ and modes of $(A_p,B_p,C_p)$, and common values of eigenvalues of $A_2$ and zeros of $(A_p,B_p,C_p)$.}. Thus, since in this case of left reset compensation the base control system of Fig. 7 is simply a feedback connection of the series connection of $(A_{r_1},B_{r_1},C_{r_1})$, $(A_{r_2},B_{r_2},C_{r_2})$, and $(\bar{A}_p,\bar{B}_p,\bar{C}_p)$, it is clear that every unobservable mode $\lambda \in \sigma_{ \bar{ \mathcal{O} } }(A,C)$ of the base control system must be a pole of $(A_{r_1},B_{r_1},C_{r_1})$ and/or $(A_{r_2},B_{r_2},C_{r_2})$, and a zero of $(\bar{A}_p,\bar{B}_p,\bar{C}_p)$. 

For an unobservable mode $\lambda$, let $m_{\lambda}$ be the algebraic multiplicity as a zero of $(\bar{A}_p,\bar{B}_p,\bar{C}_p)$; $q_{\lambda}$, the pole algebraic multiplicity in $(A_{r_1},B_{r_1},C_{r_1})$; and $r_{\lambda}$, the pole algebraic multiplicity in  $(A_{r_2},B_{r_2},C_{r_2})$. In addition, the number of cancellations is $d_\lambda = \min (q_\lambda+r_\lambda,m_\lambda)$. On the other hand, since the geometric multiplicity of the unobservable modes in the base control system is 1 (see Prop. A.1 in the Appendix), then the index of an unobservable mode of $A$ is equal to its algebraic multiplicity $ma_\lambda$, that in general will be greater or equal than $d_\lambda$. In addition,  the dimension of $\mathcal{N}\left(\mathcal{O}\right)$ is the sum of all the cancellations corresponding to the unobservables modes (see Proposition A.3 in the Appendix), that is
\begin{equation}
 dim(\mathcal{N}\left(\mathcal{O}\right)) = s:= \sum\limits_{\lambda \in \sigma_{ \bar{ \mathcal{O} } }(A,C)} d_\lambda 
\label{snrho}
\end{equation}


{\bf Proposition III.5}: Consider the reset control system $(A,C,n_\rho)$ of Fig. 7, with $P$ given by (17) and $R$ being a 
left reset compensator with base system $R_{b} = (A_{r},B_{r},C_{r})$ as given by (32). $(A,C,n_\rho)$ has well-posed reset instants if and only if
\begin{equation}
n_\rho \geq s
\label{nrho}
\end{equation}


{\bf Proof}:  
({\em if}): By using Prop. A.2 and Prop. III.2, it it is sufficient to prove that $\mathcal{M}_{RU}=\mathcal{M}_{R}\cap\mathcal{N}\left(  \mathcal{O}\right)   
=\left\{ \mathbf{0} \right\}$.
 In virtue of Proposition A.3, let $\{  \mathbf{v}_{\lambda}^{(d_\lambda-k)}\in\mathbb{R}^{n}:\lambda\in\sigma
_{\mathcal{\bar{O}}}\left(  A,C\right), k = 0,\cdots,d_\lambda-1  \}  $ be a basis of
$\mathcal{N}\left(  \mathcal{O}\right)  $ with $\mathbf{v}_{\lambda}^{(d_\lambda-k)} = ({\mathbf{v}}_{p_\lambda}^{(d_\lambda-k)},\mathbf{v}_{{\bar \rho}_\lambda}^{(d_\lambda-k)},\mathbf{v}_{\rho_\lambda}^{(d_\lambda-k)})$ being a
generalized eigenvector of $A$ corresponding to the unobservable mode $\lambda$ ($\mathbf{v}_{\lambda}^{(d_\lambda)}$ is the eigenvector). Thus,  any $\mathbf{v} = (\mathbf{v}_p,\mathbf{v}_{\bar{\rho}},\mathbf{v}_\rho)\in \mathcal{N}\left(  \mathcal{O}\right)$ can be expressed as $\mathbf{v} = \sum\limits_{\lambda \in \sigma_{ \bar{ \mathcal{O} } } (A,C)} \sum\limits_{k=0}^{d_\lambda -1}\alpha_{\lambda}^{(d_\lambda-k)} \mathbf{v}_{\lambda}^{(d_\lambda-k)}$, where $\mathbf{v}_\rho = \sum\limits_{\lambda \in \sigma_{ \bar{ \mathcal{O} } } (A,C)} \sum\limits_{k=0}^{d_\lambda -1}\alpha_{\lambda}^{(d_\lambda-k)} \mathbf{v}_{\rho_{\lambda}}^{(d_\lambda-k)}$. In the following, it will be shown that if $\mathbf{v} \in \mathcal{N}\left(  \mathcal{O}\right)\cap \mathcal{M}_R$, and thus $\mathbf{v}_\rho=\mathbf{0}$, then the scalars $\alpha_{\lambda}^{(d_\lambda-k)}$ are all zero and thus $\mathbf{v}=\mathbf{0}$. Note that, in general $\{\mathbf{v}_{\rho_{\lambda}}^{(d_\lambda-k)}:\lambda\in\sigma
_{\mathcal{\bar{O}}}\left(  A,C\right), k = 0,\cdots,d_\lambda-1   \}$ is not the set of generalized eigenvectors (including the eigenvector) of $A_{r_2}$ corresponding to the mode $\lambda$, and thus it can not be directly concluded that it is a linearly independent set. 

Consider a state transformation given by a matrix  $S = \left(
\begin{array}
[c]{cc}%
I_{n_{p}} & O\\
O & T
\end{array}
\right) \in\mathbb{R}^{n\times n} $ where $T= \left(
\begin{array}
[c]{cc}%
M\\
N%
\end{array}
\right) =
\left(
\begin{array}
[c]{cc}%
M_{1} & M_{2}\\
N_{1} & N_{2}%
\end{array}
\right)  $ with $M = \left( \begin{array}{cc}  M_1 & M_2 \end{array} \right) \in\mathbb{R}^{n_{\bar{\rho}}\times\left(  n_{\rho}%
+n_{\bar{\rho}}\right)  }$, and $N = \left( \begin{array}{cc}  N_1 & N_2 \end{array} \right) \in\mathbb{R}^{n_{{\rho}}\times\left(  n_{\rho}%
+n_{\bar{\rho}}\right)  }$. Following a procedure analogous to the Kalman
decomposition (observable/unobservable decomposition), $M$ can be selected in
such a way that $M_{1}$ is nonsingular, $\mathcal{N}\left(  M\right)  $ is
$A_{r}-$invariant, and that includes the subspace spanned by those generalized
eigenvectors of $A_{r}$ associated with unobservable modes of $\left(
A,C,n_{\rho}\right)  $\footnote{The condition of being $\mathcal{N}\left(
\mathcal{O}\right)  $ a subspace of the $A_{r}-$invariant subspace
$\mathcal{N}\left(  M\right)  $ can be achieved by putting the basis of
$\mathcal{N}\left(  \mathcal{O}\right)  $ (composed by generalized
eigenvectors of $A$) as rows of $N$, and then completing $N$ with other
generalized eigenvectors so as to obtain complete Jordan subchains of vectors
(which spans cyclic subspaces). There exists freedom in order to select $M$
(completion of a basis of row vectors) but always $M_{1}$ can be assured to be
nonsingular by making $M_{1}$ to be a large multiple of the identity.}. As a
result if $T^{-1}= \left(
\begin{array}
[c]{cc}%
X &Y%
\end{array}
\right) =
\left(
\begin{array}
[c]{cc}%
X_{1} & Y_{1}\\
X_{2} & Y_{2}%
\end{array}
\right)  $ with $X\in\mathbb{R}^{\left(  n_{\rho}+n_{\bar{\rho}}\right)
\times n_{\bar{\rho}}}$ and $Y\in\mathbb{R}^{\left(  n_{\rho}+n_{\bar{\rho}%
}\right)  \times n_{\rho}}$, the matrix of the system is transformed into%
\[
\tilde{A}=SAS^{-1}=\left(
\begin{array}
[c]{ccc}%
\bar{A}_{p} & \bar{B}_{p}C_{r_{2}}^{T}X_{2} & \bar{B}_{p}C_{r_{2}}^{T}Y_{2}\\
MB_{r}\bar{C}_{p} & \tilde{A}_{r_{11}} & O\\
NB_{r}\bar{C}_{p} & \tilde{A}_{r_{21}} & \tilde{A}_{r_{22}}%
\end{array}
\right)
\]
It is straightforward to check that the new unobservable subspace
$\mathcal{N}(  \widetilde{\mathcal{O}})  $ with $\widetilde
{\mathcal{O}}=\mathcal{O}S^{-1}$ is spanned by a set $\{  \widetilde
{\mathbf{v}}_{\lambda
}^{(d_\lambda-k)}=(   
{\mathbf{v}}_{p_{\lambda}}^{(d_\lambda-k)},\mathbf{0},\widetilde
{ \mathbf{v}}_{\rho_{\lambda}}^{(d_\lambda-k)})
:\lambda\in\sigma_{\mathcal{\bar{O}}}\left(  A,C\right), k = 0,\cdots,d_\lambda-1  \}  $ where
$\{\widetilde
{\mathbf{v}}_{\rho_{\lambda}}^{(d_\lambda-k)}\in\mathbb{R}_{\rho}^{n}:\lambda\in\sigma_{\mathcal{\bar{O}}}\left(  A,C\right), k = 0,\cdots,d_\lambda-1 \} $ is now a subset of the union of generalized eigenvectors sets of $\tilde{A}_{r_{22}}$ corresponding to the unobservable modes. Note that it is sufficient with the condition (35) for building such a subset.

 Owing to $\mathcal{N}(  \widetilde{\mathcal{O}})  $ is
mapped into $\mathcal{N}\left(  \mathcal{O}\right)  $ via $S^{-1}$ it is clear
that any $\mathbf{v}\in\mathcal{N}\left(  \mathcal{O}\right)  $ can be written as
$\mathbf{v}=\sum\limits_{\lambda\in\sigma_{\mathcal{\bar{O}}}\left(  A,C\right)  }%
\sum\limits_{k=1}^{d_\lambda-1}
\alpha_{\lambda}^{(d_\lambda-k)}S^{-1}\widetilde{\mathbf{v}}_{\lambda}^{(d_\lambda-k)}$, 
for some scalars $\alpha_{\lambda}\in\mathbb{R}$.

Note that since $S^{-1}%
\widetilde{\mathbf{v}}_{\lambda}^{(d_\lambda-k)}=\left(\mathbf{v}_{p_{\lambda}}^{(d_\lambda-k)},Y_1\widetilde{\mathbf{v}}_{\rho_{\lambda}}^{(d_\lambda-k)}, Y_2\widetilde{\mathbf{v}}_{\rho_{\lambda}}^{(d_\lambda-k)}\right)
$, it directly follows that $\mathbf{v} \in\mathcal{M}_{R} \cap \mathcal{N}(\mathcal{O})$ if and only if $\mathbf{0} =(I- A_R)\mathbf{v} = Y_{2}\sum\limits_{\lambda\in\sigma_{\mathcal{\bar{O}}}\left(  A,C\right)  }%
\sum\limits_{k=1}^{d_\lambda-1}
\alpha_{\lambda}^{(d_\lambda-k)}\widetilde{\mathbf{v}}_{\rho_{\lambda}}^{(d_\lambda-k)}$. In addition, since $M_1$ is not singular then
\begin{align*}
MY &  =M_{1}Y_{1}+M_{2}Y_{2}=O\Rightarrow Y_{1}=-M_{1}^{-1}M_{2}Y_{2}\\
NY &  =N_{1}Y_{1}+N_{2}Y_{2}=I\Rightarrow\left(  -N_{1}M_{1}^{-1}M_{2}%
+N_{2}\right)  Y_{2}=I
\end{align*}

\noindent and thus $Y_{2}$ is not singular with inverse $Y_2^{-1} = -N_{1}M_{1}^{-1}M_{2}%
+N_{2}$ . Since in addition the set of generalized eigenvectors $\widetilde{\mathbf{v}}_{\rho_{\lambda}}^{(d_\lambda-k)}$ is linearly independent then $\alpha_{\lambda}^{(d_\lambda-k)}=0$ for all $\lambda\in\sigma_{\mathcal{\bar{O}}}\left(
A,C\right) $ and $k=0,\cdots,d_\lambda-1$, and thus  $\mathcal{M}_{R} \cap \mathcal{N}(\mathcal{O}) = \{\mathbf{0}\}$. 

\vspace{0.125cm}

({\em only if}): 
Let $\left\{  \mathbf{v}_{1},\ldots, \mathbf{v}_{s}\right\}  $ be a basis of $\mathcal{N}\left(\mathcal{O}\right) $, where $\mathbf{v}_i$ is partitioned as $ \mathbf{v}_i = ( \mathbf{v}_p^{(i)}, \mathbf{v}_{\bar{\rho}}^{(i)}, \mathbf{v}_\rho^{(i)})$, for $i=1,\cdots,s$. If \eqref{nrho} is false then $n_\rho < s$,  and thus $\left\{  \mathbf{v}_\rho^{(1)},\ldots, \mathbf{v}_\rho^{(s)}\right\}  $ is a linearly dependent set. As a result, it is obtained that $\mathbf{0} =  \sum\limits_{i =1}^{s} \alpha_i \mathbf{v}_\rho^{(i)}$ for some scalars $\alpha_i$, $i = 1,\cdots,s$, not all zero. Now, using those scalars the vector 
$\mathbf{v} =  \sum\limits_{i =1}^{s} \alpha_i \mathbf{v}_i$ must be nonzero since $\sum\limits_{i =1}^{s} \alpha_i \mathbf{v}_i = \mathbf{0}$ if and only $\alpha_i = 0$ for $i=1,\cdots,s$. As a result, there exists a nonzero $\mathbf{v} = (\mathbf{v}_p, \mathbf{v}_{\bar{\rho}}, \mathbf{v}_\rho) \in \mathcal{N}\left(\mathcal{O}\right) $ with $\mathbf{v}_\rho = \mathbf{0}$, that is a nonzero $\mathbf{v} \in \mathcal{M}_{RU}$, and then by Proposition A.2 (see Appendix) and Proposition III.2 it follows that $(A,C,n_\rho)$ has ill-posed reset instants. $\Box$

\vspace{0.125cm}


{\em Example III.8 (ill-posed reset control system with left reset compensation and disturbance input)}: 
Consider the reset control system of first row in Table I, with a sinusoidal disturbance input $d(t) = sin(t)$ generated by a exosystem $\Sigma_2(s) = 1/(s^2+1)$, a system $P$ with a transfer function $P(s) = 1/(s+1)$, and a left reset compensator $R$ given by the tandem connection of a (two inputs) Clegg integrator ($n_\rho =1$), and a system $G_1$ with transfer function $G_1(s)=1/(s^2+1)$. Note that in this case the unobservable modes are the common poles of $G_1(s)$ and $\Sigma_2(s)$, that is $\sigma_{\bar{{\mathcal O}}}= \{+j,-j\}$, and in addition $q_{+j}  = 1$, $r_{+j} =0$, $m_{+j}  = 1$, and $q_{-j}  = 1$, $r_{-j} =0$, $m_{-j}  = 1$, and therefore $d_{+j} =1$, and $d_{-j}  = 1$.  Since $n_\rho =1 < d_{+j}+d_{-j}  = 2$ then the reset control system is ill-posed.

\begin{table*}[ht]
\caption{Examples of reset control systems with partial-reset compensators}
\begin{center}
\begin{tabular}{|c|c|c|}
\hline
Reset control system & Left/Right reset-Algebraic multiplicities & Reset Instants Well-posedness \\
\hline \hline
\includegraphics[width=6cm,keepaspectratio]{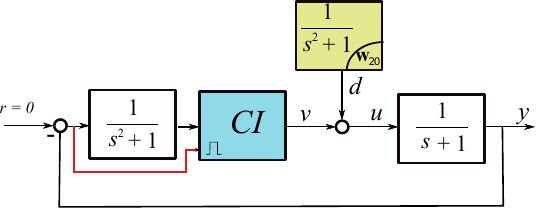} 
& 
$
\begin{array}{c}
 \text{Left-reset}   \\
   q_{+ j} = 1, r_{+ j} = 0, m_{+ j} =1 \rightarrow d_{+ j} = 1 \\
   q_{- j} = 1, r_{- j} = 0, m_{- j} =1 \rightarrow d_{- j} = 1\\
    n_\rho = 1, s = 2
\end{array}
$
&
Ill-posed \\
\hline
\includegraphics[width=6cm,keepaspectratio]{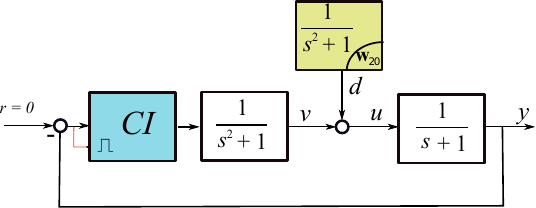} & Right-reset  & Well-posed\\
\hline
\includegraphics[width=6cm,keepaspectratio]{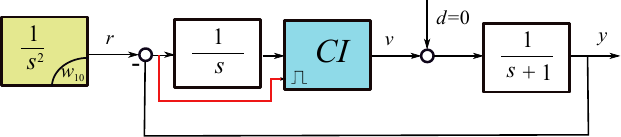} 
&
$
\begin{array}{c}
 \text{Left-reset}   \\
   q_{0} = 1, r_{0} = 1, m_{0} = 2 \rightarrow d_{0} = 2 \\
   n_\rho = 1, s = 2
\end{array}
$
& Ill-posed \\
\hline
\includegraphics[width=6cm,keepaspectratio]{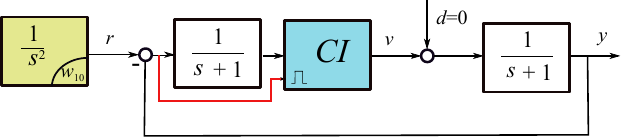} 
&
$
\begin{array}{c}
 \text{Left-reset}   \\
   q_{0} = 0, r_{0} = 1, m_{0} = 2 \rightarrow d_{0} = 1 \\
   n_\rho = 1, s = 1
\end{array}
$
& Well-posed \\
\hline
\includegraphics[width=6cm,keepaspectratio]{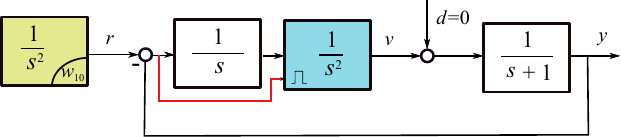} 
&
$
\begin{array}{c}
 \text{Left-reset}   \\
   q_{0} = 1, r_{0} = 2, m_{0} = 2 \rightarrow d_{0} = 2 \\
   n_\rho = 2, s = 2
\end{array}
$
& Well-posed \\
\hline
\includegraphics[width=6cm,keepaspectratio]{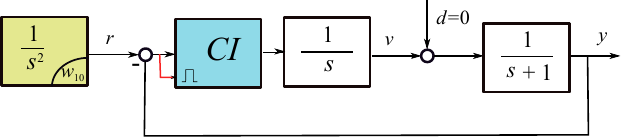} & Right-reset& Well-posed \\
\hline
\includegraphics[width=6cm,keepaspectratio]{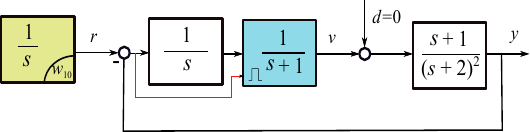} 
&
$
\begin{array}{c}
 \text{Left-reset}   \\
   q_{0} = 1, r_{0} = 0, m_{0} = 1 \rightarrow d_{0} = 1 \\
   q_{-1} = 0, r_{-1} = 1, m_{-1} = 1 \rightarrow d_{-1} = 1 \\
   n_\rho = 1, s = 2
\end{array}
$
& Ill-posed \\
\hline
\includegraphics[width=6cm,keepaspectratio]{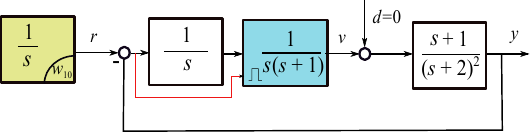} 
&
$
\begin{array}{c}
 \text{Left-reset}   \\
   q_{0} = 1, r_{0} = 1, m_{0} = 1 \rightarrow d_{0} = 1 \\
   q_{-1} = 0, r_{-1} = 1, m_{-1} = 1 \rightarrow d_{-1} = 1 \\
   n_\rho = 2, s = 2
\end{array}
$
& Well-posed \\
\hline
\end{tabular}
\end{center}
\label{default}
\end{table*}%

This can be alternatively done, with some effort, by directly checking if ${\mathcal M}_{RU} $ is $A$-invariant. The systems $P$, $R$, and the exosystem $\Sigma_2$ have the following realizations:
\begin{eqnarray}
P&:& 
   \begin{array}{llll}
     A_p = -1  &  B_p = 1 & C_p = 1
    \end{array}
 \\
 R&:&
   \begin{array}{ll}
     A_r = \left( \begin{array}{ccc}
 			0 & 1& 0   \\
 			-1 & 0 & 0\\
 			1 & 0 & 0
			\end{array}
		\right)  &
   B_r = \left(\begin{array}{c}
 			0   \\
 			1   \\
 			0
		  \end{array}
	   \right) \\
   C_r = \left(\begin{array}{ccc}
 			0  & 0 & 1
		   \end{array}
             \right) &
  \end{array}
 \\
 \Sigma_2&:& 
   \begin{array}{cc}
     A_2 = \left(\begin{array}{cc}
 			0 & 1   \\
 			-1 & 0 
		     \end{array}
	       \right) &
   C_2 = \left(\begin{array}{cc}
  			1&0        
		   \end{array}
	     \right)
    \end{array}
\end{eqnarray}

The reset control system $(A,C,1)$, with state  $\mathbf{x} = (\mathbf{w}_2,{x_p},\mathbf{x}_r)$, is given by the matrices 
\begin{equation}\small
\begin{array}{cc}
 {A} = \left(
\begin{array}{cccccc}
 0 & 1 & 0 & 0 & 0 & 0 \\
 -1 &0 &0 &0 &0 &0  \\
  1& 0  &-1& 0&0 &1\\
 0& 0  &0&0&1 &0 \\
  0& 0  &-1&-1&0 &0 \\   
   0& 0  &0&1&0 &0  
\end{array}
\right)
   &
  {C} = \left(
\begin{array}{cccccc}
 0 & 0 & 1 & 0 & 0 & 0  
\end{array}
\right)
\end{array}
\end{equation}

\noindent and it can be obtained that 
\begin{equation}
{\mathcal M}_{RU} = {\mathcal N} \left(
\begin{array}{c}
 I - {A}_R \\
 {\mathcal O}_{base}  
\end{array}
\right) = span \{ \left(  
\begin{array}{cccccc}
0, 1, 0, -1, 0, 0 
\end{array}
\right)\}
\end{equation}
and, finally, it can be easily check that ${\mathcal M}_{RU}$ is not $A$-invariant and thus the reset control system is ill-posed. Note that in this case, for any initial condition in ${\mathcal M}_{RU}$ ($\subset {\mathcal M}_{R}$), the system evolves to a subset of the unobservable subspace ${\mathcal N}({\mathcal O}_{base}) \subseteq {\mathcal N}(C)$ that does not contain after reset states, and thus it is a part of the reset set ${\mathcal M}$. 

On the other hand, consider the reset control system of the second row in Table I; in this case, the reset control system is well-posed since the compensator is a right reset compensator. This can be also concluded by checking that ${\mathcal M}_{RU}$ is $A$-invariant (note that in this case ${\mathcal M}_{RU} = span\{(0,-1,0,1,0,0),(-1,0,0,0,1,0)\} \neq \{\mathbf{0}\}$).
Besides the above reset control systems, Table I shows several examples of reset control systems with well/ill-posed reset instans, based on the direct application of Prop. III.4-III.5 (note that reset systems with ill-posed reset instants must have base systems with at least two unobservable modes).


\section{Continuous dependence on the initial condition}
Several convenient metrics has been successfully developed in the literature to represent distances between impulsive/hybrid systems solutions; for example, the Skorokhod distance \cite{canada}, and the graphical distance in the HI framework \cite{cai,GSTbook}.  In the particular case of IDSs, it will be shown that  
a convenient metric for determining the continuous dependence on the initial condition is directly the Hausdorff distance between the set of points defining the IDS trajectories; this metric has been used in \cite{ahmad} for analysis of impulsive integro-differential equations, and more recently in \cite{dishliev} for the analysis of continuous dependence of solutions of differential equations with nonfixed moments of impulses on the initial condition.  In the following, this approach will be followed to develop analogous results about reset control systems.

Given two nonempty subsets $\mathcal{A},\mathcal{B} \subset \mathds{R}^n$, the {\em Hausdorff distance} between them is 
\begin{equation}
d_H(\mathcal{A},\mathcal{B}) := max\{h(\mathcal{A},\mathcal{B}),h(\mathcal{B},\mathcal{A}) \}
\end{equation} 
where $ h(\mathcal{A},\mathcal{B}):=sup\{inf\{\|a-b\|,a \in \mathcal{A} \}, b \in \mathcal{B} \}$.
On the other hand, the euclidean distance $d_E(\mathcal{A},\mathcal{B})$ is defined as 
\begin{equation}
d_E(\mathcal{A},\mathcal{B}) := inf \{ inf \{ \|a-b\|,b \in \mathcal{B} \},a \in \mathcal{A} \}
\end{equation} 

For a function ${\mathbf f}:\mathds{R}^+ \rightarrow \mathds{R}^n$ and some scalar $T>0$, by definition ${\mathbf f}([0,T]):= \{ \mathbf{f}(t) \in \mathds{R}^n: t \in [0,T] \}$. In general, for two left-continuous functions ${\mathbf f}$ and ${\mathbf g}$, a notion of distance using directly the Hausdorff distance may be problematical. Note that the distantce $d_H({\mathbf f}([0,T]),{\mathbf g}([0,T]))$ by itself does not give a good characterization of the property; for example, two functions like $sin(t)$ and $cos(t)$ would be at a distance 0 over the interval $[0,2\pi]$, that is $d_H({sin}([0,2\pi]),{cos}([0,2\pi]))=0$ (similar examples can be easily found for left-continuous functions with jump discontinuities). In spite of this fact, it will be shown in next Section how Hausdorff distance can be successfully used. 
Note that for example $d_H({sin}([0,T]),{cos}([0,T])) > 0$ if $T\in [0,\pi/2) \cup (\pi/2,3\pi/2)$.

\subsection{Definition and motivating examples}

For a reset control system $(A,C,n_\rho)$, consider a solution $\mathbf{x}$ corresponding to a initial condition $\mathbf{x}_0$, and another solution $\mathbf{x}^\ast$ corresponding to a perturbed initial condition $\mathbf{x}^\ast_0$. With some abuse of notation, let $\mathbf{x}([0,T],\mathbf{x}_0)$ be the set of points corresponding to the trajectory of the system for $t\in [0,T]$, that is 
\begin{equation}
\mathbf{x}([0,T],\mathbf{x}_0) := \{ \mathbf{x}(t) \in \mathds{R}^n: t \in [0,T] \}
\end{equation}
where $\mathbf{x}$ is the solution of $(A,C,n_\rho)$ with initial condition $\mathbf{x}_0$.  $\mathbf{x}([0,T])$ is used if the initial condition is clear from the context, and  $\mathbf{x}^\ast([0,T],\mathbf{x}^\ast_0)$ is defined accordingly. Then, $d_H(\mathbf{x}([0,T],\mathbf{x}_0),\mathbf{x}^\ast([0,T],\mathbf{x}^\ast_0))$, or simply $d_H(\mathbf{x}([0,T]),\mathbf{x}^\ast([0,T]))$,  will be referred to as the {\em Hausdorff distance} between the two trajectories.

Continuous dependence will be characterized by the property that for {\em almost all} $T\geq 0$, it is possible to find two trajectories $\mathbf{x}([0,T],\mathbf{x}_0)$ and $\mathbf{x}^\ast([0,T],\mathbf{x}^\ast_0)$ arbitrarily close, in the sense that $d_H(\mathbf{x}([0,T],\mathbf{x}_0),\mathbf{x}^\ast([0,T],\mathbf{x}^\ast_0))$ is arbitrarily small, by choosing close enough initial conditions $\mathbf{x}_0$ and  $\mathbf{x}^\ast_0$.\\

{\bf Definition IV.1 (continuous dependence)}: {\em For a reset control system $(A,C,n_\rho)$ with well-posed reset instants, the solution depends continuously on the initial condition at $\mathbf{x}_0 \in \mathds{R}^n$ if for any $\epsilon > 0$, $T  \in \mathds{R}^+\setminus {\mathds{T}}$, there exist $\delta >0$ 
such that for any $\mathbf{x}^\ast_0 \in \mathds{R}^n$, with $\| \mathbf{x}_0  - \mathbf{x}^\ast_0\| < \delta$, it is true that 
\begin{equation}
d_H(\mathbf{x}^\ast([0,T],\mathbf{x}^\ast_0), \mathbf{x}([0,T],\mathbf{x}_0)) < \epsilon
\end{equation}}

A first analysis shows that continuous dependence on the initial condition fails at ${\mathbf x}_0 \in \mathcal{M}$. 
The following example describes this behavior, and also an analysis of the necessity of removing reset instants when checking the Hausdorff distance (31). \\

{\em Example IV.1}: Consider the reset control system $(A,C,1)$ with
\begin{equation}
\begin{array}{cc}
A=\left( \begin{array}{cc}
0&1 \\
-1&0
\end{array} \right) &
C=\left( \begin{array}{cc}
1&0
\end{array} \right)
\end{array}
\end{equation}
The reset set is $\mathcal{M} = \{(x_1,x_2) \in \mathds{R}^2: x_1 = 0\} \setminus \{(0,0)\}$. For any $\mathbf{x}_0 \neq \mathbf{0}$ there is only one reset instant, and thus $\mathds{T}=(t_1,\infty)$. By simplicity, consider $\mathbf{x}_0 = (1,0)$ that results in $t_1 = \frac{\pi}{2}$, and the perturbed initial condition $\mathbf{x}^\ast_0 = (1+\delta \cos(\phi),\delta \sin(\phi))$, for some $\delta>0$ and $\phi \in[0,2\pi)$, satisfying $\|\mathbf{x}_0 -\mathbf{x}^\ast_0 \| < \delta$, that results in a resetting instant $t_1^\ast \in [\frac{\pi}{2}- \text{atan }\delta,\frac{\pi}{2}+ \text{atan }\delta]$.
\begin{figure}[t]
\centerline{\includegraphics[scale=1.25]{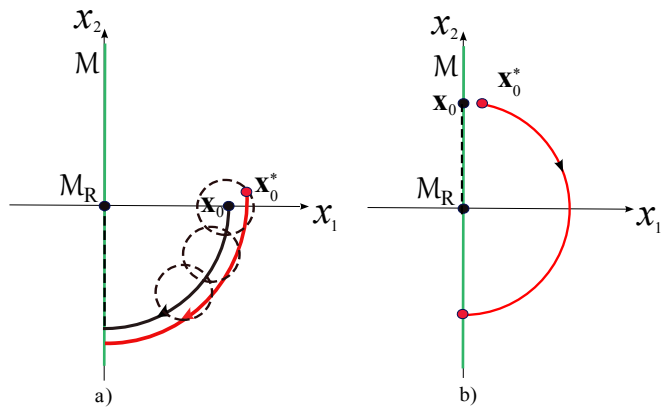}}
\caption{$\mathbf{x}$ (black) and $\mathbf{x}^\star$ (red): a) $\mathbf{x}_0 \in \mathds{R}^n\setminus {\mathcal M}$, b) $\mathbf{x}_0 \in {\mathcal M}$.} 
\label{fig:controlsetup3}
\end{figure}
Two cases are  separetaly analyzed: $T<t_1$ and $T>t_1$. For $T<t_1=\frac{\pi}{2}$ it always possible to find some $\delta_1 >0$ such as $T <   \frac{\pi}{2}-\text{atan } \delta_1$ meaning that both the solution and the perturbed solution always reset after the instant $T$; thus, both solutions correspond to the base system solutions and directly $d_H(\mathbf{x}^\ast([0,T],\mathbf{x}^\ast_0), \mathbf{x}([0,T],\mathbf{x}_0)) \leq \delta_1$. For the case $T> t_1$, it is also possible to find some $\delta_2$ such as $\frac{\pi}{2}+ \text{atan } \delta_2 < T$, that is both the solution and the perturbed solution always reset before the instant $T$, and again $d_H(\mathbf{x}^\ast([0,T],\mathbf{x}^\ast_0), \mathbf{x}([0,T],\mathbf{x}_0)) \leq \delta_2$. 
On the other hand, note that for $T = t_1 = \frac{\pi}{2}$ it is not possible to choose any $\delta>0$ such as  $T = \frac{\pi}{2} <  \frac{\pi}{2}-\text{atan } \delta$ or $T = \frac{\pi}{2} >  \frac{\pi}{2}+\text{atan } \delta$, meaning that for $T = t_1 = \frac{\pi}{2}$, there always exist perturbed solutions that have performed resets at instants $t_1^\ast < T$ and thus $\mathbf{0} \in \mathbf{x}^\ast([0,\frac{\pi}{2}],\mathbf{x}^\ast_0)$, while on the other hand $\mathbf{0} \notin \mathbf{x}([0,\frac{\pi}{2}],\mathbf{x}_0)$ and thus there exists perturbed solutions for which $d_H(\mathbf{x}^\ast([0,\frac{\pi}{2}],\mathbf{x}^\ast_0), \mathbf{x}([0,\frac{\pi}{2}],\mathbf{x}_0)) > 1$. This is the reason why resetting instants corresponding to $\mathbf{x}_0$ needs to be removed when checking Hausdorff distances in (31) according to Def IV.1. 
On the other hand, for $\mathbf{x}_0 = (0,1) \in {\mathcal M}$ (see Fig. \ref{fig:controlsetup3}.b) the perturbed initial condition $\mathbf{x}^\ast_0 = (\delta,1)$, and $T = t_1^\ast$ (the first crossing instant corresponding to $\mathbf{x}^\ast_0$), it is true that $\| \mathbf{x}_0  - \mathbf{x}^\ast_0\| = \delta$, and $d_H(\mathbf{x}([0,t_1^\ast],\mathbf{x}_0),\mathbf{x}^\ast([0,t_1^\ast],\mathbf{x}^\ast_0)) \geq d_H(\{\mathbf{0}\},\mathbf{x}^\ast([0,t_1^\ast],\mathbf{x}^\ast_0)) = \sqrt{1+\delta^2} > 1$ for any $\delta > 0$. As a result, the solution does not depends continuously on the initial condition at any $\mathbf{x}_0 \in {\mathcal M}$.

\vspace{0.125cm}

Another source of problems arises when there are reset system trajectories that are {\em tangential} to the reset set. More specifically,  for a reset control system $(A,C,n_\rho)$ there exist a {\em tangential crossing} of a solution $\mathbf{x}$ with initial condition $\mathbf{x}_0$, at the instant $t >0$, if $\mathbf{x}(t) \in \mathcal{M}$ and $C\dot{\mathbf{x}}(t) = 0$. In addition, a crossing that it is not tangential will be referred to as a {\em transversal crossing}.

\vspace{0.125cm}

{\em Example IV.2}: The reset control system $(A,C,1)$, with 

\begin{equation}
\begin{array}{cc}
A= \left(
\begin{array}{ccccc}
 0 & -3 & 1 \\
 1&  -1& 0 \\
0& -1 &   -1\\
\end{array}
\right)
	& \hspace{-0.2cm}
  C= \left(
\begin{array}{ccccc}
  0 &  1 &  0 
 \end{array}
\right)
 \end{array}
\end{equation}

\noindent has a tangential crossing with $\mathcal{M}$ at the first crossing for ${\mathbf x}_0 = (x_{01},0.2,1)$, where $x_{01}\approx -0.3794$, and for $t_1 \approx 0.7926$. These values can be obtained by numerically solving $Ce^{At_1}{\mathbf x}_0 = 0$ and $CAe^{At_1}{\mathbf x}_0 = 0$ for $t_1$. Note that $\mathbf{x}_1 = e^{At_1}{\mathbf x}_0 \in \mathcal{M}$, whereas $\mathbf{x}_1 \approx (0,0,0.4258)$. Fig. \ref{t} shows the zero-level curves of both $Ce^{At_1}{\mathbf x}_0$ and $CAe^{At_1}{\mathbf x}_0$ in the plane $(x_{01},t_1)$, clearly showing that there is only one solution for  ${\mathbf x}_0 = (x_{01},0.2,1)$ corresponding to $x_{01}\approx -0.3794$. In Fig. \ref{t2} two solutions of the reset system have been plotted for the initial condition  ${\mathbf x}_0$ and a perturbed initial condition ${\mathbf x}_0^\ast= (x_{01}+\delta,0.2,1)$ for some small value $\delta>0$. Note that ${\mathbf x}_0^\ast$ does not produce a crossing and thus no reset action is performed at the instant $t_1$. The two upper plots show the system output $y(t) = C\mathbf{x}(t)$ and the reset state $x_3(t)$, while the bottom plot shows the Hausdorff distance $d_H(\mathbf{x}([0,t],\mathbf{x}_0),\mathbf{x}^\ast([0,t],\mathbf{x}^\ast_0))$ versus $t$. It turns out that for $t > t_1$, $d_H(\mathbf{x}([0,t],\mathbf{x}_0),\mathbf{x}^\ast([0,t],\mathbf{x}^\ast_0))$ can not be made arbitrarily small by making $\delta$ small enough, and thus the solution does not depends continuously on the initial condition at ${\mathbf x}_0 = (x_{01},0.2,1)$. 


\begin{figure}[t]
\centering
\includegraphics[scale=0.4]{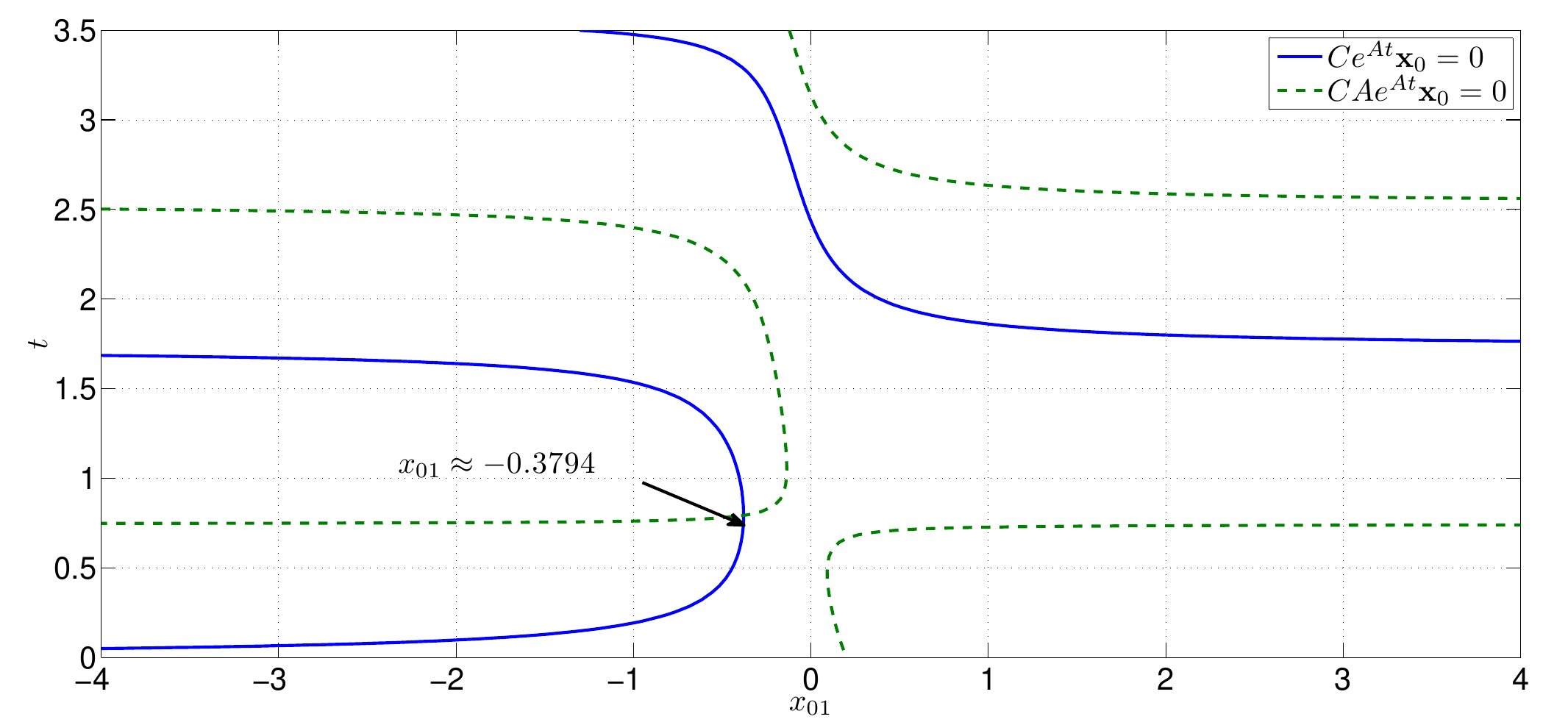}
\caption{Zero-level curves and tangential crossing at $x_{01} \approx -0.3794$.}
\label{t}
\end{figure}

\begin{figure}[t]
\centering
{\includegraphics[scale=0.4]{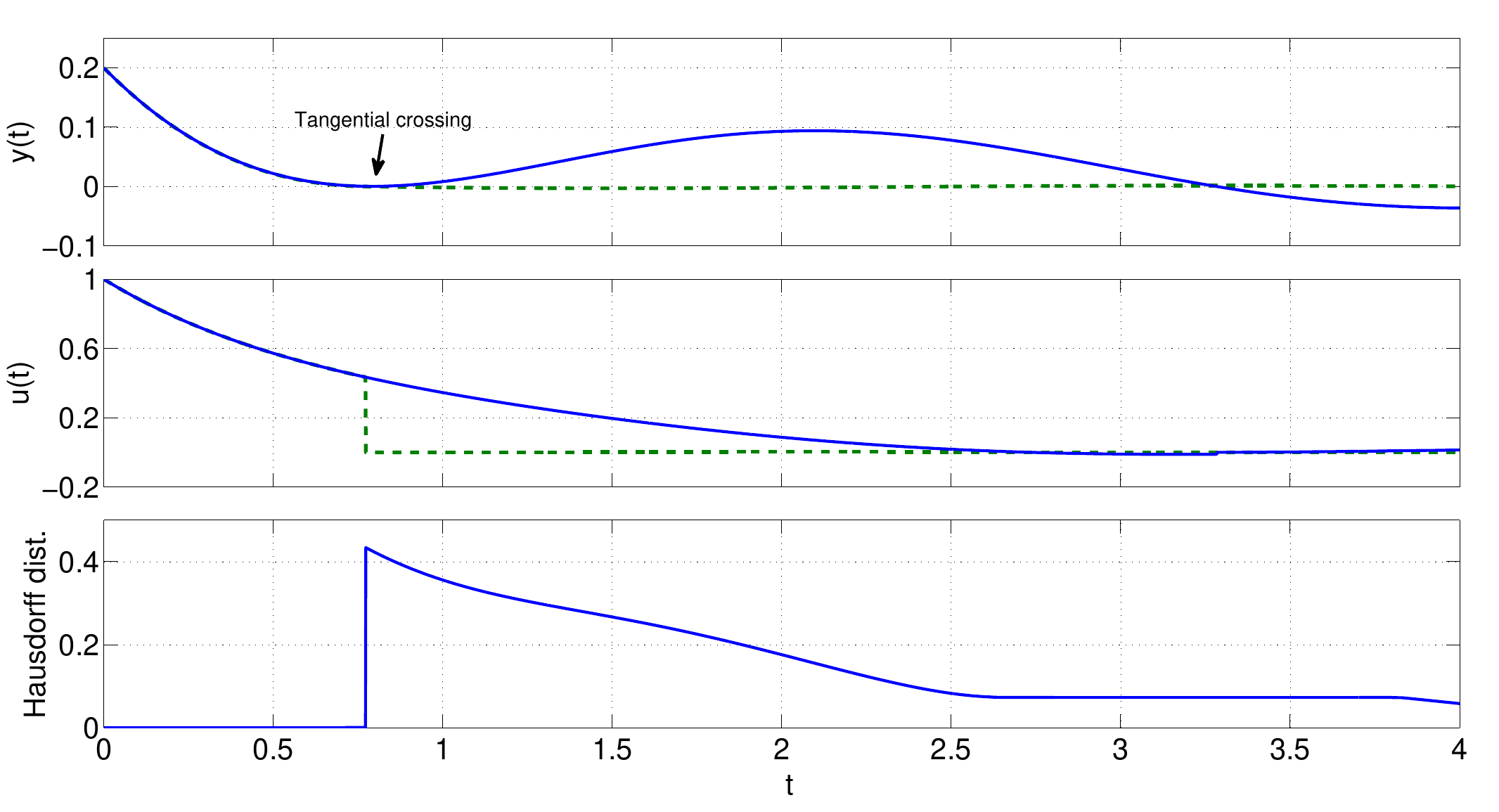}}  
\caption{Time simulation and Hausdorff distance: (dotted) tangential crossing at $t_1\approx 0.7926$, (solid) no crossing at $t_1\approx 0.7926$. }
\label{t2}
\end{figure}

\subsection{Crossing instants versus reset instants}
Besides reset instants, crossings of $\mathcal{H}_C$ ($= \bar{\mathcal{M}}$) play an important role in the  continuous dependence analysis of reset control systems.

{\bf Definition IV.2 (well-posed crossing instants)}: {\em A reset control system $(A,C,n_\rho)$ has {well-posed crossing instants} if for any $\mathbf{{x}}_0 \in \mathds{R}^n$ there exists a sequence $\bar{\mathds{T}}:\{1,2,\cdots\} \rightarrow \mathds{R}^+ \cup \{\infty\}$, denoted by $\bar{\mathds{T}}= (\bar{t}_1,\bar{t}_2,\cdots)$,  and given by (being  $\bar{t}_0 = 0$):

\begin{itemize}
\item 
$\small
	\bar{t}_1 = \left \{ 
		\begin{array}{ccc}
		0  &,\mathbf{{x}}_{0} \in \bar{\mathcal{M}}    \\
		\min \{ \Delta \in \mathds{R}^+ : e^{A\Delta}\mathbf{{x}}_{0} \in \bar{\mathcal{M}} \}  &,\mathbf{{x}}_{0}  \in \mathds{R}^n\setminus\ \bar{\mathcal{M}} 
  	\end{array} \right. \\
\nonumber
\nonumber
$

\item 
\begin{algorithmic}
\STATE $i = 1$
\WHILE {$\bar{t}_i \neq \infty$}  
	\IF{$\mathbf{x}_{i-1} \in \mathcal{M}_{RU}$} 
	\STATE $\bar{t}_{i+1} = \infty$
	\ELSE
	\STATE $
		\begin{array}{l}
 		 	\mathbf{{x}}_i = A_Re^{A(\bar{t}_{i}-\bar{t}_{i-1})}\mathbf{{x}}_{i-1} \\
			\bar{t}_{i+1} = \bar{t}_{i} + \min \{ \Delta >0 : e^{A\Delta}\mathbf{{x}}_{i} \in 									\bar{\mathcal{M}} \}\\
			i \leftarrow i+1\\
			N = i
		\end{array}$
	\ENDIF
\ENDWHILE
\end{algorithmic}
\end{itemize}
}

\vspace{0.125cm}

Note that $\bar{\mathds{T}} = (\infty)$ corresponds to a solution without crossings of $\bar{\mathcal{M}}$, $\bar{\mathds{T}} = (\bar{t}_1,\infty)$ to a solution with a unique crossing at the instant $\bar{t}_1$, $\cdots$. Note that since $\mathcal{M} \subset \bar{\mathcal{M}}$ then $\mathds{T}$ is a subsequence of $\bar{\mathds{T}}$; and that $\bar{\mathds{T}} = (0,\infty)$  for $\mathbf{x}_0 \in \mathcal{M}_{RU} \subset \bar{\mathcal{M}}$. 

{\em Example IV.3}: The reset control system of Example IV.1 (Fig. 8) has the crossing instants sequence $\bar{\mathds{T}} = \mathds{T} = ({t_1},\infty)$, with $\bar{t}_1 = t_1$, for $\mathbf{x}_0 \neq 0$; and $\bar{\mathds{T}} = (0,\infty)$, $\mathds{T} = (\infty)$, for $\mathbf{x}_0 = \mathbf{0}$ (note that $\mathcal{M}_{RU} = \{\mathbf{0}\}$).  

{\em Example IV.4}: For the reset control system of Example IV.2,  consider an initial condition $(1,x_{02},0)$, where $x_{02} \in \mathds{R}$. 
In Fig. \ref{t3}, the solutions (crossing instants) of $Ce^{A\bar{t}}(1,x_{02},0) =0$ for $\bar{t}$ has been plotted vs. $x_{02}$. To decide if a crossing instant is also a reset instant it has to be checked if the state corresponding to that instant belong to $\mathcal{M}_R$ or not. For $x_{02} = 0$, it result that 
$\bar{t}_1 = 0$, $\bar{t}_2 = t_1$, $\bar{t}_3 = t_2$, $\cdots$; and for $x_{20}<0$, 
$\bar{t}_1^\ast =t_1^\ast$, $\bar{t}_2^\ast =t_2^\ast$, $\cdots$ ($t_1$, $t_2$, ${t}_1^\ast$, and ${t}_2^\ast$ are shown in Fig.11).
\begin{figure}[t]
\centering
\includegraphics[scale=0.6]{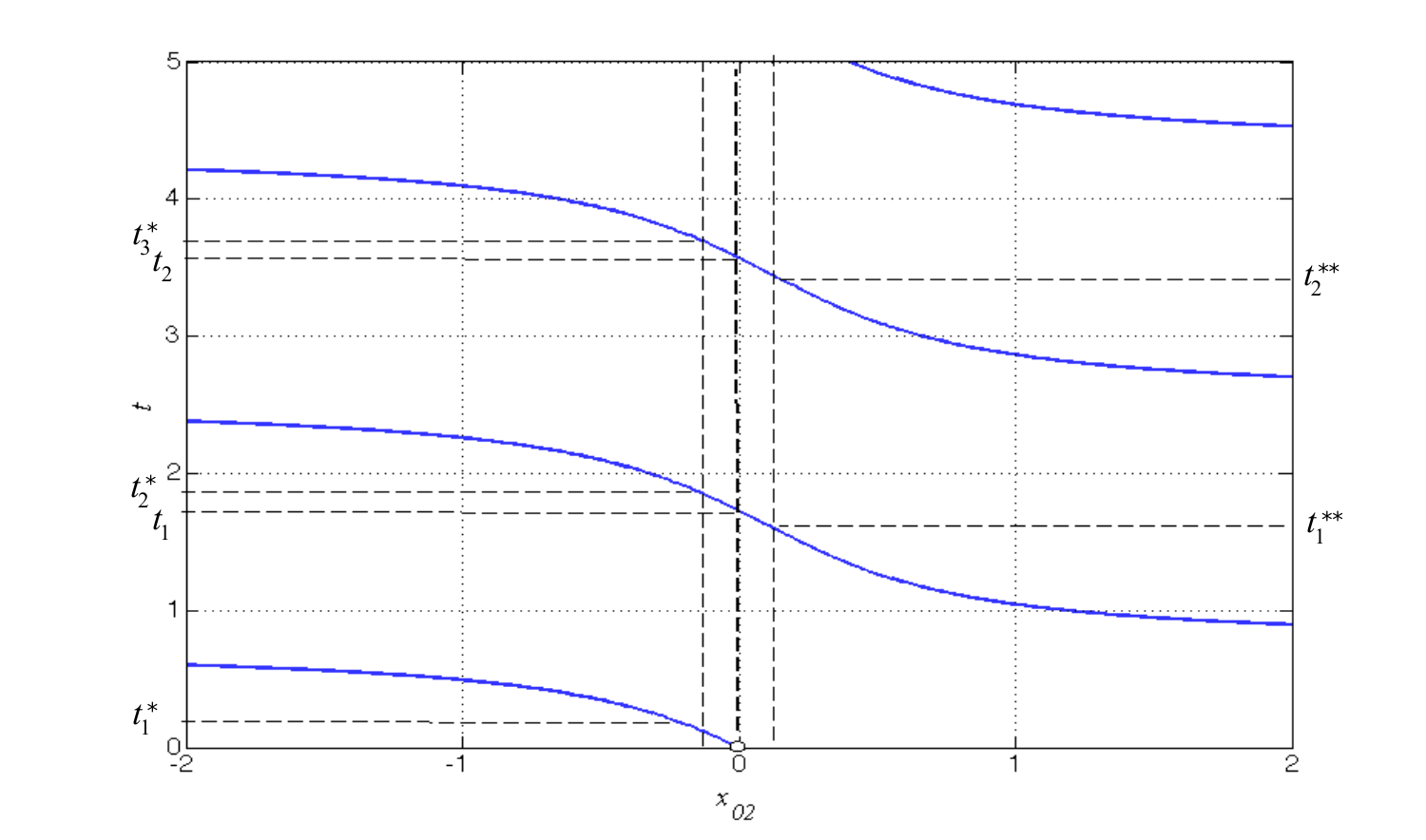}
\caption{Zero-level curves and reset instants: ($x_{02} = 0$) $\mathds{T} = (t_1,t_2,\cdots)$,  ($x_{02} < 0$) $\mathds{T}^\ast = (t_1^\ast,t_2\ast,\cdots)$, ($x_{02} > 0$) $\mathds{T}^{\ast \ast} = (t_1^{\ast \ast},t_2^{\ast \ast},\cdots)$.}\label{t3}
\end{figure}

{\bf Proposition IV.1}: {\em A reset control system $(A,C,n_\rho)$ with well-posed reset instants has well-posed crossing instants, that is there exists $\bar{\mathds{T}} = (\bar{t}_1,\bar{t}_2, \cdots)$ for any $\mathbf{x}_0\in \mathds{R}^n$, and in addition $0 \leq \bar{t}_1 < \bar{t}_2 < \cdots$, and $\bar{t}_k \rightarrow \infty$ as $k \rightarrow \infty$ if the sequence is infinite.}

\vspace{0.125cm}
{\bf Proof}: It follows similar arguments to proof of Prop. III.2-III.3 and is omitted by brevity. 
$\Box$

\vspace{0.125 cm}

Besides functions $\tau_i$, $i=1,2,\cdots$ mapping $\mathbf{x}_0$ to the $i^{th}$ reset instant,  functions $\bar{\tau_i}:\mathds{R}^n \rightarrow \mathds{R}^+$ may be defined such as $\bar{t}_i = \bar{\tau}_i(\mathbf{x}_0)$, $i=1,2,\cdots$. Even in simple cases, like the third order reset control system of Example IV.2,  it is true that they have jump discontinuities. In that example (see Fig. 11), for an $x_{20}$ arbitrary close to zero and $x_{20}<0$, note that $\mathds{T} = (t_1^\ast,t_2^\ast, t_3^\ast, \cdots)$, where $t_1^\ast >0 $  is arbitrarily close to $0$, $t_2^\ast$  is arbitrarily close to $t_1$,  $t_3^\ast$  is arbitrarily close to $t_2$, $\cdots$; and for $x_{20}>0$, $\mathds{T} = (t_1^{\ast \ast},t_2^{\ast \ast},\cdots)$ (see Fig. 11), with $t_1^{\ast \ast}$ arbitrarily close to $t_1$,  $t_2^{\ast \ast}$ arbitrarily close to $t_2$, $\cdots$). Thus any $\tau_i$, $i = 1,2, \cdots$ is discontinuous at $(1,0,0)$. Obviously, since the reset instants sequence is a subsequence of the crossing instants sequence, then  this is also the case for $\bar{\tau}_i$. In spite of the lack of regularity of the reset instants pattern, some simple properties of the crossing instants sequences have been discovered. These properties will be key  to derive a sufficient condition for continuous dependence on the initial condition.

\vspace{0.125cm}

{\bf Proposition IV.2}: {\em Let $(A,C,n_\rho)$ be a a reset control system with well-posed reset instants and $\mathbf{x}_0 \in \mathds{R}^n\setminus \bar{\mathcal{M}}$. If 
$\bar{t}_{1}< \infty$ and $CA\mathbf{x}\left(  \bar{t}_{1},\mathbf{x}_{0}\right)  \neq0$ then 
$\bar{\tau}_{1}$ is continuous at $\mathbf{x}_{0}$.}
 
{\bf Proof}: It is a direct consequence of the implicit function theorem. Consider the continuously differentiable function $f:\mathds{R}^n \times \mathds{R} \rightarrow \mathds{R}$ given by $f(\mathbf{x},t) = Ce^{At}\mathbf{x}$, where $f(\mathbf{x}_0,t_1)= 0$ and in addition $\left ( \frac{\partial f}{\partial t} \right )(\mathbf{x}_0,t_1) = CAe^{At_1}\mathbf{x}_0 = CA\mathbf{x}(t_1,\mathbf{x}_0) \neq 0$. Then, there exists open sets $U \subset \mathds{R}^n$ and $I \subset{\mathds{R}}$, with $\mathbf{x}_0 \in U$ and $t_1 \in I$,  and a (unique) continuously differentiable function $\bar{\tau}_1:U \rightarrow I$ such as $0=f(\mathbf{x}_0,\bar{\tau}_1(\mathbf{x}_0))= Ce^{A\bar{\tau}_1(\mathbf{x}_0)}\mathbf{x}_0$. $\Box$

\vspace{0.125cm}

{\bf Proposition IV.3}: {\em Let $(A,C,n_\rho)$ be a a reset control system with well-posed reset instants and $\mathbf{x}_0 \in \mathcal{M}_{R}\setminus \mathcal{M}_{RU}$ (and thus $\bar{t}_{1}=0$). If 
$\bar{t}_{2}< \infty$ and $CA\mathbf{x}\left(  \bar{t}_{k},\mathbf{x}_{0}\right)  \neq0$, $k = 1,2$ then 
 there exist $d_{0},d_{I}>0$ such that $\mathbf{x}^{\ast}\left(  \left[  0,\bar{t}_2+d_I \right]  ,\mathbf{x}_{0}^{\ast}\right)  $, with $\left\Vert \mathbf{x}_{0}-\mathbf{x}_{0}^{\ast}\right\Vert <d_{0}$,
intersects\ the set $\mathcal{\bar{M}}$ either:
\begin{itemize}
\item at the instant $\bar{t_1}^\ast \in (\bar{t_2}-d_I,\bar{t}_2 + d_I)$, and for any $\alpha >0$ there exists some $\delta >0$ such as if $\|\mathbf{x}_0^\ast - \mathbf{x}_0\|<\delta$ then  $|\bar{t}_1^\ast - \bar{t}_2| < \alpha$,  or
 \item at the instants $\bar{t_1}^\ast \in [0,d_I)$ and $\bar{t}_2^\ast \in (\bar{t_2}-d_I,\bar{t}_2 + d_I)$, and for any $\alpha >0$ there exists some $\delta >0$ such as if $\|\mathbf{x}_0^\ast - \mathbf{x}_0\|<\delta$ then $\bar{t}_1^\ast < \alpha$ and $|\bar{t}_2^\ast - \bar{t}_2| < \alpha$.
\end{itemize}}

{\bf Proof}: Consider again the continuously differentiable function $f$ such as $f\left(\mathbf{x},t\right):=Ce^{At}\mathbf{x}$, and thus $f\left(\mathbf{x}_{0},0\right)  =f\left( 
\mathbf{x}_{0}, \bar{t}_{2}\right)  =0$.  Assume that $f\left(\mathbf{x}_{0},t\right)  >0$ for all $t\in\left( 0,\bar{t}_{2}\right)$, otherwise a similar case is obtained.
In addition, it is possible to find constants $\epsilon>0$, and $\alpha
_{1}\left(  \epsilon\right), \alpha_{2}\left(  \epsilon\right) > 0  $ such that:

\noindent (i) $f\left(\mathbf{x}_0, \alpha_{1}\right)  = f\left(\mathbf{x}_0, \bar{t}%
_{2}-\alpha_{2}\right)  =\epsilon$ with $\alpha_{1}+\alpha_{2}<\bar{t}_{2}$.

\noindent (ii) $f\left(\mathbf{x}_0, t\right)$ is monotone increasing with respect to $t \in\left(0,\alpha_{1}\right)$  and
monotone decreasing with respect to $t \in \left(  \bar{t}_{2}-\alpha_{2},\bar{t}_{2}\right)$.

\noindent (iii) $f\left(\mathbf{x}_0,t\right) >\epsilon$ for all $t\in\left(\alpha_{1},\bar{t}_{2}-\alpha_{2}\right)  $.

Also, since $f$ is continuous at $(\mathbf{x}_0,t) \in \mathds{R}^n \times \mathds{R}^+$ then there exists $\delta\left(  \epsilon,\bar{t}_{2}\right)  $
such that for all $\mathbf{x}_{0}^{\ast}\in \mathds{B}\left(  \mathbf{x}_{0},\delta\right)  $ and
$t\in\left(\alpha_{1},\bar{t}_{2}-\alpha_{2}\right)  $: %
$
0<f\left(\mathbf{x}_0,t \right)  -\epsilon<f\left(\mathbf{x}_0^\ast,t \right)  <f\left(
\mathbf{x}_0,t \right)  +\epsilon.
$
Now, since $\frac{\partial f}{\partial t}(\mathbf{x}(\bar{t_2},\mathbf{x}_0),\bar{t_2}) = CA\mathbf{x}(\bar{t_2},\mathbf{x}_0) \neq 0$ then, in virtue of the
implicit function theorem, there exist constants $\delta_{1}>0$ and $\alpha>0$
such that $\mathbf{x}^\ast\left(\left(  \bar{t}_{2}-\alpha,\bar{t}_{2}+\alpha\right),\mathbf{x}_{0}^{\ast}\right)  $ intersects $\mathcal{\bar{M}}$ for any $\mathbf{x}_{0}^{\ast}\in \mathds{B}\left(  \mathbf{x}_{0},\delta_{1}\right)  $, and as a consequence.
$\bar{t}_{1}^{\ast}<\bar{t}_{2}+\alpha$

Property (i) can be rewritten by defining a constant $\alpha_{3}\left(
\epsilon\right)  $ such that $f\left(\mathbf{x}_0,\bar{t}_{2}+\alpha_{3}\right)
=-\epsilon$. 
From (i), (ii) and (iii) it is concluded that
$\bar{t}_{1}^{\ast}\in\left(0,\alpha_{1}\right)
{\cup}\left(  \bar{t}_{2}-\alpha_{2},\bar{t}_{2}+\alpha
_{3}\right)  $. Also due to $\lim_{\epsilon\rightarrow0}A_{R}e^{At_{1}^{\ast}%
}\mathbf{x}_{0}=\mathbf{x}_{0}$, it is possible to find $\epsilon_{0}$ such that for all
$\epsilon<\epsilon_{0}$, $\left\Vert A_{R}e^{At_{1}^{\ast}}\mathbf{x}_{0}%
-\mathbf{x}_{0}\right\Vert <\delta$ and then $\bar{t}%
_{2}^{\ast}\in\left(0,\alpha_{1}\right)
\overset{\cdot}{\cup}\left(  \bar{t}_{2}-\alpha_{2},\bar{t}_{2}+\alpha
_{3}\right)  $. 

Now, it is proven that if $\bar{t}_{1}^{\ast}\in\left(0,\alpha_{1}\right)  $ then $\bar{t}_{2}^{\ast}\notin\left(0,\alpha_{1}\right)  $. By contradiction, 
assume that $\bar{t}_{1}^{\ast}\in\left(0,\alpha_{1}\right)$ and $\bar{t}_{2}^{\ast}\in\left(0,\alpha_{1}\right)$, then in virtue of the
Rolle theorem there exists $s\in\left(0,\bar{t}%
_{2}\right)  $ such that 
 $\frac{\partial f}{\partial t}(\mathbf{x}(s,\mathbf{x}_0),s) = 0$.
It is obvious that $\mathbf{x}_{0}^{\ast}\rightarrow \mathbf{x}_{0}$ as $\epsilon\rightarrow 0$ and then from (ii) $\alpha_1  \rightarrow 0$ as $\epsilon \rightarrow 0$ , which
means that $s \rightarrow 0$ as $\epsilon\rightarrow 0$ and finally $CA\mathbf{x}_{0}=0$, which is in contradiction with one of the assumptions in the Proposition.

Finally, (ii) implies that $\alpha_{k} \rightarrow 0$ as $\epsilon\rightarrow 0$, $k=1,2$. For an arbitrary $\alpha>0$ there exists
 $\epsilon_{1}<\epsilon_{0}$ such that $\alpha_{k}\left(  \epsilon\right)
<\alpha$ for $\epsilon<\epsilon_{1}$ and such that (iii) holds. Then the result follows immediately since
$\bar{t}_{1}^{\ast}\in\left(  0,\alpha\right)  {\cup}\left(
\bar{t}_{2}-\alpha,\bar{t}_{2}+\alpha\right)  $, and if $\bar{t}_{1}^{\ast}%
\in\left(  0,\alpha\right)  $ then $\bar{t}_{2}^{\ast}\in\left( \bar{t}%
_{2}-\alpha,\bar{t}_{2}+\alpha\right)  $. In particular there exist constants
$d_{I}=\alpha$ and $d_{0}=\delta\left(  \epsilon\right)  $ satisfying the Proposition.
 $\Box$
 
\vspace{0.125cm}

\subsection{ A sufficient condition for continuous dependence on the initial condition}

Since continuous dependence is not possible to obtain for any arbitrary initial condition (for example, initial conditions in the reset set $\mathcal{M}$), the problem is to characterize the set $D$ of initial conditions that have the property. As a result, $D \subset \mathds{R}^n\setminus \mathcal{M}$. In addition,
initial conditions that produce tangential crossings (see Example IV.2) must be excluded. 
 

In the following, several basic results to be used in the next Proposition will be derived. Firstly, consider a solution $\mathbf{x}(t)$ on $[s_0,s_1]$, such that $t_k \notin [s_0,s_1)$, $k=1,2,\cdots$,  and with $\|\mathbf{x}(s_0)\| < M$ for some constant $M>0$, then:
\begin{itemize}
\item $\|\mathbf{x}(t) - \mathbf{x}(s_0)\| \leq \|e^{A(t -s_0)}\mathbf{x}(s_0)-\mathbf{x}(s_0)\| \leq (e^{L(s_1-s_0)}-1)áM$, where $L$ is the spectral norm of $A$. Thus, $\beta_1(\cdot,M):\mathds{R}^+ \rightarrow \mathds{R}^+$ is defined as 
\begin{equation}
 \mathcal{\beta}_1(\epsilon,M) := \frac{1}{L}\text{ln}(1+\frac{\epsilon}{M})
\end{equation}
and $s_1-s_0 < \alpha = \mathcal{\beta}_1(\epsilon,M) \Rightarrow \|\mathbf{x}(t) - \mathbf{x}(s_0)\| <\epsilon$ for $t \in [s_0,s_1]$.
%
\item (CDBS property)  $\|\mathbf{x}^\ast(t) - \mathbf{x}(t)\| \leq e^{L(s_1-s_0)} \|\mathbf{x}^\ast(s_0) - \mathbf{x}(s_0)\|$. $\beta_2(\cdot,s_0,s_1):\mathds{R}^+ \rightarrow \mathds{R}^+$ is defined as 
\begin{equation}
\beta_2(\epsilon,s_0,s_1) := \frac{\epsilon}{e^{L(s_1-s_0)}}  
\end{equation}
Thus, if $\|\mathbf{x}^\ast(s_0) - \mathbf{x}(s_0)\| < \delta = \beta_2(\epsilon,s_0,s_1)$ then $\|\mathbf{x}^\ast(t) - \mathbf{x}(t)\| \leq \epsilon$ for $t \in [s_0,s_1]$
\end{itemize}

In addition, if conditions of Prop. IV.2 are satisfied then it is possible to make $|t_1^\ast -t_1| < \alpha$ by doing  $\|\mathbf{x}^\ast_0 - \mathbf{x}_0\| \leq \delta$, and $\delta >0$ small enough. Thus, a map $\beta_3:[0,d_I) \rightarrow \mathds{R}^+$ is defined such as $\delta = \beta_3(\alpha)$ is the largest $\delta$ with that property. And finally, if Prop. IV. 3 applies then by doing $\|\mathbf{x}^\ast_0 - \mathbf{x}_0\| \leq \delta$, and $\delta >0$ small enough, in the cases there are intersections of $\bar{\mathcal{M}}$ either $|\bar{t}_1^\ast - \bar{t}_2| < \alpha$ or  $\bar{t}_1^\ast < \alpha$ and $|\bar{t}_2^\ast - \bar{t}_2| < \alpha$; now, a map $\beta_4:[0,d_I) \rightarrow \mathds{R}^+$ is defined such as $\delta = \beta_4(\alpha)$ is the largest $\delta$ with that property.
\\
\\

{\bf Proposition IV.4}: {\em For a reset control system $(A,C,n_\rho)$ with well-posed reset instants, the solution depends continuously on the initial condition at $\mathbf{x}_0 \in \mathds{R}^n\setminus \mathcal{M}$ if $CA\mathbf{x}(\bar{t}_k,\mathbf{x}_0) \neq 0$, $k=1,2, \cdots$. }


\vspace{0.125cm}
{\bf Proof}: Since $(A,C,n_\rho)$ has well-posed reset instants and Prop. IV.1 applies, then there exists reset and crossing instants sequences ${\mathds T} = (t_i)$ and $\bar{\mathds T} = (\bar{t}_i)$, $i=1,2\cdots$, with $0 \leq t_1 < t_2 < \cdots$ and $0 \leq \bar{t}_1 < \bar{t}_2 < \cdots$, and also there exists a unique solution $\mathbf{x}(t)$ for any $t\geq0$.
Two cases, $\mathbf{x}_0 \notin \mathcal{M}_R$ (that is $\mathbf{x}_0 \notin \bar{\mathcal{M}}$ since $\mathbf{x}_0 \in \mathds{R}^n\setminus \mathcal{M}$ by asumption) and $\mathbf{x}_0 \in \mathcal{M}_R$, will be separately treated. In any case, there exist a constant $M_0>0$ such as $\|\mathbf{x}_0\| \leq M_0$.

{\em Case A} ($\mathbf{x}_0  \notin \mathcal{M}_R$). It will be assumed that $\bar{t}_1 < \infty$, otherwise the result directly follows from the CDBS property.
If $A_R\mathbf{x}(\bar{t}_{1}) \notin \mathcal{M}_{RU}$, then ${\mathbf x}({t}) \notin \bar{\mathcal{M}}$ for any ${t}\in (\bar{t}_1,\bar{t}_2)$ 
and in this case the proof will be based on checking if $d_H(\mathbf{x}^\ast([0,T]),\mathbf{x}([0,T]))$ is arbitrarily small 
when $\mathbf{x}^\ast_0$ and $\mathbf{x}_0$ are arbitrarily close, for $0 \leq T < \bar{t}_{12}$ and some intermediate instant $\bar{t}_{12} \in (\bar{t}_1,\bar{t}_2)$.
 If $\bar{t}_2 = \infty$ then the proof will be finished; otherwise, note that since ${\mathbf x}(\bar{t}_{12}) \notin \mathcal{M}_R$ (in fact ${\mathbf x}(\bar{t}_{12}) \notin \bar{\mathcal{M}}$) then ${\mathbf x}(\bar{t}_{12})$ can be redefined as a new initial condition belonging to this Case A, and thus a similar argument may be used to analyze $d_H(\mathbf{x}^\ast([0,T]),\mathbf{x}([0,T]))$ for $\bar{t}_{12} \leq T < \bar{t}_{23}$, and some $\bar{t}_{23} \in (\bar{t}_2,\bar{t}_3)$, etc. 
 The cases  ${\mathbf x}(\bar{t}_1) \in \mathcal{M}$ and  ${\mathbf x}(\bar{t}_1) \notin \mathcal{M}$ (that is ${\mathbf x}(\bar{t}_1) \in \mathcal{M}_R$), will be separately treated as Cases A.1 and A.2. Finally, if $A_R\mathbf{x}(\bar{t}_{1}) \in \mathcal{M}_{RU}$ then $\mathbf{x}({t}) \in \mathcal{M}_{RU}$ for any $t\in(\bar{t}_1,\bar{t}_2) = (\bar{t}_1,\infty)$, since $\mathcal{M}_{RU}$ is $A$-invariant by well-posedness of the reset instants; this will be the Case A.3.

 {\em Case A.1} (${\mathbf x}_0 \notin \mathcal{M}_R$, ${\mathbf x}(\bar{t}_1) \in \mathcal{M}$, $A_R{\mathbf x}(\bar{t}_1) \notin \mathcal{M}_{RU}$): Here $0 < \bar{t}_1 = t_1 < \bar{t}_{2}$, and by Prop. IV.2 it is true that the perturbed trajectory ${\mathbf x}^\ast([0,{t}_1+d_I],{\mathbf x}_0^\ast)$ intersects the set ${\mathcal{M}}$ at some instant $\bar{t}_1^\ast={t}_1^\ast \in ({t}_1 -\alpha, {t}_1+\alpha)$, where $\alpha \in [0,d_I)$ for $\|\mathbf{x}_0^\ast - \mathbf{x}_0\| < \delta$ and $\delta = \beta_3(\alpha)$. 
Now,  $d_H(\mathbf{x}^\ast([0,T],\mathbf{x}^\ast_0), \mathbf{x}([0,T],\mathbf{x}_0))$ is checked for $T \in [0,\bar{t}_2)$ and $T\neq t_1$.
\\
{\em -- Case A.1.1}: For $0 \leq T< {t}_1 (= \bar{t}_1)$ (Fig \ref{fig:caso1}.a), choose $\alpha > 0$ such as $T < {t}_1 -\alpha$, meaning that the solution and any perturbed solution intersects the reset set after the instant $T$, and all the trajectories are given by the base system. Now, for $\delta = \min \{\beta_2(\epsilon,0,T),\beta_3(\alpha) \}$ it is true that $\|\mathbf{x}^\ast(t) - \mathbf{x}(t)\| \leq \epsilon$ for $t \in [0,T]$, and thus it directly follows that $\|\mathbf{x}_0-\mathbf{x}_0^\ast\|< \delta \Rightarrow d_H(\mathbf{x}^\ast([0,T],\mathbf{x}^\ast_0), \mathbf{x}([0,T],\mathbf{x}_0)) < \epsilon$. 
\\
{\em -- Case A.1.2}: For ${t}_1 < T < \bar{t}_{12}$ (Fig. \ref{fig:caso1}.b), choose $\alpha_1>0$ such as ${t}_1 + \alpha_1 < T$, and thus any perturbed trajectory also intersects the reset set at an instant ${t}_1^\ast < T$.  Now, assume that ${t}_1 <  t_1^\ast$ (otherwise a similar reasoning may be applied), thus ${t}_1 < {t}_1^\ast < T$. Firstly, distances between trajectories and perturbed trajectories will be bounded in some intervals, then these bounds will be used to obtain bounds of the Hausdorff distance.
\begin{itemize}
 \item $0 \leq {t} \leq {t}_1$: Directly, by the CDBS property $\|\mathbf{x}_0^\ast - \mathbf{x}_0\| \leq \delta_1 = \min \{\beta_2(\epsilon_1,0,t_1),\beta_3(\alpha_1)\} \Rightarrow  \|\mathbf{x}^\ast(t) - \mathbf{x}(t)\| \leq \epsilon_1$. In addition, $\|\mathbf{x}(t_1)\| = \|e^{At_1}\mathbf{x}_0\| \leq e^{Lt_1}M_0 =:M_1$, and $\|\mathbf{x}^\ast(t_1)\| \leq \|\mathbf{x}^\ast(t_1) - \mathbf{x}(t_1)\| + \|\mathbf{x}(t_1)\| \leq \epsilon_1 + M_1 =: M_1^\ast$.
\item ${t}_1 < {t} \leq {t}_1^\ast$: By making $\delta_2 = \min\{\beta_3(\beta_1(\frac{\epsilon_2}{2},M_1^\ast)),\delta_1\}$ it is true that $\|\mathbf{x}_0^\ast - \mathbf{x}_0\| \leq \delta_2 \Rightarrow$
\begin{equation}
\|\mathbf{x}^\ast(t) - \mathbf{x}({t}_1)\| \leq \|\mathbf{x}^\ast(t) - \mathbf{x}^\ast({t}_1)\| +
\nonumber
\end{equation}
\begin{equation}
+\|\mathbf{x}^\ast({t}_1) - \mathbf{x}({t}_1)\| \leq \frac{\epsilon_2}{2} + \epsilon_1 \leq \epsilon_2
\end{equation}
\noindent by doing $\epsilon_1 \leq \epsilon_2/2$. 
%
%
In addition, $\|\mathbf{x}_0^\ast - \mathbf{x}_0\| \leq \delta_3 = \min\{\beta_3(\beta_1(\frac{\epsilon_3}{2},M_1)),\delta_2\} \Rightarrow$
\begin{equation}
\|\mathbf{x}(t) - \mathbf{x}^\ast(t_1^{\ast +})\| \leq \|\mathbf{x}(t)- \mathbf{x}(t_1^+) \| + \| \mathbf{x}(t_1^+)-\mathbf{x}^\ast(t_1^{\ast +})\|
\nonumber
\end{equation}
\begin{equation}
<  \frac{\epsilon_3}{2}+ \|A_R\mathbf{x}({t}_1) - A_R\mathbf{x}^\ast({t}_1^\ast)  \|
\nonumber
\end{equation}
\begin{equation}
< \frac{\epsilon_3}{2} + \|\mathbf{x}({t}_1) - \mathbf{x}^\ast({t}_1^\ast)  \| < \frac{\epsilon_3}{2}+ \epsilon_2 \leq \epsilon_3
\end{equation}
where it has been used the fact that $\|A_R\|=1$, and chosen $\epsilon_2 \leq \epsilon_3/2$. 
\item ${t}_1^\ast < t \leq T$: By doing $\epsilon_3 = \beta_2(\epsilon_4,t_1^{\ast +},T)$, and since $\mathbf{x}(t_1^{\ast +}) = \mathbf{x}(t_1^\ast)$ then by using the CDBS property,  $\|\mathbf{x}^\ast(t_1^{\ast +}) - \mathbf{x}(t_1^{\ast +})\| < \epsilon_3 \Rightarrow \|\mathbf{x}^\ast(t) - \mathbf{x}(t)\| < \epsilon_4$.
\end{itemize}
\begin{figure}[t] 
   \centering
   \includegraphics[scale=0.75]{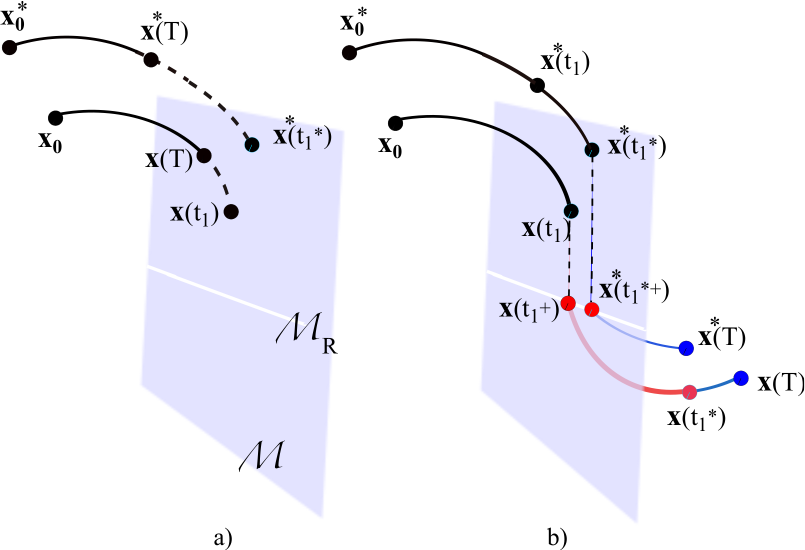}
   \caption{Case A.1 ($t_1 < t_1^\ast$): a) A.1.1: $T < t_1 (< t_1^\ast)$, b) A.1.2: $t_1 (< t_1^\ast)< T < \bar{t}_2$)}
   \label{fig:caso1}
\end{figure}
Now, the Hausdorff distance between the sets $\mathbf{x}^\ast([0,T])$ and $\mathbf{x}([0,T])$ will be bounded using the above bounds. Firstly, 
since
\begin{equation}\small 
d_E(\mathbf{x}^\ast(t),\mathbf{x}([0,T])) \leq \left \{
\begin{aligned}
\small
&d_E(\mathbf{x}^\ast(t),\mathbf{x}([0,t_1])) \leq \epsilon_1, &0 \leq t \leq t_1\\
&d_E(\mathbf{x}^\ast(t),\mathbf{x}([t_1,t_1^\ast])) \leq \epsilon_3 , &t_1 \leq t \leq t_1^\ast\\
&d_E(\mathbf{x}^\ast(t),\mathbf{x}((t_1^\ast,T]))\leq \epsilon_4, &t_1^\ast \leq t \leq T 
\end{aligned}
\right.
\end{equation}
then $\|\mathbf{x}_0^\ast - \mathbf{x}_0\| \leq \delta_3 \Rightarrow$
\begin{equation} 
h(\mathbf{x}([0,T]),\mathbf{x}^\ast([0,T])) \leq \epsilon_4 
\end{equation}
Finally, since 
\begin{equation} \small
d_E(\mathbf{x}(t),\mathbf{x}^\ast([0,T]) \leq \left \{
\begin{aligned}
\small
&d_E(\mathbf{x}(t),\mathbf{x}^\ast([0,t_1]) \leq \epsilon_1, &0 \leq t \leq t_1\\
&d_E(\mathbf{x}(t),\mathbf{x}^\ast((t_1,t_1^\ast]) \leq \epsilon_2, &t_1 \leq t \leq t_1^\ast\\
&d_E(\mathbf{x}(t),\mathbf{x}^\ast((t_1^\ast,T]) \leq \epsilon_4, &t_1^\ast \leq t \leq T 
\end{aligned}
\right.
\end{equation}
then $\|\mathbf{x}_0^\ast - \mathbf{x}_0\| \leq \delta_3 \Rightarrow$
\begin{equation} 
h(\mathbf{x}^\ast([0,T]),\mathbf{x}([0,T])) \leq \epsilon_4 
\end{equation}
and using (12) it is concluded that for a given $\epsilon := \epsilon_4$ there exist $\delta:= \delta_3$ such as
\begin{equation}
\|\mathbf{x}_0^\ast - \mathbf{x}_0\| \leq \delta \Rightarrow d_H(\mathbf{x}^\ast([0,T]),\mathbf{x}([0,T])) \leq \epsilon
\end{equation}
%
%

 {\em Case A.2} (${\mathbf x}_0 \notin \mathcal{M}_R$, ${\mathbf x}(\bar{t}_1) \in \mathcal{M}_R$): Here $0 < \bar{t}_1 < \bar{t}_2 \leq t_1$, and by Prop. IV.2 the perturbed trajectory ${\mathbf x}^\ast([0,\bar{t_1}+d_I),{\mathbf x}_0^\ast)$ intersects the set $\bar{\mathcal{M}}$ at instant $\bar{t}_1^\ast \in (\bar{t}_1 -\alpha, \bar{t}_1+\alpha)$, where $\alpha>0$ is arbitrarily small for $\|\mathbf{x}_0^\ast - \mathbf{x}_0\| < \delta$ and $\delta > 0$ arbitrarily small. 
 \\
{\em -- Case A.2.1}: For $0 < T < \bar{t}_1$, choose $\alpha > 0$ such as $T < \bar{t}_1 -\alpha < \bar{t}_1^\ast$, and thus  $\|\mathbf{x}^\ast_0 - \mathbf{x}_0\| < \delta = \min \{\beta_2(\epsilon,0,\bar{t}_1),\beta_3(\alpha)\} \Rightarrow d_H(\mathbf{x}^\ast([0,T],\mathbf{x}^\ast_0), \mathbf{x}([0,T],\mathbf{x}_0)) < \epsilon$ 
directly follows from the CDBS property.
\\
{\em -- Case A.2.2}: For $\bar{t}_1 \leq T < \bar{t}_{12}$ assume that $\bar{t}_1^\ast < \bar{t}_1$, otherwise a similar reasoning may be applied; thus, 
\begin{itemize}
\item $0 \leq t \leq \bar{t}_1^\ast$: Again choose $\alpha_1 > 0$ such as $T < \bar{t}_1 -\alpha_1 < \bar{t}_1^\ast$ and thus  $\|\mathbf{x}^\ast_0 - \mathbf{x}_0\| < \delta_1 = \min \{\beta_2(\epsilon_1,0,\bar{t}_1),\beta_3(\alpha_1)\} \Rightarrow \|\mathbf{x}^\ast(t) - \mathbf{x}(t)\| \leq \epsilon_1$. In addition, $\|\mathbf{x}(\bar{t}_1^\ast)\| = \|e^{A\bar{t}_1^\ast}\mathbf{x}_0\| \leq e^{L\bar{t}_1^\ast}M_0 \leq e^{L\bar{t}_1}M_0 =:M_1$, and $\|\mathbf{x}^\ast(\bar{t}_1^\ast)\| \leq \|\mathbf{x}^\ast(\bar{t}_1^\ast) - \mathbf{x}(\bar{t}_1^\ast)\| + \|\mathbf{x}(\bar{t}_1^\ast)\| \leq \epsilon_1 + M_1 =: M_1^\ast$.
\item $\bar{t}_1^\ast < t \leq \bar{t}_1$: For some $\epsilon_2 > 0$ and $\epsilon_1 = \epsilon_2/2$, it is true that $\|\mathbf{x}^\ast_0 - \mathbf{x}_0\| < \delta_2 = \min \{\beta_4(\beta_1(\frac{\epsilon_2}{2},M_1)),\delta_1\} \Rightarrow$ 
\begin{equation}
\|\mathbf{x}(t) - \mathbf{x}^\ast(\bar{t}_1^\ast)\| \leq \|\mathbf{x}(t) - \mathbf{x}(\bar{t}_1^\ast)\|
+\|\mathbf{x}(\bar{t}_1^\ast) - \mathbf{x}^\ast(\bar{t}_1^\ast)\|
\nonumber
\end{equation}
\begin{equation}
\leq \frac{\epsilon_2}{2} + {\epsilon_1} = \epsilon_2
\end{equation}
%
In addition, since $\mathbf{x}(\bar{t}_1)\in \mathcal{M}_R$, and thus $\mathbf{x}(\bar{t}_1) = A_R\mathbf{x}(\bar{t}_1)$, and $\|A_R\|=1$, then for some $\epsilon_3 > 0$ and $\epsilon_2 = \epsilon_3/2$, it is true that $\|\mathbf{x}^\ast_0 - \mathbf{x}_0\| < \delta_3 = \min \{\beta_4(\beta_1(\frac{\epsilon_3}{2},M_1^\ast)),\delta_2\} \Rightarrow$ 
\begin{equation}
\|\mathbf{x}^\ast(t) - \mathbf{x}(\bar{t}_1)\| \leq \|\mathbf{x}^\ast(t)- \mathbf{x}^\ast(\bar{t}_1^{\ast+})\|  
\end{equation} 
\begin{equation}
 + \| \mathbf{x}^\ast(\bar{t}_1^{\ast +}) - \mathbf{x}(\bar{t}_1)\| \leq \|\mathbf{x}^\ast(t)- A_R\mathbf{x}^\ast(\bar{t}_1^{\ast})\|  
\nonumber
\end{equation}
\begin{equation}
+ \|A_R \mathbf{x}^\ast(\bar{t}_1^\ast) - A_R\mathbf{x}(\bar{t}_1)\| \leq \frac{\epsilon_3}{2} + \|\mathbf{x}^\ast({t}_1^\ast) - \mathbf{x}(\bar{t}_1)\|
\end{equation} 
\begin{equation}
\leq \frac{\epsilon_3}{2} + \epsilon_2 = \epsilon_3
\nonumber
\end{equation}
\item $\bar{t}_1 < t \leq T < \bar{t}_2$: By the CDBS property,  $\|\mathbf{x}^\ast(\bar{t}_1) - \mathbf{x}(\bar{t}_1)\| < \epsilon_3 = \beta_2(\epsilon_4,\bar{t}_1,T) \Rightarrow \|\mathbf{x}^\ast(t) - \mathbf{x}(t)\| < \epsilon_4$.
\end{itemize}
Finally, making a reasoning similar to Case A.1.2, it again follows that for a given $\epsilon =: \epsilon_4$ there exist some $\delta =: \delta_3$ such as 
\begin{equation}
\|\mathbf{x}_0^\ast - \mathbf{x}_0\| \leq \delta \Rightarrow d_H(\mathbf{x}^\ast([0,T]),\mathbf{x}([0,T])) \leq \epsilon
\end{equation}

{\em Case A.3} (${\mathbf x}_0 \notin \mathcal{M}_R$, ${\mathbf x}(\bar{t}_1) \in \mathcal{M}$, $A_R{\mathbf x}(\bar{t}_1) \in \mathcal{M}_{RU}$): Here $0 < \bar{t}_1 = t_1 < \bar{t}_{2} = \infty$, and again by Prop. IV.2 it results that the perturbed trajectory ${\mathbf x}^\ast([0,{t}_1+d_I],{\mathbf x}_0^\ast)$ intersects the set ${\mathcal{M}}$ at some instant $\bar{t}_1^\ast={t}_1^\ast \in ({t}_1 -\alpha, {t}_1+\alpha)$, where $\alpha \in [0,d_I)$ for $\|\mathbf{x}_0^\ast - \mathbf{x}_0\| < \delta$ and $\delta = \beta_3(\alpha)$. For $T\in [0,t_1)$ the case is identical to Case A.1.1; for $T\in (t_1,\infty)$
the case is similar to Case A.1.2 (assume for example that $t_1 < t^\ast_1 < T$), the difference is that now $t_2 = \infty$ and there may exist (infinitely) many reset instants $t^\ast_k \in (t^\ast_1, \infty)$, $k= 2, 3, \cdots$. In this case, the result follows from application of the CDBS property and the continuity of the map $A_R$ (note that $\|A_R\| = 1$).

\vspace{0.125cm}
\noindent  {\em Case B} (${\mathbf x}_0 \in \mathcal{M}_R$). If ${\mathbf x}_0 \in \mathcal{M}_{RU}$ a case similar to case A.3 is obtained. Otherwise, if ${\mathbf x}_0 \in\mathcal{M}_R\setminus\mathcal{M}_{RU}$ then Prop. IV.3 applies and $d_H(\mathbf{x}^\ast([0,T]),\mathbf{x}([0,T]))$ will be checked to be arbitrarily small, and 
$\mathbf{x}^\ast(T)$ and $\mathbf{x}(T)$ arbitrarily close, when $\mathbf{x}^\ast_0$ and $\mathbf{x}_0$ are arbitrarily close, for $0 = \bar{t}_1 \leq T < \bar{t}_2$. Once again, the argument is that if that property holds for $T \in [0,\bar{t}_{12})$,  where $\bar{t}_{12} \in (0,\bar{t}_2)$, then since ${\mathbf x}(\bar{t}_{12}) \notin \mathcal{M}_R$, in fact ${\mathbf x}(\bar{t}_{12}) \notin \bar{\mathcal{M}}$, then ${\mathbf x}(\bar{t}_{12})$ can be redefined as a new initial condition belonging to Case A, and thus a similar argument may be used to analyze $d_H(\mathbf{x}^\ast([0,T]),\mathbf{x}([0,T]))$ for $\bar{t}_{12} \leq T < \bar{t}_{23}$, and some $\bar{t}_{23} \in (\bar{t}_2,\bar{t}_3)$, etc. 
The proof of this case is somehow sketched since the reasoning is similar to Case A.

By Prop. IV. 3, the perturbed trajectory $\mathbf{x}^\ast([0,\bar{t}_2+d_I),\mathbf{x}_0^\ast)$ either intersects the set $\bar{\mathcal{M}}$ at an instant $\bar{t}_1^\ast$ with $|\bar{t}_1^\ast - \bar{t}_2| < \alpha$ or intersects $\bar{\mathcal{M}}$ at the instants $\bar{t}_1^\ast$ and $\bar{t}_2^\ast$, with $\bar{t}_1^\ast < \alpha$, and $|\bar{t}_2^\ast - \bar{t}_2| < \alpha$. 
Now, for $0< T<\bar{t}_2$, choose $\alpha_1>0$ such as $\alpha_1 < T < \bar{t}_2 - \alpha_1$, and analyze the two possibilities: i) if  $|\bar{t}_1^\ast - \bar{t}_2| < \alpha_1$ then $T < \bar{t}_1^\ast$ and a case similar to Cases A.1.1 and A.2.1 is obtained; that is,  $\|\mathbf{x}^\ast_0 - \mathbf{x}_0\| < \delta_1 = \min \{\beta_2(\epsilon_0,0,T),\beta_4(\alpha_1)\} \Rightarrow \|\mathbf{x}^\ast(t) - \mathbf{x}(t)\| \leq \epsilon_0$; ii) if $\bar{t}_1^\ast < \alpha_1$ and $|\bar{t}_2^\ast - \bar{t}_2| < \alpha_1$ then $0 \leq \bar{t}_1^\ast < T < \bar{t}_2^\ast$:
\begin{itemize}
\item $0 \leq t \leq \bar{t}_1^\ast$: $\|\mathbf{x}^\ast_0 - \mathbf{x}_0\| \leq \delta_1 \Rightarrow \|\mathbf{x}^\ast(t) - \mathbf{x}(t)\| \leq \epsilon_0$.
\item $\bar{t}_1^\ast< t \leq T$: Here $\|\mathbf{x}^\ast(\bar{t}_1^{\ast +}) - \mathbf{x}(\bar{t}_1^{\ast +})\| =  \| A_R\mathbf{x}^\ast(\bar{t}_1^{\ast}) - \mathbf{x}(\bar{t}_1^{\ast})\| \leq \| \mathbf{x}^\ast(\bar{t}_1^{\ast}) - \mathbf{x}(\bar{t}_1^{\ast})\| \leq \epsilon_0 = \beta_2(\epsilon_1,0,T) \Rightarrow \|\mathbf{x}^\ast(t) - \mathbf{x}(t)\| \leq \epsilon_1$
\end{itemize}

As a result, in any case $\|\mathbf{x}^\ast_0 - \mathbf{x}_0\| \leq \delta_1 \Rightarrow \|\mathbf{x}^\ast(t) - \mathbf{x}(t)\| \leq \epsilon_1$, and following a reasoning similar to Case A it results that  $\|\mathbf{x}^\ast_0 - \mathbf{x}_0\| \leq \delta_1 \Rightarrow d_H(\mathbf{x}^\ast([0,T]),\mathbf{x}([0,T])) < \epsilon_1$, for $T \in [0,\bar{t}_2)$.
\\
$\Box$

\subsection{Relaxed conditions for continuous dependence on the initial condition}

In general, it may be hard to check for $\mathbf{x}_0$ the condition $CA\mathbf{x}(\bar{t}_k,\mathbf{x}_0) \neq 0$, $k=1,2, \cdots$, since except in the case of low order reset systems (e. g. Example III.1) crossing/reset instants are hard to compute. Some relaxed conditions that have been found to be useful are developed in the following. Consider the set $\bar{\mathcal{M}}_T$, defined as the set of all points in $\bar{\mathcal{M}}\setminus\mathcal{M}_{RU}$ that produce a tangential crossing: 
\begin{equation}
\bar{\mathcal{M}}_T := \{ \mathbf{x} \in \bar{\mathcal{M}}\setminus \mathcal{M}_{RU} :  \text{  } CA \mathbf{x} = 0 \}
\end{equation}

In addition, a notion of set backward reachability is also needed for the base system. The backward reachable set from $\bar{{\mathcal{M}}}_T$, $\mathcal{B}(\bar{{\mathcal{M}}}_T )$, is defined as
\begin{equation}
\mathcal{B}(\bar{\mathcal{M}}_T ) := \{ \mathbf{x} \in \mathds{R}^n:  \exists t \in \mathds{R}^+, e^{At} \mathbf{x} \in \bar{\mathcal{M}}_T  \}
\end{equation}

\begin{figure}[t] 
   \centering
   \includegraphics[scale=0.8]{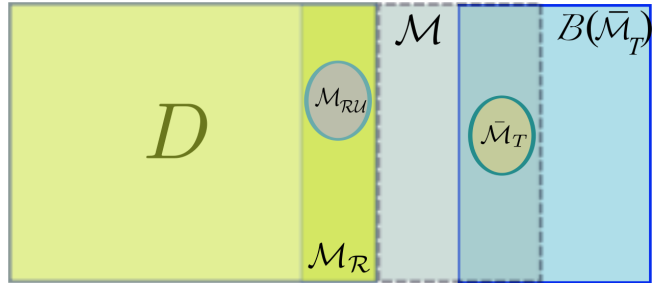} 
   \caption{Set Diagrams for  $\mathcal{M_R}$, $\mathcal{M_{RU}}$, $\mathcal{M}$, $\bar{\mathcal{M}}_T$, $\mathcal{B}(\bar{\mathcal{M}}_T)$, and $D = \mathds{R}^n\setminus(\mathcal{M} \cup\mathcal{B}(\bar{\mathcal{M}}_T))$. }
   \label{fig:conjuntos}
\end{figure}

As result, it is easy to check (see Fig. \ref{fig:conjuntos} for a representation of the involved sets) that $CA\mathbf{x}(\bar{t}_k,\mathbf{x}_0) \neq 0$, $k=1,2, \cdots$ if 

\begin{equation}
\mathbf{x}_0 \in D:=\mathds{R}^n \setminus (\mathcal{M} \cup \mathcal{B}(\bar{\mathcal{M}}_T )  )
\end{equation}
 
\noindent and 
\begin{equation}
\mathcal{M}_R \cap \mathcal{B}(\bar{\mathcal{M}}_T )  ) = \varnothing
\end{equation}

Moreover, a more relaxed and conservative condition, but even easier to evaluate, is simply that 
\begin{equation}
 CA\mathbf{x} \neq 0 \text{ for any } \mathbf{x} \in \bar{\mathcal{M}}
 \end{equation}

\noindent note that in this case $ {{\mathcal M}}_T = \mathcal{B}({{\mathcal M}}_T) =\varnothing$, and thus from (49) it directly follows that $D = \mathds{R}^n \setminus \mathcal{M}$; and, in addition, (50) is trivially satisfied.


{\em Example IV.5}: The reset control system of Example IV.1 depends continuously on the initial condition at $\mathbf{x}_0 \in  \mathds{R}^2\setminus \mathcal{M}$, since  ${\bar{\mathcal M}}_T= {\mathcal N} ( \left ( \begin{array}{cc} 1 &0\\0& 1  \end{array} \right )) = \varnothing$. Note that for second order reset control systems with observable base system, it always turns out that $D =\mathds{R}^2\setminus \mathcal{M}$. 

\vspace{0.125cm}
{\em Example IV.6}: For the reset control system of Example IV.2, $\bar{\mathcal{M}}_T = span\{(0,0,1)\} \setminus \{ {\bf 0} \} $ and thus a reachability analysis is necessary to determine an initial set $D$ for the reset control system to depend continuously on the initial condition. Note that the more relaxed  condition $\bar{\mathcal{M}}_T = \varnothing$ do not apply in this example. Here, the backward reachable set is   
\begin{equation}
\mathcal{B}(\bar{\mathcal M}_T)= \bigcup_{t\geq0} span\{ e^{-At} (0, 0, 1)\}
\end{equation}
\noindent Fig. \ref{fig:surface} shows $\mathcal{B}(\bar{\mathcal M}_T)$, $\bar{\mathcal{M}}_T$ and the after-reset set $\mathcal{M}_R$. It results that $\mathcal{B}(\bar{{\mathcal M}}_T) \cap \mathcal{M}_R = \varnothing$, and then $D = \mathds{R}^3\setminus (\mathcal{M} \cup \mathcal{B}(\bar{{\mathcal M}}_T))$.
Although it is not possible to exactly compute the set $\mathcal{B}(\bar{{\mathcal M}}_T)$, a superset of $\mathcal{B}(\bar{{\mathcal M}}_T)$ may be obtained 
by the union of two polytopes $P$ and $\hat{P}$, that may be computed by the following method: 
for some constant $N>0$, let us define a set of row vectors $\mathbf{n}_i$, $i=1,2,\cdots,N$, using
\begin{equation}
\mathbf{n}_i(\theta_i):=\left( sin(\phi_i)\ \ cos(\phi_i)\ \ cos(\theta_i) \right)
\end{equation}
where $\phi_i=\frac{i \pi }{2 N}$, and $\theta_i$ is a constant to be determined, $i=1,\cdots,N$. Then, the polytope $P$ is given by
\begin{equation}
P=\bigcap^N_{i=1} \left\{ {\bf x} \in \mathds{R}^3 : \mathbf{n}_i(\theta_i) {\bf x} \leq 0 \right\}
\end{equation}
and $\hat{P}$ is similarly defined using $-\mathbf{n}_i$ instead of $\mathbf{n}_i$. In order to achieve
$ \mathcal{B}(\bar{\mathcal{M}}_T) \subset P \cup \hat{P} $ with a tight enclosing, $\theta_i$ is maximized subject to: 
\begin{equation}
\left \{
\begin{array}{l}
 \mathbf{n}_i(\theta_i) e^{-At} \left( \begin{array}{c}0\\0\\1 \end{array} \right)\neq0, \text{ }  \forall t\geq0  \\
 0\leq \theta_i \leq \frac{\pi}{2}
 \end{array}
\right.
\end{equation}
 
\noindent Fig. \ref{fig:surface} shows a solution for $N=64$.
\begin{figure}[t] 
   \centering
   \includegraphics[scale=0.5]{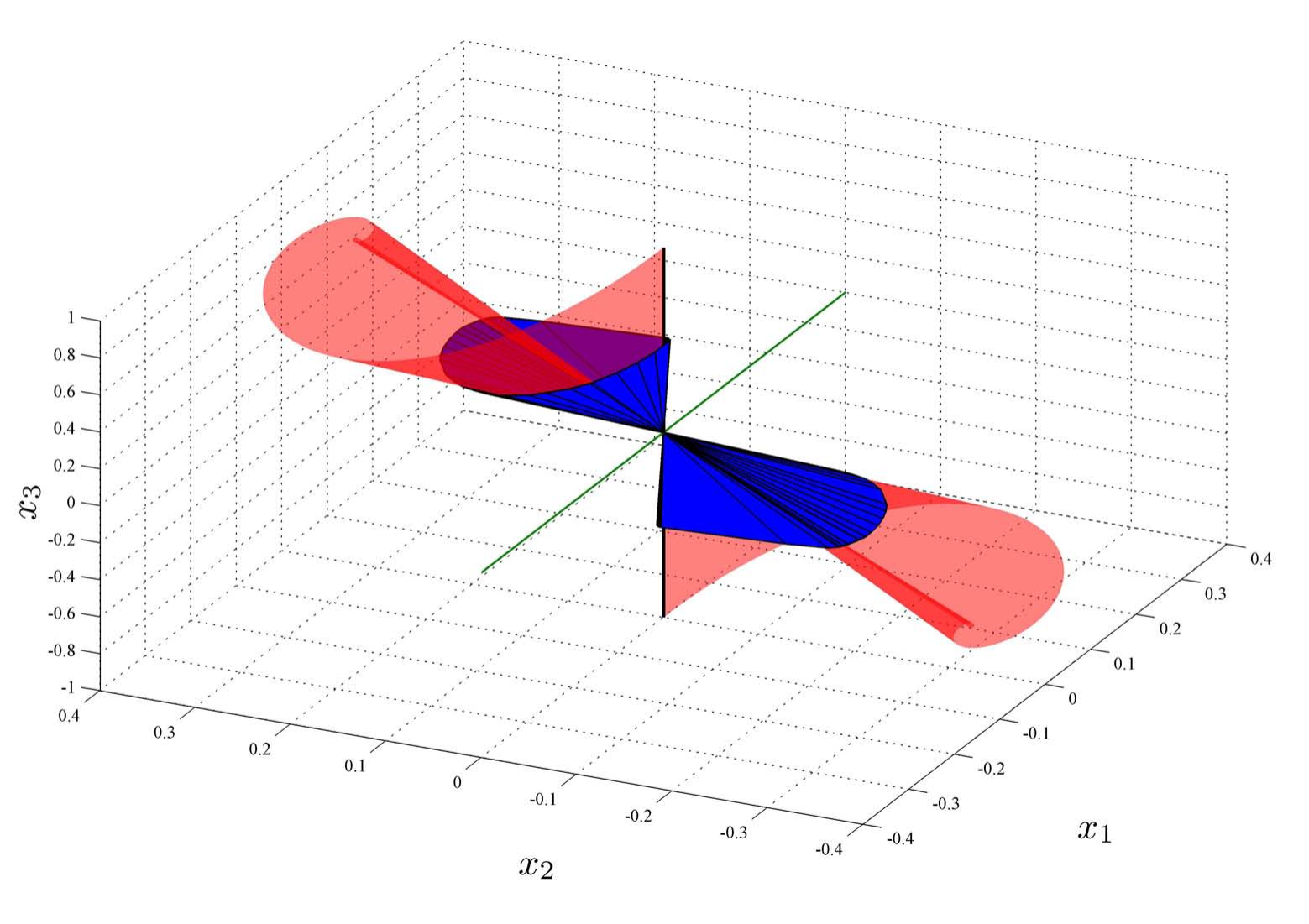} 
   \caption{  $\mathcal{M_R}$ (green), $\bar{\mathcal{M}}_T$ (black), $\mathcal{B}(\bar{\mathcal{M}}_T)$ (red) , enclosing polytopes $P$ and $\hat{P}$ (blue). }
   \label{fig:surface}
\end{figure}

\subsection{Sensitivity of reset control systems to sensor noise}

In control practice, the sensitivity to sensor noise is an important issue. It is expected that a reset control system would produce close closed-loop output responses with and without sensor noise, as  the sensor noise becomes smaller in some sense. In general, for impulsive and hybrid systems this is a hard issue, and has been one of the main motivation for HI framework \cite{GSTbook}. In the following, it will be shown how the property of continuous dependence on the initial condition can be used to analyze sensitivity of a reset control system to sensor noise in the IDS framework, without introducing nondeterminism. It will be assumed that the  sensor noise (as well as the other exogenous signals) is a Bohl function, a not overly restrictive condition in practice. 
 
For a reset control system $(A,C,n_{\rho})$, with $A$ and $C$  given by (21), and with state $\mathbf{x}=(\mathbf{w}_1,\mathbf{w}_2,\mathbf{x}_p,\mathbf{x}_r)$, a perturbed extended state $\mathbf{z_n}$ is defined as $\mathbf{z_n}= (\mathbf{x},\mathbf{n})$, where sensor noise is generated by the exosystem (see Fig. \ref{fig:noise})
\begin{equation}
  \Sigma_n: \left \{
   \begin{array}{llll}
     \mathbf{\dot{n}}(t) &= A_{n}\mathbf{n}(t), \hspace{0.5cm} \mathbf{n}(0)=\mathbf{n}_{0} \\
     n(t) & = C_{n}\mathbf{n}(t) \hspace{0.5cm}  \\
    \end{array}
   \right.
    \label{noise}
\end{equation}
\noindent  where $\mathbf{n} \in \mathds{R}^{m_n}$, $n \in \mathds{R}$, and $A_n, C_n$ are matrices with appropriate dimensions. In this way, a noisy solution $\mathbf{x}^\ast$ is recovered as a projection of  the noisy extended solution $\mathbf{z_n}$ with initial condition $(\mathbf{x}_0,\mathbf{n}_0)$, for some $\mathbf{n}_0 \neq \mathbf{0}$, that is $\mathbf{x}^\ast = \Pi \mathbf{z_n}$.  Moreover, the noisy control system will be referred to as $(A_z,C_z,n_\rho)$, where matrices $A_z$ and $C_z$ are can be easily obtained. And the noise-free solution $\mathbf{x}$ is simply $\mathbf{x}= \Pi \mathbf{z}$, where $\mathbf{z}$ is the extended solution with initial condition $(\mathbf{x}_0,\mathbf{0})$.

\vspace{0.125cm}
\begin{figure}[t] 
   \centering
   \includegraphics[scale=0.85]{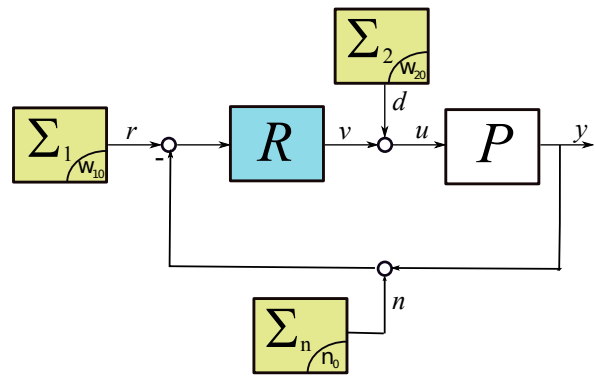} 
   \caption{Reset control system with added sensor noise in the feedback loop.}
   \label{fig:noise}
\end{figure}

{\bf Definition IV.2}: {\em A reset control system $(A,C,n_{\rho})$ with well-posed reset instants, and with initial condition $\mathbf{x}_0$, is not sensitive to noise if for any exosystem  $\Sigma_n$, any 
$ \epsilon > 0$, and for almost any $T > 0$ there exist $\delta > 0$ such that $\|{\mathbf n}_0 \| < \delta \Rightarrow d_H(\Pi\mathbf{z_n}([0,T],(\mathbf{x}_0,\mathbf{n}_0)), \mathbf{x}([0,T],\mathbf{x}_0)) < \epsilon$.}\\

In the following, it will be shown how continuous dependence on the initial condition results in that a reset control system is not sensitive to noise. Since an easily checkable condition is wanted, the result is particularized for full-reset/right-reset compensation and the relaxed condition (49)-(50) (obviously a more relaxed and conservative condition is (51)).\\

{\bf Proposition IV.5}: {\em A reset control system $(A,C,n_{\rho})$, with a full reset or right reset compensator, and with initial condition $\mathbf{x}_0$, is not sensitive to noise if $\mathbf{x}_0 \in D = \mathds{R}^n \setminus (\mathcal{M}\cup \mathcal{B}(\bar{\mathcal{M}}_T))$ and $\mathcal{M}_R \cap \mathcal{B}(\bar{\mathcal{M}}_T)) = \varnothing$. } 

\vspace{0.125cm}
{\bf Proof}: Since the reset compensator is full reset or right reset, then by Prop. III.4 and Prop. IV.1 both the noise-free and the noisy reset control systems have well-posed crossing and reset instants. 
Let $(\bar{t}_k)$ and $(\bar{t}^n_k)$, $k = 1,2,\cdots$ be the crossing instants sequences corresponding to the noise-free and the noisy reset control systems, respectively. It turns out that for the initial conditions $\mathbf{x}_0$ and $(\mathbf{x}_0,\mathbf{0})$, $\bar{t}^n_k=\bar{t}_k$, for $k = 1,2,\cdots$. Now, from (49)-(50) it directly follows that $CA\mathbf{x}(\bar{t}_k,\mathbf{x}_0) \neq 0$, $k=1,2, \cdots$; and thus for the noisy system $(A_z,C_z,n_{\rho})$ (see Fig. 15) it results that 
\begin{equation} \small
C_zA_z \mathbf{z}(\bar{t}^n_k,(\mathbf{x}_0,\mathbf{0})) = 
\left(
\begin{array}{ccc}
C  & -C_n    
\end{array}
\right)
\left(
\begin{array}{ccc}
A  & \star\\
O & A_n    
\end{array}
\right)
\left(
\begin{array}{c}
\mathbf{x}(\bar{t_k},\mathbf{x}_0)\\
\mathbf{0}     
\end{array}
\right) \nonumber
\end{equation} 
\begin{equation}
= CA\mathbf{x}(\bar{t}_k,\mathbf{x}_0) \neq 0
\end{equation}
and then directly by Prop. IV.4 the solution of the noisy reset control system $(A_z,C_z,n_{\rho})$ depends continuously on the initial condition at  $\mathbf{z}_0 = (\mathbf{x}_0,\mathbf{0})$.  As a result, it is true that for any $\epsilon>0$ and almost any $T\in \mathds{R}^+$ there exist a $\delta >0$ such as  for any perturbed initial condition $(\mathbf{x}_0,\mathbf{n}_0) \in \mathds{R}^{n+m_n}$, with $\| \mathbf{n}_0\| < \delta$  it is satisfied that $d_H(\mathbf{z}_n([0,T],(\mathbf{x}_0,\mathbf{n})), \mathbf{z}([0,T],(\mathbf{x}_0,\mathbf{0}))) < \epsilon$ and the result directly follows. 
 $\Box$

\vspace{0.125cm}

{\em Example IV.7 }: Consider a reset control system $(A,C,1)$ as given by Fig. 15, Where $R$ is a P+CI compensator, a parallel connection of a proportional compensator and a CI (see \cite{libro} for a detailed definition), and the plant $P$ is an integrator. P+CI is given by matrices $A_r=0$, $B_r=1$, $C_r=K_{CI}$, $D_r=K_P$, $A_\rho=0$, where $K_P = 2$ and $K_{CI} = 1$ for this example; and, in addition, the exogenous signal $r$ is a step of height $w_{10}$ (no disturbance is considered in this example). In this case, $A$ and $C$ are given by
\begin{equation}
\begin{array}{cc}
A = \left(
\begin{array}{ccc}
 0 & 0  & 0  \\
 2 & -2  &1   \\
 1 & -1  &0   
\end{array}
\right)
  &
C =\left(
\begin{array}{ccc}
 1 & -1  &0   
\end{array}
\right)
\end{array}
\end{equation}

\noindent and the after-reset and reset sets are  $\mathcal{M}_R = span\{(1,1,0)\}$ and $\mathcal{M} = \mathcal{H}_C \setminus \mathcal{M}_R$, respectively, where $ \mathcal{H}_C$ is the hyperplane $ \mathcal{H}_C = span\{(1,1,0),(0,0,1) \}$. Note that P+CI is a full reset compensator, and in addition $\mathcal{M}_{RU} =\mathcal{M}_{R}= span\{(1,1,0)\}$, and
\begin{equation}
\bar{\mathcal{M}}_T= \mathcal{N} (\left( \begin{array}{ccc} 1 & -1 & 0 \\ -2 &2 &1 \end{array} \right)) \setminus span\{(1,1,0)\} =\varnothing
\end{equation}

\begin{figure}[t] 
   \centering
   \includegraphics[scale=0.6]{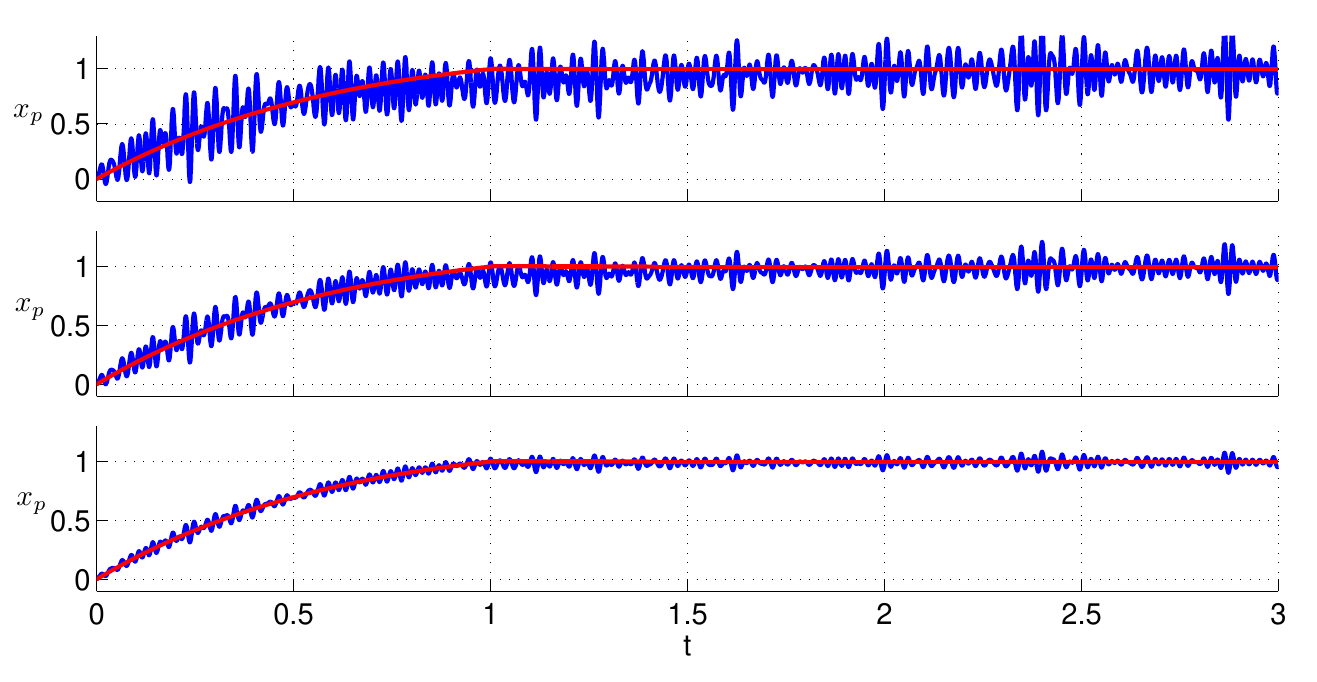} 
   \caption{Closed-loop output $(x_p)$: (blue) noisy case for different values of $\|\mathbf{n}_0\|$, (red) Noise-free case.}
   \label{fig:noiseexamplesignals}
\end{figure}
 
\begin{figure}[t] 
   \centering
   \includegraphics[scale=0.6]{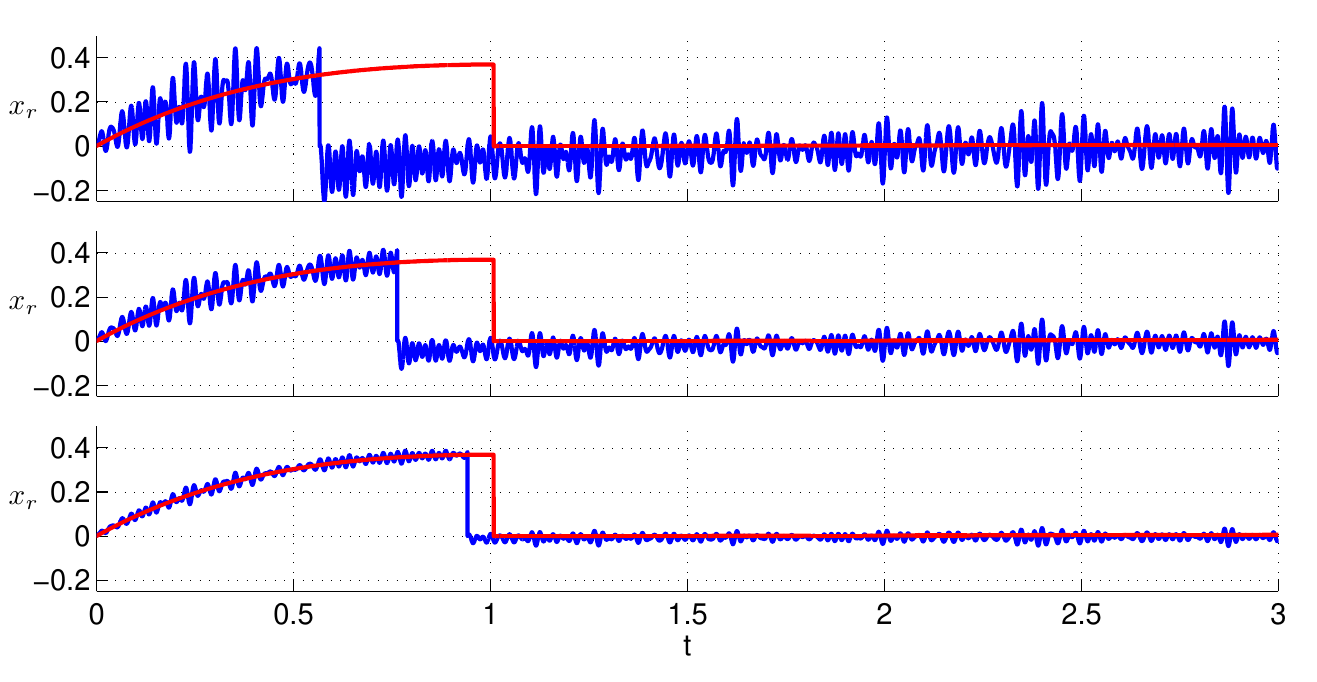} 
   \caption{CI state $(x_r)$: (blue) noisy case for different values of $\|\mathbf{n}_0\|$-there are many reset actions, with the first reset getting closer to $t=1$ as $\|\mathbf{n}_0\|$ is getting smaller, (red) Noise-free case-there is only a reset action at $t=1$.}
   \label{fig:noiseexamplesignalsControlador}
\end{figure}

Thus, by Prop. IV.5, $(A,C,1)$ is not sensitive to noise for any initial condition ${\bf x}_0 \in D =\mathds{R}^n \setminus \mathcal{M}$. Fig. \ref{fig:noiseexamplesignals}-\ref{fig:noiseexamplesignalsControlador} show a time simulation, including closed-loop output $y ( = x_p)$ and the reset compensator state $x_r$,  for a noise signal generated by an exosystem $\Sigma_n$ with different values of  $\mathbf{n_0}$: it is given by the sum of  $20$ sinusoidal signals with frequencies greater than $200$ rad/s. The reference is a unit step and the $CI$ and the plant are initially at rest, that is $\mathbf{x}_0 = (1, 0, 0) \in D$.  

  \begin{figure}[t] 
\centering
   \includegraphics[scale=0.6]{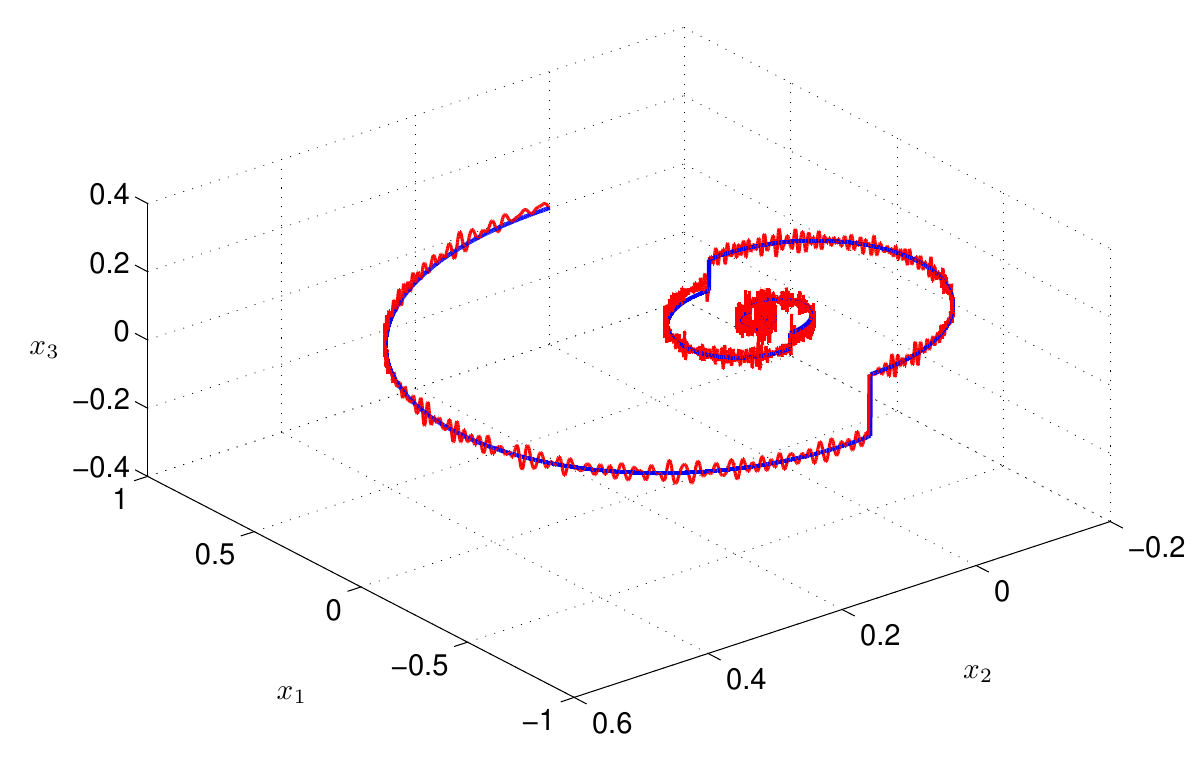} 
   \caption{Solutions of the reset control system-Ex. IV.8: (blue) noise-free case, (red) noisy case.} 
   \label{fig:noiseexamplesignals2}
\end{figure}

{\em Example IV.8 }: Consider the reset control system of Example IV.6, in which $\mathcal{M}_R \cap \mathcal{B}(\bar{\mathcal{M}}_T) =\varnothing$ (see Fig. 14). The after-reset and reset sets are  $\mathcal{M}_R = span\{(1,0,0)\}$ and $\mathcal{M} =  span\{(1,0,0), (0,0,1)\} \setminus \mathcal{M}_R$, respectively. Thus, Prop. IV.5 applies and this reset control system is not sensitive to noise for any initial condition ${\bf x}_0 \in D =\mathds{R}^n \setminus \mathcal{M} \cup \mathcal{B}(\bar{\mathcal{M}}_T)$, where the set $\mathcal{B}(\bar{\mathcal{M}}_T)$ can be bounded by two polytopes. 
Fig. \ref{fig:noiseexamplesignals2} shows the solution of the reset control system for $\mathbf{x}_0=(1,0,0) \in D$ and with a noise signal $n$ generated as in Example IV.7. Again, this simulation (jointly with many others) reflects the property of reset control systems to be not sensitive to noise according to Def. IV.2 and Prop. IV.5.

\section{Conclusions}
Well-posedness of reset control systems, that is existence and uniqueness of solutions and continuous dependence on the initial condition, has been investigated in an impulsive dynamical systems framework. Necessary and sufficient conditions for existence and uniquenesss of solutions, and a sufficient condition for continuous dependence on the initial condition have been obtained. It turns out that reset compensators that have been successfuly used in practice (full reset and right reset compensators) result in well-posed reset control systems, as far as the plant is strictly proper and exogenous signals are represented by Bohl functions. An immediate consequence is that time regulation is not needed for avoiding Zeno solutions (in fact there is no Zeno solutions), and that the reset control system is not sensitive to sensor noise once the continuous dependence property is satisfied. This work has been centered in reset control system with a zero-crossing resetting law. It is believed that the different concepts and methods that have been developed will provide a solid IDS framework to analyze several others resetting laws that has been found useful in practice.


\appendix

In the following, some technical results will be derived for the left reset control system of Section IV.C (LRC system in the following). In general, including single-input single-ouput systems, it is not true that for an arbitrary system $(A,B,C)$ the geometric multiplicity of unobservable modes is 1. However, for the LRC base system this is indeed the case. On the other hand, in this left compensation case, the subspace of after reset and unobservable states,  $\mathcal{M}_{RU}$, is $A$-invariant only in the simple case in which it consists of the zero state. In addition, the unobservable subspace shows a particular structure, it can be expressed through the root spaces of the unobservable modes.  

\vspace{0.125cm}

{\em Notation and Background}: $\mathcal{R}_{\lambda}\left(  A\right)$ represents the root space of $A$ associated to  $\lambda\in\sigma\left( A\right) $.  $\oplus$ stands for direct sum; $V$ is a A-cyclic subspace of $\mathds{R}^n$ generated by $\mathbf{v}$ if $V = span \{\mathbf{v}, A\mathbf{v}, \cdots \}$. Assume that $\lambda$ has algebraic multiplicity $ma_\lambda$ and geometric multiplicity 1, and let $\{\mathbf{v_1},\mathbf{v_2}, \cdots,\mathbf{v_{ma_\lambda}}\}$ be the set of generalized eigenvectors (including the eigenvector $\mathbf{v}_1$), then $\mathcal{R}_{\lambda}\left(  A\right) = \mathcal{N}(\lambda I - A)^{ma_\lambda} = span \{\mathbf{v_1},\mathbf{v_2}, \cdots,\mathbf{v_{ma_\lambda}} \}$; therefore, if the mode $\lambda\in\sigma\left( A\right)$ is observable then $\mathbf{v}_i \notin \mathcal{N}(\mathcal{O})$, for $i = 1, \cdots,n$ (note that $C\mathbf{v}_1 \neq 0$ by the PBH test $\Rightarrow \mathbf{v}_1 \notin \mathcal{N}(\mathcal{O})$; $CA\mathbf{v}_2 = \lambda C \mathbf{v}_2 + C\mathbf{v}_1  \Rightarrow  C\mathbf{v}_2 \neq 0$ or $CA\mathbf{v}_2 \neq 0 \Rightarrow \mathbf{v}_2 \notin \mathcal{N}(\mathcal{O})$, $\cdots$).

\vspace{0.125cm}

{\em Proposition A.1}: For the LRC system, the geometric multiplicity of
any $\lambda\in\sigma_{\bar{O}}\left( {A},{C}\right)$, as an eigenvalue of $A$, is $1$.
 
{\em Proof}:
By using the PBH test, for any $\lambda \in \sigma_{ \bar{ \mathcal{O} } }(A,C)$ it is true that 
\begin{equation}
\small
\left(
\begin{array}{ccc}
 \lambda I - \bar{A}_p & O &  -\bar{B}_p C_{r_2}\\
  B_{r_1} \bar{C}_p &  \lambda I - A_{r_{1}}   &  O \\
 O &-B_{r_2}C_{r_1}   &    \lambda I - A_{r_{2}} \\
 \bar{C}_p  & O & O
\end{array}
\right)     
\left(
\begin{array}{c}
\bar{\mathbf{v}}_p  \\ \mathbf{v}_1 \\ \mathbf{v}_2   
\end{array}
\right) =
\left(
\begin{array}{c}
\mathbf{0}  \\ \mathbf{0} \\ \mathbf{0} \\ 0   
\end{array}
\right)       
\label{PBHA1}
\end{equation}
\noindent for some nonzero state $(\bar{\mathbf{v}}_p, \mathbf{v}_1, \mathbf{v}_2)$ and thus
\begin{equation}
\begin{array}{l}
\left(
\begin{array}{cc}
 \lambda I - \bar{A}_p &  -\bar{B}_p \\
  \bar{C}_p  & O  
  \end{array}
\right) 
\left(
\begin{array}{c}
 \bar{\mathbf{v}}_p   \\
   C_{r_2}\mathbf{v}_2  
  \end{array}
\right) =
\left(
\begin{array}{c}
\mathbf{0} \\ 0   
\end{array}
\right)    \\   
(\lambda I - A_{r_{1}})\mathbf{v}_1 = \mathbf{0} \\

(\lambda I - A_{r_{2}})\mathbf{v}_2 = B_{r_2}C_{r_1}\mathbf{v}_1   

\end{array}\label{PBHA2}
\end{equation}

\noindent Now, $\lambda$ is a zero of $\bar{P}$ and an eigenvalue of $A_{r_1}$ and/or $A_{r_2}$; and in any case, the geometric multiplicity of the zero and the eigenvalues is 1, since the realizations are minimal. As a result, two cases are possible: i) $\lambda$ is not an eigenvalue of $A_{r_1}$ and is an eigenvalue of $A_{r_2}$, thus $\mathbf{v}_1  =  \mathbf{0}$ and $\mathbf{v}_2  \neq \mathbf{0}$ with $dim(\mathcal{N}(\lambda I - A_{r_{2}}))= 1$, ii) $\lambda$ is an eigenvalue of $A_{r_1}$, thus $\mathbf{v}_1  \neq \mathbf{0}$ with $dim(\mathcal{N}(\lambda I - A_{r_{1}}))= 1$, and  $\mathbf{v}_2 = \mathbf{0}$ or $\mathbf{v}_2  \neq \mathbf{0}$ with $dim(\mathcal{N}(\lambda I - A_{r_{2}}))= 1$. As a result, the geometric multiplicity of $\lambda$ as an eigenvalue of $A$ is 1. $\Box$ 

\vspace{0.125cm}


{\em Proposition A.2}:
For the LRC system, $\mathcal{M}_{RU}$ is $A-$invariant if and only if $\mathcal{M}_{RU}=\left\{\mathbf{0}\right\}  $.
 
{\em Proof}: ({\em if}) If $\mathcal{M}_{RU}=\left\{\mathbf{0}\right\}  $ then $\mathcal{M}_{RU}$ is trivially $A$-invariant.
 ({\em only if})  If $\mathcal{M}_{RU}$ is $A$-invariant then $E_{\lambda}\left(  A\right)  \cap\mathcal{M}_{RU}\neq \left\{  \mathbf{0} \right \}$ for some $\lambda \in \sigma_{ \bar{ \mathcal{O} } }(A,C)$, that is $\mathcal{M}_{RU}$ contains at least one eigenvector, or $\mathcal{M}_{RU} =  \left\{\mathbf{0}\right\}$. By using the PBH test for observability  \eqref{PBHA1}, for any eigenvector  $\mathbf{v} = (\bar{\mathbf{v}}_p,\mathbf{v}_{\bar{\rho}} ,\mathbf{v}_\rho)$ with $\mathbf{v}_{{\rho}} = \mathbf{0}$ it results that $(\bar{A}_p - \lambda I)\bar{\mathbf{v}}_p = \mathbf{0}$ and thus $\bar{\mathbf{v}}_p = \mathbf{0}$ since $\lambda$ is not an eigenvalue of $\bar{A}_p$ (otherwise, the realization $(\bar{A}_p,\bar{B}_p,\bar{C}_p)$ would not be minimal since $\lambda$ is a zero). In addition, from \eqref{PBHA2} it results that $(A_{r_1} - \lambda I)\mathbf{v}_{\bar{\rho}} = \mathbf{0}$ and $B_{r_2}C_{r_1}\mathbf{v}_{\bar{\rho}} = \mathbf{0}$. Now, since both $(A_{r_1},B_{r_1},C_{r_1})$ and $(A_{r_2},B_{r_2},C_{r_2})$ are minimal it is true that $B_{r_2} \neq \mathbf{0}$ and  $C_{r_1}\mathbf{v}_{\bar{\rho}} \neq {0}$ (scalar) for $\mathbf{v}_{\bar{\rho}} \neq \mathbf{0}$, then it must be true that $\mathbf{v}_{\bar{\rho}} = \mathbf{0}$ and thus $E_{\lambda}\left(  A\right)  \cap\mathcal{M}_{RU} = \left\{  \mathbf{0} \right \}$ 
for any $\lambda \in \sigma_{ \bar{ \mathcal{O} } }(A,C)$. As a result, $\mathcal{M}_{RU}$ does not contain any eigenvector and thus $\mathcal{M}_{RU} =  \left\{\mathbf{0}\right\}$. $\Box$
 
\vspace{0.125cm}

\color{black}{ 
 
{\em Proposition A.3}:  The unobservable subspace of the LRC system is given by
\begin{equation}
\mathcal{N}(\mathcal{O}) = \bigoplus\limits_{\lambda \in \sigma_{ \bar{ \mathcal{O} } }(A,C)} \mathcal{N}(\lambda I - A)^{d_\lambda}
\end{equation}

{\em Proof}: Consider some $\lambda\in\sigma_{\bar{O}}\left( {A},{C}\right)$. Firstly, since $\lambda$ is a zero of  $(\bar{A}_p,\bar{B}_p,\bar{C}_p)$ with algebraic multiplicity $m_\lambda \geq d_\lambda$ ($d_\lambda$ is the number of cancellations) then $\bar{C}_p(\lambda I -\bar{A}_p)^{-j}\bar{B}_p = 0$, for $j = 0, \cdots, d_{\lambda}$. In addition, $\lambda$ is an eigenvalue of $A $ with geometric multiplicity 1, and algebraic multiplicity $ma_\lambda \geq d_\lambda$.  Thus, it is possible to choose a subset  $\{\mathbf{v}_\lambda^{(d_\lambda - k)} = (\mathbf{v}_{p_\lambda}^{(d_\lambda - k)},\mathbf{v}_{\bar{\rho}_
\lambda}^{(d_\lambda - k)},\mathbf{v}_{\rho_\lambda}^{(d_\lambda - k))}), k = 0,\cdots,d_\lambda-1\}$ of its generalized eigenvalues (including the eigenvector corresponding to $k = 0$), that it is a basis of $\mathcal{N}(\lambda I - A)^{d_\lambda}$.  By using again the PBH test, it is clear from (42)-(43) that  $\mathbf{v}_{p_\lambda}^{(d_\lambda)} =   (\lambda I - \bar{A}_p)^{-1}\bar{B}_pC_{r_2}\mathbf{v}_{{\rho}_\lambda}^{(d_\lambda)} $, and in general $\mathbf{v}_{p_\lambda}^{(d_\lambda - k)} =  \sum_{j=1}^{k+1} (\lambda I - \bar{A}_p)^{-j}\bar{B}_pC_{r_2}\mathbf{v}_{{\rho}_\lambda}^{(d_\lambda+j-(k+1))}$ for $k = 0,\cdots,d_\lambda-1$. Note that the inverse is well-defined in all cases since $(\bar{A}_p,\bar{B}_p,\bar{C}_p)$ is observable and $\lambda$ is a zero of $(\bar{A}_p,\bar{B}_p,\bar{C}_p)$ with multiplicity at least $d_\lambda$. For the eigenvector $\mathbf{v}_\lambda^{(d_\lambda)} = (\mathbf{v}_{p_\lambda}^{(d_\lambda)},\mathbf{v}_{\bar{\rho}_\lambda}^{(d_\lambda)}, \mathbf{v}_{\rho_\lambda}^{(d_\lambda))})$ it has been shown that the scalar $C_{r_2}\mathbf{v}_{\rho_\lambda}$ is not equal to zero, 

As a result, it is true that $C\mathbf{v}_\lambda^{(d_\lambda - k)} =  \sum_{j=1}^{k+1} \bar{C}_p(\lambda I - \bar{A}_p)^{-j}\bar{B}_pC_{r_2}\mathbf{v}_{{\rho}_\lambda}^{(d_\lambda+j-(k+1))} = 0$, for any $k = 0,\cdots,d_\lambda-1$. Thus, since by construction the subspace $\mathcal{N}(\lambda I - A)^{d_\lambda}$ is $A$-invariant it follows that $CA^j\mathbf{v}_\lambda^{(d_\lambda - k)} = 0$ for $j = 0,1,\cdots$, and $k = 0,1,\cdots, d_\lambda-1$; in other words, $\mathcal{N}(\lambda I - A)^{d_\lambda} \subset \mathcal{N}(\mathcal{O})$.

In addition, since the algebraic multiplicity of $\lambda$, as eigenvalue of $A$, is greater than $d_\lambda$ only when $m_\lambda = d_\lambda$ then it follows that in this case  $\bar{C}_p(\lambda I -\bar{A}_p)^{-j}\bar{B}_p \neq 0$, for $j > d_\lambda$; and thus it is not difficult to see that  $\mathcal{N}(\lambda I - A)^{d_\lambda + n} \cap \mathcal{N}(\mathcal{O}) = \mathcal{N}(\lambda I - A)^{d_\lambda}$, for any integer $n > 0$; and, in particular $
\mathcal{R}_\lambda(A) \cap \mathcal{N}(\mathcal{O}) = \mathcal{N}(\lambda I - A)^{d_\lambda}$. 

Now, consider an observable mode $\lambda\in\sigma_{{O}}\left( {A},{C}\right)$ with algebraic multiplicity $ma_\lambda$. 
Its root space  $\mathcal{R}_\lambda(A)$ is spanned by the set of generalized eigenvectors $\{\mathbf{v}_\lambda^{(d_\lambda - k)} = (\mathbf{v}_{p_
\lambda}^{(d_\lambda - k)},\mathbf{v}_{\bar{\rho}_\lambda}^{(d_\lambda - k)},\mathbf{v}_{\rho_\lambda}^{(d_\lambda - k))}), k = 0,\cdots,ma_\lambda-1\}$, and it is true that $\mathbf{v}_\lambda^{(d_\lambda - k)}  \notin \mathcal{N}(\mathcal{O}) $
, for any $k = 0,\cdots,ma_\lambda-1$. As a result, since no generalized eigenvector is an element of the unobservable subspace, then it follows that  no $A$-cyclic subspace must be in the unobservable subspace, and thus $\mathcal{R}_\lambda(A) \cap \mathcal{N}(\mathcal{O}) = \{ \mathbf{0}\}$ for observable modes. 

Finally, since $\mathcal{N}(\mathcal{O})$ is $A$-invariant, it may be obtained as the direct sum of its intersection with the the root spaces, and thus (43) directly follows: 

\begin{equation}
\mathcal{N}(\mathcal{O}) = \bigoplus\limits_{\lambda \in \sigma(A)} \mathcal{R}_\lambda(A) \cap \mathcal{N}(\mathcal{O}) = 
\bigoplus\limits_{\lambda \in \sigma_{ \bar{ \mathcal{O} } }(A,C)} \mathcal{N}(\lambda I - A)^{d_\lambda}
\nonumber
\end{equation}

$\Box$

\end{document}